\documentclass{aa}

\usepackage[colorlinks,linkcolor=blue,citecolor=blue,urlcolor=blue,bookmarks=false,hypertexnames=true]{hyperref}
\usepackage{txfonts}
\usepackage{xcolor}
\usepackage{amsmath}
\usepackage{soul}
\usepackage{float}
\usepackage{subcaption}
\usepackage[version=4]{mhchem}
\usepackage[labelformat=parens,labelsep=quad,skip=3pt]{caption}
\usepackage{graphicx}
\usepackage[toc,page]{appendix}
\definecolor{yellowhl}{rgb}{0,1,1}
\definecolor{greyhl}{rgb}{0.8,0.8,0.8}
\definecolor{redhl}{rgb}{1,0.5,0.5}
\definecolor{bluehl}{rgb}{0.75,0.75,1}
\definecolor{greenhl}{rgb}{0.5,1.0,0.5}
\definecolor{purplehl}{rgb}{0.75,0.5,0.9}
\definecolor{orangehl}{rgb}{1,0.4,0.3}
\definecolor{pinkhl}{rgb}{0.9,0.4,0.9}

\begin{document}

   \title{The prebiotic molecular inventory of Serpens SMM1}

   \subtitle{I. An investigation of the isomers CH$_{3}$NCO and HOCH$_{2}$CN}

   \author{N.F.W. Ligterink
          \inst{1}
          \and
          A. Ahmadi
          \inst{2} 
          \and
          A. Coutens
          \inst{3}    
          \and
          \L{}. Tychoniec
          \inst{2}  
          H. Calcutt
          \inst{4,5}
          \and
          E.F. van Dishoeck 
          \inst{2,6}
          \and
          H. Linnartz
          \inst{7}
          \and
          J.K. J{\o}rgensen
          \inst{8}
          \and
          R.T. Garrod
          \inst{9}
          \and
          J. Bouwman
          \inst{7}
          }

   \institute{Physics Institute, University of Bern, Sidlerstrasse 5, 3012 Bern, Switzerland \\
              \email{niels.ligterink@csh.unibe.ch}
         \and
             Leiden Observatory, Leiden University, PO Box 9513, 2300 RA Leiden, The Netherlands  
         \and         
		Laboratoire d'Astrophysique de Bordeaux, Univ. Bordeaux, CNRS, B18N, all\'{e}e Geoffroy Saint-Hilaire, 33615 Pessac, France
         \and
         	Department of Space, Earth and Environment, Chalmers University of Technology, 41296, Gothenburg, Sweden
         \and
         	Institute of Astronomy, Faculty of Physics, Astronomy and Informatics, Nicolaus Copernicus University, Grudziadzka 5, 87-100 Torun, Poland
         \and
         	 Max-Planck Institut f\"{u}r Extraterrestrische Physik (MPE), Giessenbachstr. 1, 85748 Garching, Germany
         \and
             Laboratory for Astrophysics, Leiden Observatory, Leiden University, PO Box 9513, 2300 RA Leiden, The Netherlands
		 \and
         Centre for Star and Planet Formation, Niels Bohr Institute \& Natural History Museum of Denmark, University of Copenhagen, {\O}ster Voldgade 5--7, 1350 Copenhagen K., Denmark         
         \and
             Departments of Chemistry and Astronomy, University of Virginia, Charlottesville, VA 22904, USA              
             }

   \date{Received October 8, 2020; accepted December 22, 2020}

 
  \abstract
   {}
   {Methyl isocyanate (CH$_{3}$NCO) and glycolonitrile (HOCH$_{2}$CN) are isomers and prebiotic molecules that are involved in the formation of peptide structures and the nucleobase adenine, respectively. These two species are investigated to study the interstellar chemistry of cyanides (CN) and isocyanates (NCO) and to gain insight into the reservoir of interstellar prebiotic molecules.}
   {ALMA observations of the intermediate-mass Class 0 protostar Serpens SMM1-a and ALMA-PILS data of the low-mass Class 0 protostar IRAS~16293B are used. Spectra are analysed with the CASSIS line analysis software package in order to identify and characterise molecules.}
   {CH$_{3}$NCO, HOCH$_{2}$CN, and various other molecules are detected towards SMM1-a. HOCH$_{2}$CN is identified in the PILS data towards IRAS~16293B in a spectrum extracted at a half-beam offset position from the peak continuum. CH$_{3}$NCO and HOCH$_{2}$CN are equally abundant in SMM1-a at [X]/[CH$_{3}$OH] of 5.3$\times$10$^{-4}$ and 6.2$\times$10$^{-4}$, respectively. A comparison between SMM1-a and IRAS~16293B shows that HOCH$_{2}$CN and HNCO are more abundant in the former source, but CH$_{3}$NCO abundances do not differ significantly. Data from other sources are used to show that the [CH$_{3}$NCO]/[HNCO] ratio is similar in all these sources within $\sim$10\%.}
   {The new detections of CH$_{3}$NCO and HOCH$_{2}$CN are additional evidence for a large interstellar reservoir of prebiotic molecules that can contribute to the formation of biomolecules on terrestrial planets. The equal abundances of these molecules in SMM1-a indicate that their formation is driven by kinetic processes instead of thermodynamic equilibrium, which would drive the chemistry to one product. HOCH$_{2}$CN is found to be much more abundant in SMM1-a than in IRAS~16293B. From the observational data, it is difficult to indicate a formation pathway for HOCH$_{2}$CN, but the thermal Strecker-like reaction of CN$^{-}$ with H$_{2}$CO is a possibility. The similar [CH$_{3}$NCO]/[HNCO] ratios found in the available sample of studied interstellar sources indicate that these two species either are chemically related or their formation is affected by physical conditions in the same way. Both species likely form early during star-formation, presumably via ice mantle reactions taking place in the dark cloud or when ice mantles are being heated in the hot core. The relatively high abundances of HOCH$_{2}$CN and HNCO in SMM1-a may be explained by a prolonged stage of relatively warm ice mantles, where thermal and energetic processing of HCN in the ice results in the efficient formation of both species.}


   \maketitle
%

\section{Introduction}

Observations of molecules towards star-forming regions give insight into the kind of species that end up in planet-forming discs. These molecules not only aid planet formation but can also seed newly formed planets with a cocktail of molecules from which larger organic molecules can be formed. Prebiotic molecules are of particular interest, as they are involved in the formation of biomolecules, such as amino acids, nucleobases, proteins, and lipids \citep[][]{sandford2020}. In the interstellar medium (ISM) and on planets, prebiotic molecules are the building blocks from which biomolecules are made.

\begin{figure}
\centering
   \includegraphics[width=0.9\hsize]{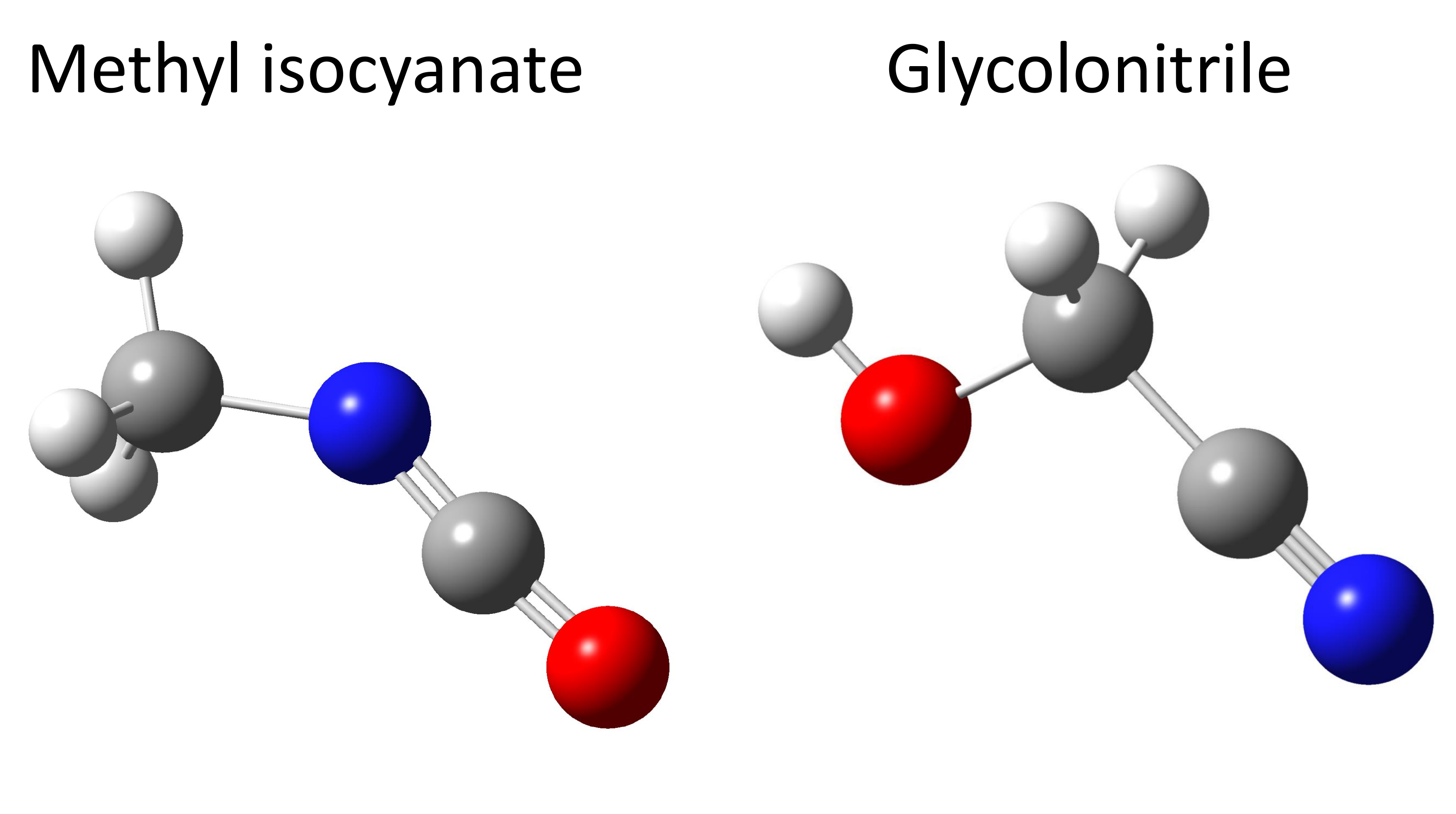}
\caption{Structures of the C$_{2}$H$_{3}$NO isomers methyl isocyanate (CH$_{3}$NCO, left) and glycolonitrile (HOCH$_{2}$CN, right).}
\label{fig:c2h3no}
\end{figure}

In the ISM, several prebiotic molecules have been detected. Examples are formamide \citep[NH$_{2}$CHO,][]{rubin1971} a precursor to nucleobases and amino acids \citep{saladino2012}, the simplest sugar-like molecule glycolaldehyde \citep[HOCH$_{2}$CHO,][]{hollis2000,jorgensen2012}, methylamine \citep[CH$_{3}$NH$_{2}$,][]{kaifu1974,bogelund2019a} and aminoacetonitrile \citep[NH$_{2}$CH$_{2}$CN,][]{belloche2008}, building blocks of the amino acid glycine \citep{holtom2005,lee2009}, the peptide building blocks acetamide (CH$_{3}$C(O)NH$_{2}$) and N-methylformamide \citep[CH$_{3}$NHCHO,][]{hollis2006,halfen2011,belloche2017,belloche2019,ligterink2020}, the chiral molecule propylene oxide \citep[CH$_{3}$CHCH$_{2}$O,][]{mcguire2016}, cyanomethanimine (NHCHCN), which can oligomerise to form adenine \citep{zaleski2013,rivilla2019}, and the nucleobase precursors cyanamide \citep[NH$_{2}$CN,][]{turner1975,coutens2018}, hydroxylamine \citep[NH$_{2}$OH,][]{rivilla2020} and carbamide \citep[also known as urea, NH$_{2}$C(O)NH$_{2}$,][]{belloche2019}. Over the past years, methyl isocyanate \citep[CH$_{3}$NCO,][]{halfen2015,cernicharo2016,ligterink2017} has been detected in several interstellar sources and recently its isomer glycolonitrile \citep[also known as hydroxy acetonitrile, HOCH$_{2}$CN,][]{zeng2019} was identified for the first time in the ISM towards the low-mass protostar IRAS~16293--2422B (hereafter IRAS~16293B). Both these isomers are prebiotic molecules. CH$_{3}$NCO can engage in reactions that form peptide-like structures, while HOCH$_{2}$CN is known to accelerate the oligomerisation of HCN in liquids and ice under terrestrial conditions, forming the nucleobase adenine \citep{schwartz1982a,schwartz1982b}. 

Besides their relevance to prebiotic chemistry, CH$_{3}$NCO and HOCH$_{2}$CN are also interesting molecules to gain insight into interstellar nitrogen chemistry. While these molecules, being isomers, have the same elemental composition (C$_{2}$H$_{3}$NO), their chemical structures differ significantly, see Fig. \ref{fig:c2h3no}. Recent quantum chemical calculations of the stability of C$_{2}$H$_{3}$NO isomers in general, also reveal that CH$_{3}$NCO is the most stable species of those, followed by HOCH$_{2}$CN \citep{fourre2020}. Therefore, observations of this isomer couple provide information on interstellar reactions involving cyanides (-CN) and isocyanates (-NCO), two important nitrogen-bearing chemical groups, and the physical conditions that steer or prohibit this chemistry. 

After the first detection of CH$_{3}$NCO \citep{halfen2015,cernicharo2016}, HNCO was suggested to be involved in its formation due to their structural similarity and the large HNCO abundances in interstellar gas and ice \citep[in the form of OCN$^{-}$][]{boogert2015}. Various gas-phase and solid-state methylation (the addition of a CH$_{3}$ functional group to a molecule) reactions of HNCO, the OCN radical, and the OCN$^{-}$ anion have been proposed as possible reaction pathways. Experimental investigations \citep{ligterink2017,mate2018} and modeling studies \citep{martin-domenech2017,quenard2018,majumdar2018} indicate solid-state methylation in the ice mantle as the main pathway to form CH$_{3}$NCO:
\begin{equation}
    \ce{CH3 + NCO -> CH3NCO.}
    \label{eq:nco_ch3}
\end{equation}
Variations of this pathway, such as the methylation of HNCO or OCN$^{-}$ may be possible as well, while completely different reactions, such as the hydrogenation of HCN$\ldots$CO may form CH$_{3}$NCO as well.

No gas-phase formation pathways are known for glycolonitrile, but solid-state production routes have been studied theoretically \citep{woon2001a} and experimentally \citep{danger2012,danger2014}. Laboratory work indicates that the thermally activated reaction between a cyanide anion (CN$^{-}$) and formaldehyde (H$_{2}$CO) forms HOCH$_{2}$CN:  
\begin{equation}
    \ce{[XH+CN^-] + H2CO -> HOCH2CN + X,}
    \label{eq:cn-h2co}
\end{equation}
where X is a molecule that can act as a base, such as ammonia (NH$_{3}$) or water (H$_{2}$O). This reaction is the solid-state equivalent of the Strecker synthesis, which is a sequence of chemical reactions that produce amino acids. The Strecker-like formation of HOCH$_{2}$CN is linked to the formation of aminomethanol \citep[HOCH$_{2}$NH$_{2}$,][]{bossa2009} and aminoacetonitrile \citep[NH$_{2}$CH$_{2}$CN,][]{danger2011}. The latter of these species is detected in the ISM and known as a possible intermediate in the formation of the amino acid glycine \citep{belloche2008}. Irradiation of HOCH$_{2}$CN results in the photoproducts formylcyanide (HC(O)CN) and ketenimine (CH$_{2}$CNH). Although not investigated, hydrogenation and oxygen additions of these two species may provide pathways to form HOCH$_{2}$CN in the ISM. The solid-state radical-radical reactions HO + CH$_{2}$CN and HOCH$_{2}$ + CN can also form glycolonitrile, but neither of these reactions has been investigated \citep{margules2017}. However, precursor species to these reactions can be present in ice mantles, in particular when methanol (CH$_{3}$OH) or acetonitrile (CH$_{3}$CN) are energetically processed \citep{allamandola1988,hudson2004,bulak2020}.

Methyl isocyanate and glycolonitrile can thus be used as tracers of reactions involving CN and NCO and investigating their interstellar abundances reveals information about the chemical and physical processes that drive these reactions and interstellar nitrogen chemistry as a whole. CH$_{3}$NCO and HOCH$_{2}$CN have only been detected simultaneously in the low-mass protostar IRAS~16293B \citep[][]{ligterink2017,martin-domenech2017,zeng2019}, albeit in different observational data sets. Due to this limited sample size, it is difficult to derive correlations or variations in the C$_{2}$H$_{3}$NO isomer chemistry and therefore simultaneous identifications in other sources are required. Here, deep ALMA observations of the intermediate-mass Class 0 protostar Serpens SMM1-a (hereafter SMM1-a) are presented to derive additional constraints on CH$_{3}$NCO and HOCH$_{2}$CN chemistry.

The Serpens star-forming region contains multiple deeply embedded sources, of which SMM1 is the brightest \citep{casali1993}. The Serpens region contains multiple outflows and jets, some of which originate from SMM1 \citep{dionatos2013,hull2016}. The chemistry of the SMM1 hot corino, its outflows, and the Serpens core have been characterised in various studies \citep[e.g.,][]{white1995,hogerheijde1999,kristensen2010,oberg2011,goicoechea2012,tychoniec2019}. High-resolution continuum jet observations have shown that SMM1 consists of multiple sources, of which SMM1-a is the main one \citep{choi2009,dionatos2014,hull2017}. SMM1-a has SMM1-b as a close neighbor at $\sim$500 au, while two other sources, SMM1-c and -d, are located further away to its north. Recent distance measurements place the Serpens core, and therefore SMM1, at a distance of 436.0$\pm$10 pc \citep[][]{ortiz-leon2017}, resulting in a luminosity estimate of the entire SMM1 source of $\sim$100 $L_{\odot}$. SMM1-a is considered to be an intermediate-mass protostar \citep{hull2017,tychoniec2019}.

In this work, the detection and analysis of the isomers HOCH$_{2}$CN and CH$_{3}$NCO towards SMM1-a are presented and compared with literature results of IRAS~16293B and those of other sources. In section \ref{sec:methods} the observations towards SMM1 and the analysis method are presented. The detections of HOCH$_{2}$CN, CH$_{3}$NCO, and various other molecules are presented in section \ref{sec:results}. Section \ref{sec:discussion} discusses these detections and their likely formation pathways. The conclusions of this work are presented in section \ref{sec:conclusion}.

\begin{figure*}[h]
\centering
   \includegraphics[width=0.49\hsize]{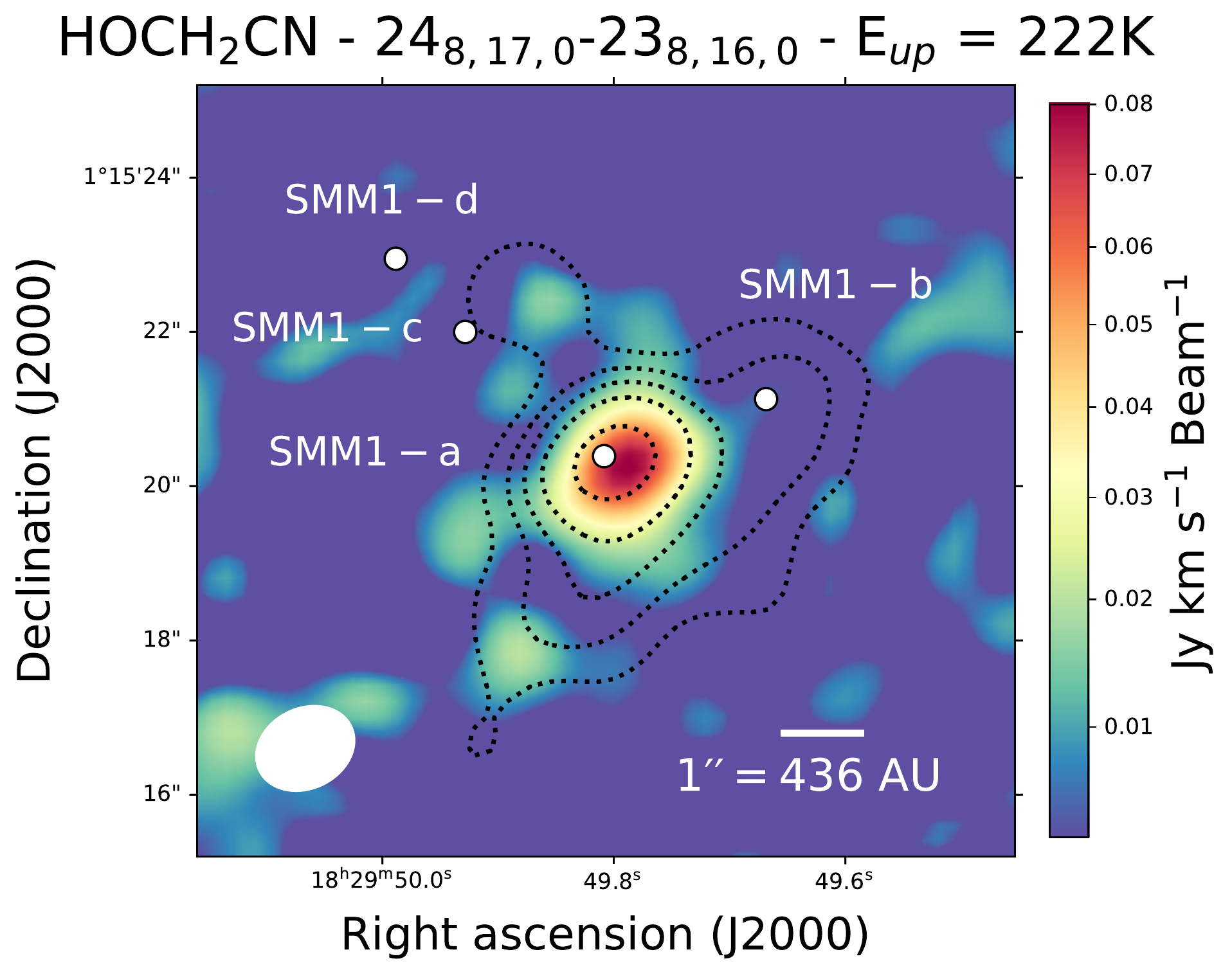}
   \includegraphics[width=0.49\hsize]{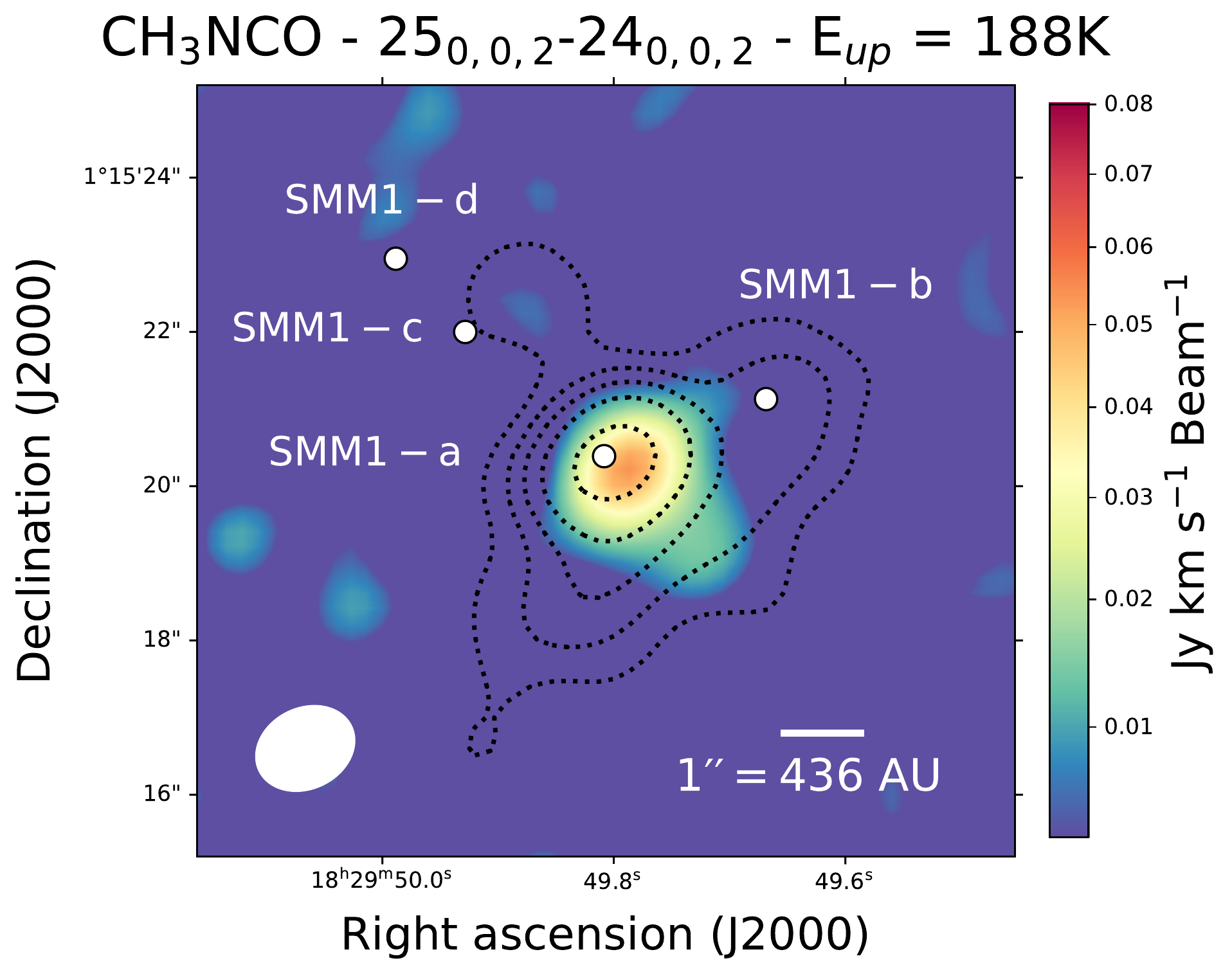} 
\caption{Moment 0 maps of the HOCH$_{2}$CN 24$_{8,17/16}$--23$_{8,16/15}$, E$_{\rm up}$ = 222~K line and the CH$_{3}$NCO $\nu$=0 25$_{0,0}$--24$_{0,0}$, E$_{\rm up}$ = 188~K line towards SMM1. Both lines are integrated over 8 velocity bins, centered on the peak frequency of each line as determined towards SMM1-a. Positions of protostars in the SMM1 region are indicated and the beam size (1$\farcs$32$\times$1$\farcs$04) is visualised in the bottom left corner. Dust continuum contours are given by the black dotted line at the levels of 0.02, 0.05, 0.1, 0.2, 0.5 Jy Beam$^{-1}$.}
\label{fig:mom0_smm1}
\end{figure*}

\begin{figure}[h]
\centering
   \includegraphics[width=1.0\hsize]{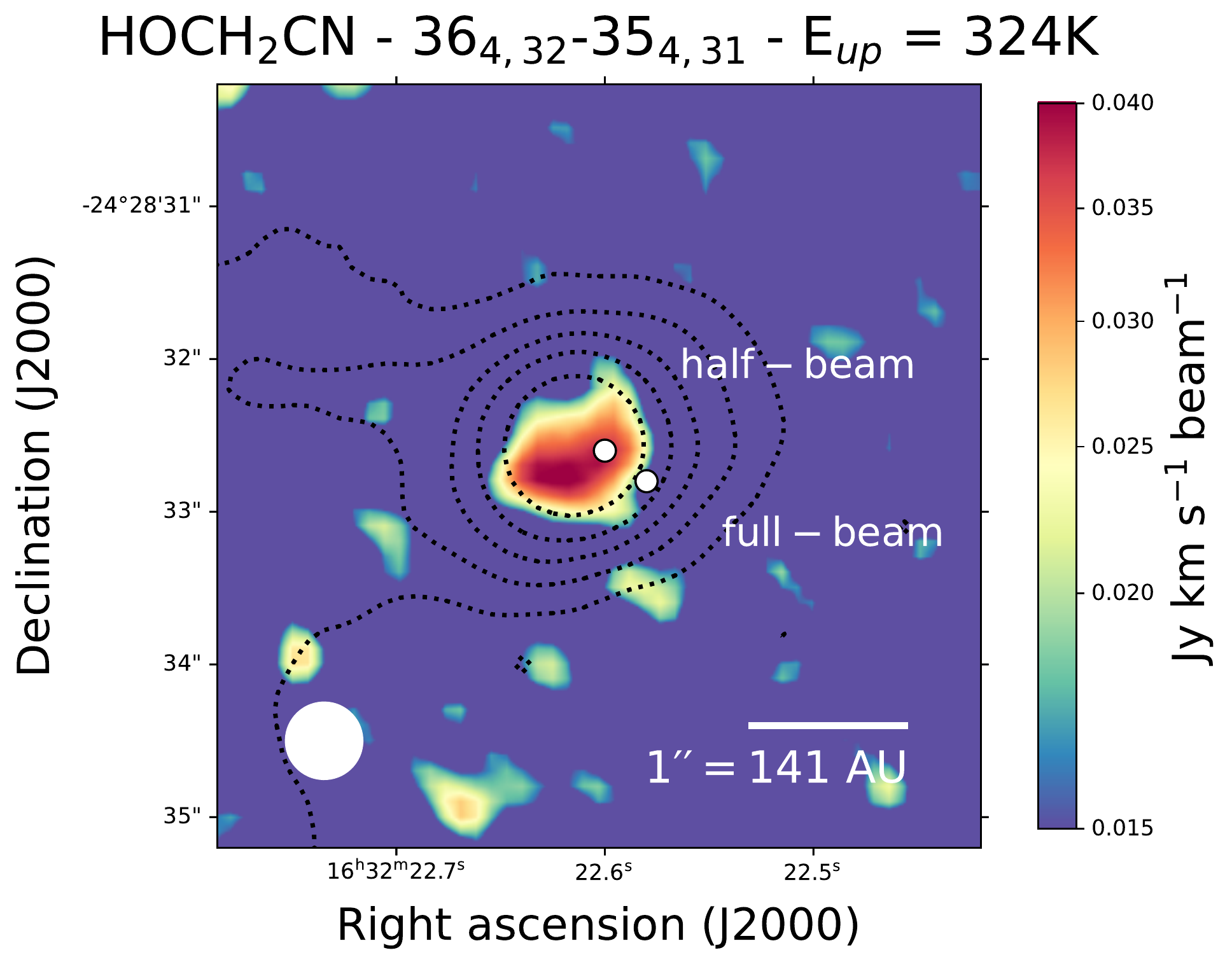}
\caption{Moment 0 map of the HOCH$_{2}$CN 36$_{4,32}$--35$_{4,31}$, E$_{\rm up}$ = 324~K line towards IRAS~16293B. The line is integrated over 8 velocity bins, centered on the peak frequency of the line. The positions of the half-beam and full-beam offset positions around IRAS~16293B are indicated and the beam size (0$\farcs$5$\times$0$\farcs$5) is visualised in the bottom left corner. Dust continuum contours are given by the black dotted line at the levels of 0.02, 0.05, 0.1, 0.2, 0.5 Jy Beam$^{-1}$.}
\label{fig:mom0_16293B}
\end{figure}

\section{Data \& Methods}
\label{sec:methods}

\subsection{Observations and spectra of Serpens SMM1}

SMM1 was observed on 27-March-2019 during ALMA cycle 6, as part of
project \#2018.1.00836.S (PI: N.F.W. Ligterink). The region was observed
using a total of 42 antennae with baselines spanning 15 -- 332 meters in
configuration C43-5. The on-source integration time was 50 minutes,
towards the phase centre $\alpha_{\rm J2000}$ = 18:29:49.80 $\delta_{\rm
J2000}$ = +01:15:20.6. Spectra were recorded in select frequency windows
between 217.59 and 235.93 GHz, at resolutions of 488.21 kHz (0.33 km~s$^{-2}$) and 1952.84 kHz (1.25 km~s$^{-2}$) for the continuum window, see Table \ref{tab:obs_freq}. The data were
calibrated and imaged with version 5.4.0-70 of the Common Astronomy
Software Applications (CASA). Bandpass and flux calibration was
conducted on J2000--1748, while phase calibration was performed on
J1851+0035. The flux uncertainty was $\leq$20\%. To reach the desired
sensitivity, the measurement sets were cleaned using the Hogbom
algorithm \citep{hogbom1974} and Briggs weighting with a robust
parameter of 0.5. This resulted in an angular resolution of
1$\farcs$32$\times$1$\farcs$04 and an rms noise of 2.6
mJy\,beam$^{-1}$\,km\,s$^{-1}$ in the final spectral data cubes. The primary beam of the observations was 26$\arcsec$.

\begin{table}[h]
\caption{Frequency settings of ALMA SMM1 observations}             
\label{tab:obs_freq}      
\centering          
\begin{tabular}{c c c c}     
\hline\hline                           
Frequency range & Bandwidth & \multicolumn{2}{c}{Resolution} \\
(GHz) & (GHz) & (kHz) & (km~s$^{-1}$) \\ 
\hline                    
217.59 -- 217.70 & 0.117 & 488.21 & 0.33 \\	
217.97 -- 218.09 & 0.117 & 488.21 & 0.33 \\	
218.43 -- 218.55 & 0.117 & 488.21 & 0.33 \\	
218.92 -- 219.03 & 0.117 & 488.21 & 0.33 \\	
219.71 -- 219.82 & 0.117 & 488.21 & 0.33 \\	
221.30 -- 221.42 & 0.117 & 488.21 & 0.33 \\	 
221.42 -- 221.53 & 0.117 & 488.21 & 0.33 \\	
221.53 -- 221.65 & 0.117 & 488.21 & 0.33 \\	
231.74 -- 231.97 & 0.234 & 488.21 & 0.33 \\	
233.41 -- 233.65 & 0.234 & 488.21 & 0.33 \\	
234.06 -- 235.93 & 1.875 & 1952.84 & 1.25 \\
\hline                  
\end{tabular}
\end{table}

Due to the line-richness of the source, the following procedure was
followed to properly subtract the continuum from the line observations.
We imaged all spectral windows without the continuum removed and used
the corrected sigma-clipping method of the STATCONT
package\footnote{\url{https://hera.ph1.uni-koeln.de/~sanchez/statcont}} 
\citep{sanchez-monge2018} to extract a continuum-subtracted line cube. STATCONT can only subtract zeroth-order polynomials, while in this dataset non-zeroth-order baselines are visible. Furthermore, continuum subtraction in the uv-plane is more desirable since the deconvolution of the line emission is more robust when it is not subjected to the deconvolution errors of the brighter continuum. Therefore, the STATCONT outputs were used to identify the line-free channels in the spectra and the continuum was subtracted in the uv-plane with the \textit{uvcontsub} task in CASA. Line-free channels are sparse, but for most spectral windows, at least 20\% of the bandwidth was given as input to the \textit{uvcontsub} task, with the exception of two spectral windows, where only 10\% of the bandwidth was line-free. From the
resulting datacube, the SMM1-a hot core spectrum was extracted towards the
peak continuum position $\alpha_{\rm J2000}$ = 18:29:49.793,
$\delta_{\rm J2000}$ = +1.15.20.200. From the average continuum flux density
(0.41 mJy beam$^{-1}$), the background temperature was determined to be
$\sim$5.2~K.

\subsection{PILS observations and spectra of IRAS~16293B}

In this work, the column density of HOCH$_{2}$CN is determined independently from the detection presented by \citep{zeng2019} by analysing data from the Protostellar Interferometric Line Survey (PILS). Other species relevant to this work are also searched for in the PILS data set. The observational details of the PILS survey have been presented in various other publications \citep[e.g.,][]{jorgensen2016} and here only the most relevant information is presented. In short, the PILS survey makes use of ALMA band 7 observations, covering a frequency range from 329 to 363 GHz at a spatial resolution of 0$\farcs$5. To investigate the chemical inventory of IRAS~16293B, spectra are extracted at several positions. These positions are on the peak continuum, a half-beam offset from the peak continuum, and a full-beam offset from the peak continuum. Most PILS analyses of IRAS~16293B make use of the spectrum at the full-beam offset position \citet[e.g.,][]{jorgensen2016,coutens2016,ligterink2017,coutens2018,persson2018,calcutt2018,jorgensen2018}. In this work, this is the main position for which molecular ratios with HOCH$_{2}$CN are determined, but the spectra of other positions are also analysed. The systemic velocity towards these positions is $V_{\rm LSR}$ = 2.7~km~s$^{-1}$ and the line width is approximately $\Delta V$ = 1.0~km~s$^{-1}$. Due to the dense dust around IRAS~16293B, the background temperatures ($T_{\rm BG}$) at these positions are higher than the cosmic microwave background radiation temperature of 2.7~K and couple with the molecular line emission \citep[see][]{ligterink2018b}. At the full-beam offset position $T_{\rm BG}$ = 21~K, while at the half-beam offset position it is $T_{\rm BG}$ = 52~K.

\begin{figure*}[h]
\includegraphics[width=\hsize]{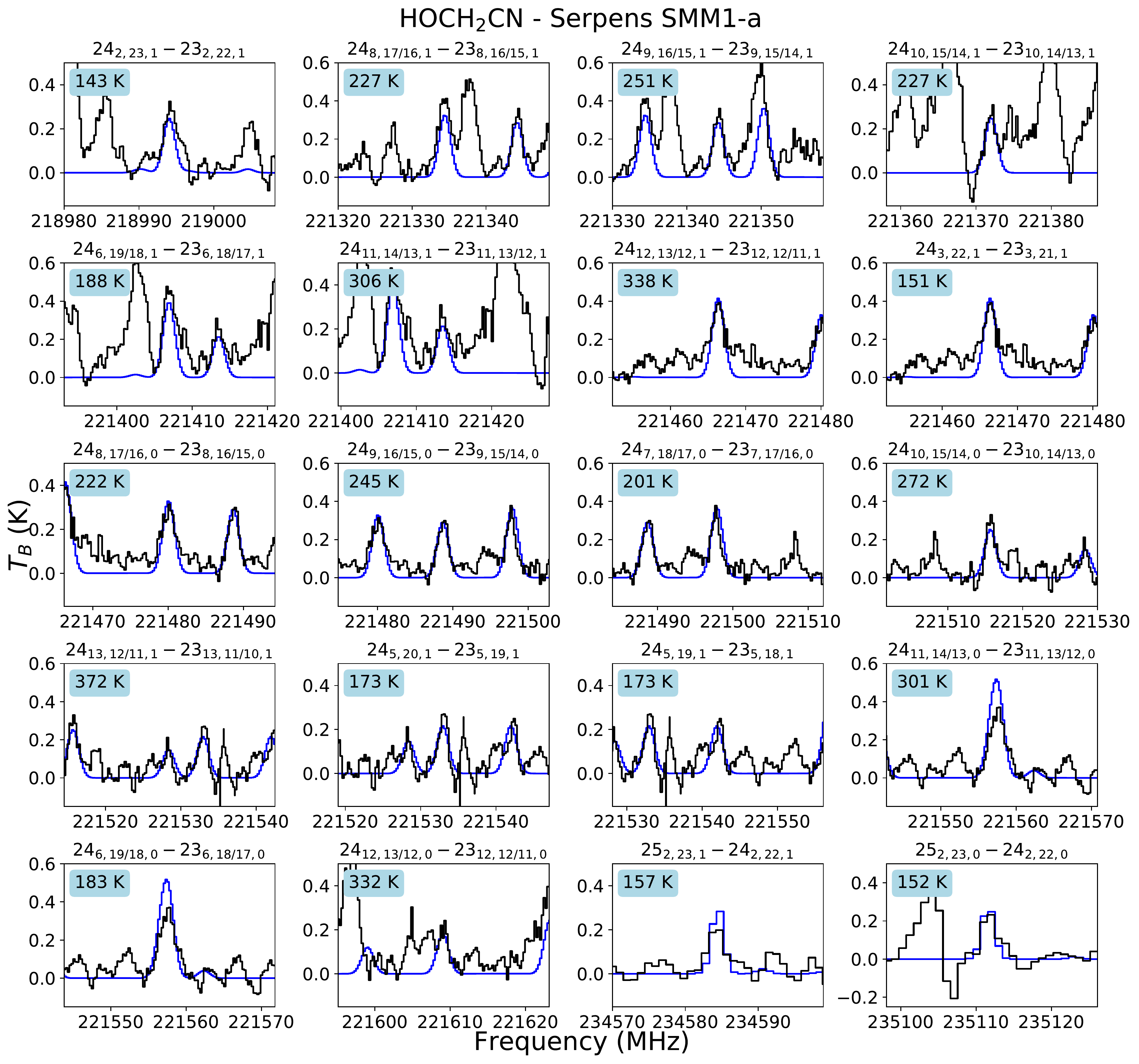}
\caption{Identified lines of HOCH$_{2}$CN towards SMM1. The observed spectrum is plotted in black, with the best-fit synthetic spectrum overplotted in blue ($N_{\rm T}$ = (7.4$\pm$0.9)$\times$10$^{14}$ cm$^{-2}$, $T_{\rm ex}$ = 260$\pm$45~K). The transition is indicated at the top of each panel and the upper state energy is given in the top left of each panel.}
\label{fig:lines_HOCH2CN}
\end{figure*}

\begin{figure*}[h]
\includegraphics[width=\hsize]{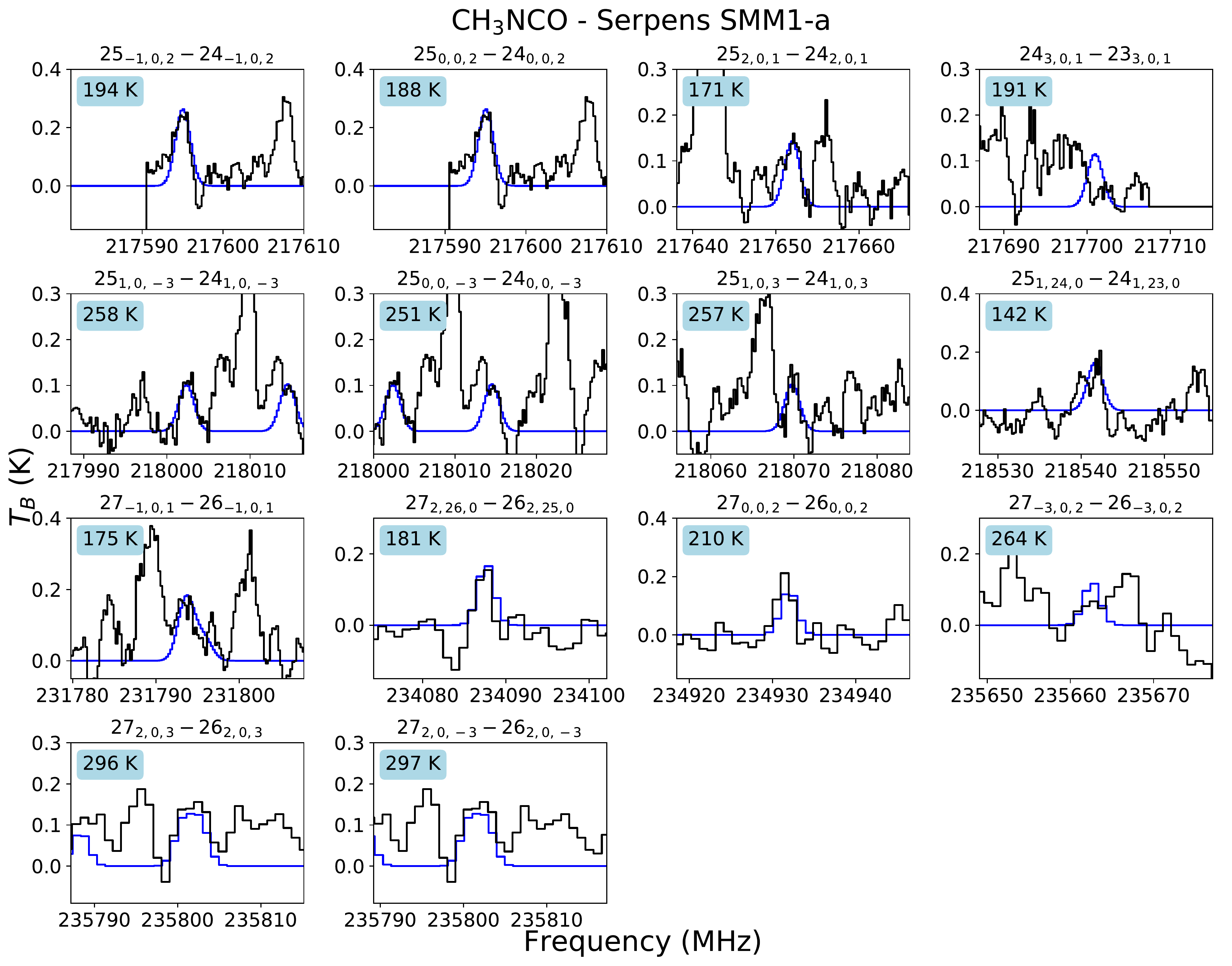}
\caption{Identified lines of CH$_{3}$NCO, $\nu$=0 towards SMM1. The observed spectrum is plotted in black, with the best-fit synthetic spectrum overplotted in blue ($N_{\rm T}$ = (6.4$\pm$1.9)$\times$10$^{14}$ cm$^{-2}$, $T_{\rm ex}$ = 240$\pm$60~K). The transition is indicated at the top of each panel and the upper state energy is given in the top left of each panel.}
\label{fig:lines_CH3NCO_v=0}
\end{figure*}

\begin{table*}[h]
\caption{Best-fit parameters for molecules detected towards SMM1-a in a 1$\farcs$2 beam.}             
\label{tab:best_fits}      
\centering          
\begin{tabular}{l c c c c c c c}     
\hline\hline  
Molecule & Lines & $N_{\rm T}$ & $T_{\rm ex}$ & $V_{\rm LSR}$ & $\Delta V$ & [X] / [CH$_{3}$OH] & [X] / [HNCO] \\ 
 & \# & (cm$^{-2}$) & (K) & km s$^{-1}$ & km s$^{-1}$ &  \\ 
\hline                    
   HOCH$_{2}$CN & 18 & (7.4$\pm$0.9)$\times$10$^{14}$ & 260$\pm$45 & 6.8$\pm$0.2 & 2.5$\pm$0.3 & 6.2$\times$10$^{-4}$ & 7.4$\times$10$^{-2}$ \\ 
   CH$_{3}$NCO & 12 & (6.4$\pm$1.9)$\times$10$^{14}$ & 240$\pm$60 & 7.0$\pm$0.4 & 3.0$\pm$0.6 & 5.3$\times$10$^{-4}$ & 6.4$\times$10$^{-2}$ \\ 
\hline
    D$_{2}$CO & 3 & (5.4$\pm$2.5)$\times$10$^{14}$ & [200] & 7.4$\pm$0.4 & 3.9$\pm$0.2 & 4.5$\times$10$^{-4}$ & 5.4$\times$10$^{-2}$ \\ 
    CH$_{3}^{18}$OH & 4 & (2.0$\pm$1.1)$\times$10$^{15}$ & 250$\pm$60 & 7.1$\pm$0.2 & 2.8$\pm$0.3 & -- & -- \\ 
   $^{12}$CH$_{3}$OH$^{a}$ & 5$^{b}$ & 1.1$\times$10$^{18}$ & -- & -- & -- & 1.0 & 110 \\ 
   CH$_{3}$CN$^{c}$ & 6 & (1.3$\pm$0.3)$\times$10$^{15}$ & 190$\pm$25 & 7.5$\pm$0.2 & 3.4$\pm$0.3 & 1.1$\times$10$^{-3}$ & 0.1 \\ 
   NH$_{2}$CN & 3 & (5.1$\pm$1.3)$\times$10$^{13}$ & 190$\pm$40 & 7.1$\pm$0.2 & 3.2$\pm$0.4 & 4.3$\times$10$^{-5}$ & 5.1$\times$10$^{-3}$ \\ 
   HN$^{13}$CO & 3 & (1.9$\pm$0.3)$\times$10$^{14}$ & 190$\pm$30 & 7.6$\pm$0.2 & 3.5$\pm$0.3 & -- & -- \\ 
   HN$^{12}$CO$^{a}$ & 5$^{b}$ & 1.0$\times$10$^{16}$ & -- & -- & -- & 1.1$\times$10$^{-2}$ & 1.0  \\ 
   CH$_{3}$CH$_{2}$OH & 14 & (4.1$\pm$0.9)$\times$10$^{15}$ & 210$\pm$25 & 7.3$\pm$0.2 & 2.8$\pm$0.6 & 4.1$\times$10$^{-3}$ & 0.3 \\ 
   CH$_{3}$OCHO & 24 & (7.4$\pm$0.7)$\times$10$^{15}$ & 215$\pm$20 & 7.3$\pm$0.2 & 3.1$\pm$0.3 & 7.4$\times$10$^{-3}$ & 0.6 \\
   a-(CH$_{2}$OH)$_{2}$ & 14 & (1.7$\pm$0.5)$\times$10$^{15}$ & 195$\pm$70 & 7.2$\pm$0.2 & 2.7$\pm$0.3 & 1.7$\times$10$^{-3}$ & 0.1 \\ 
   \hline
   CH$_{3}$CNO & 0 & $\leq$1.0$\times$10$^{13}$ & [200] & [7.0] & [3.5] & $\leq$9.1$\times$10$^{-6}$ & $\leq$1.0$\times$10$^{-3}$ \\
   CH$_{3}$OCN & 0 & $\leq$5.0$\times$10$^{13}$ & [200] & [7.0] & [3.5] & $\leq$4.6$\times$10$^{-6}$ & $\leq$5.0$\times$10$^{-3}$ \\
   CH$_{2}$CNH & 0 & $\leq$1.0$\times$10$^{15}$ & [200] & [7.0] & [3.5] & $\leq$8.3$\times$10$^{-4}$ & 0.1 \\
   CH(O)CN & 0 & $\leq$2.0$\times$10$^{14}$ & [200] & [7.0] & [3.5] & $\leq$1.7$\times$10$^{-4}$ & $\leq$2.0$\times$10$^{-2}$  \\
   NH$_{2}$CH$_{2}$CN & 0 & $\leq$1.0$\times$10$^{14}$ & [200] & [7.0] & [3.5] & $\leq$8.3$\times$10$^{-5}$ & $\leq$1.0$\times$10$^{-2}$ \\ 
\hline             
\end{tabular}
\tablefoot{Values in brackets are assumed. $^{a}$Main isotopologue column densities are determined by applying the ratios $^{16}$O/$^{18}$O = 560 and $^{12}$C/$^{13}$C = 52.5 to the minor isotopologue column densities. $^{b}$The number of lines identified of the main isotopologue in this data set. $^{c}$The CH$_{3}$CN best-fit parameters are determined from its vibrationally excited state $\nu_{8}$=1.}
\end{table*}

\begin{table*}[h]
\caption{Best-fit parameters of HOCH$_{2}$CN and related species towards IRAS~16923B in the PILS data set in a 0$\farcs$5 beam}             
\label{tab:PILS_species}      
\centering          
\begin{tabular}{l c c c c c c c}     
\hline\hline  
Molecule & $N_{\rm T}$ & $T_{\rm ex}$ & [X] / [CH$_{3}$OH]$^{a}$ & [X] / [HNCO]$^{a}$ \\
 & (cm$^{-2}$) & (K) & & \\ 
\hline                    
   HOCH$_{2}$CN & $\leq$1.0$\times$10$^{15}$ & [150] & $\leq$1.0$\times$10$^{-4}$ & $\leq$3.3$\times$10$^{-2}$ \\ 
   HOCH$_{2}$CN & $\leq$1.0$\times$10$^{15}$ & [300] & $\leq$1.0$\times$10$^{-4}$ & $\leq$3.3$\times$10$^{-2}$ \\ 
   CH$_{2}$CNH &  $\leq$1.0$\times$10$^{15}$ & [100] & $\leq$1.0$\times$10$^{-4}$ & $\leq$3.3$\times$10$^{-2}$ \\
   CH$_{2}$CNH &  $\leq$2.0$\times$10$^{15}$ & [300] & $\leq$2.0$\times$10$^{-4}$ & $\leq$6.7$\times$10$^{-2}$ \\
   CH(O)CN &  $\leq$5.0$\times$10$^{14}$ & [100] & $\leq$5.0$\times$10$^{-5}$ & $\leq$1.7$\times$10$^{-2}$ \\
   CH(O)CN &  $\leq$5.0$\times$10$^{14}$ & [300] & $\leq$5.0$\times$10$^{-5}$ & $\leq$1.7$\times$10$^{-2}$ \\
   NH$_{2}$CH$_{2}$CN & $\leq$1.0$\times$10$^{15}$ & [100] & $\leq$1.0$\times$10$^{-4}$ & $\leq$3.3$\times$10$^{-2}$ \\ 
   NH$_{2}$CH$_{2}$CN & $\leq$5.0$\times$10$^{14}$ & [300] & $\leq$5.0$\times$10$^{-5}$ & $\leq$1.7$\times$10$^{-2}$ \\ 
\hline             
\end{tabular}
\tablefoot{Values in brackets are assumed. $V_{\rm LSR}$ = 2.7 km~s$^{-1}$ and $\Delta V$ = 1.0 km~s$^{-1}$. Note that some upper limit column densities are similar because the upper state energies of the lines cover only a narrow range of energies. $^{a}$CH$_{3}$OH and HNCO column densities were adopted from \citet{jorgensen2018} and \citet{coutens2016}, respectively.}
\end{table*}

\subsection{Analysis method}
\label{sec:analysis_method}

The spectra were analyzed with the CASSIS\footnote{CASSIS has been developed by IRAP-UPS/CNRS (http://cassis.irap.omp.eu)} line analysis software. Spectral line lists were obtained from the JPL database for molecular spectroscopy \citep{pickett1998}, the Cologne Database for Molecular Spectroscopy \citep[CDMS,][]{muller2001,muller2005}, and from literature. An overview of the spectroscopic line lists used in this work and the laboratory works they are based on is given in Appendix \ref{ap:spec_data}. Given a spectroscopic line list as input, CASSIS can produce synthetic spectra of a molecule based on parameters such as column density ($N_{\rm T}$), excitation temperature ($T_{\rm ex}$), peak gas velocity ($V_{\rm LSR}$), line width at half maximum ($\Delta V$), and source size ($\theta_{\rm source}$). These parameters were given as free parameters to a Monte-Carlo Markov Chain (MCMC) algorithm and $\chi^{2}$ minimisation routine. This routine finds the best-fit of a synthetic spectrum to an observed spectrum over a specified parameter space, thereby determining the best-fit parameters and thus column density and excitation temperature of a molecule. For the analysis, optically thin lines ($\tau$ $\ll$ 1.0) were used. The $\tau$-value was approximated from the by-eye synthetic fit (see below) of the observed rotational lines with the CASSIS software. The molecules were assumed to be in local thermodynamic equilibrium (LTE). Errors on physical parameters take the uncertainty of the fit and the flux uncertainty as input and are calculated from the spread in $\chi^{2}$ values around the minimum to a 3$\sigma$ confidence level.

In this work, spectral lines of a molecule were identified in the observed spectra and a by-eye synthetic fit of the lines was made. For the by-eye fit, line width and peak velocity are determined from prominent spectral lines of molecules such as HNCO and CH$_{3}$OH and used as a first approximation for other molecules. An excitation temperature of 200~K is taken as an initial guess and followed by a round of adjusting column density and rotational temperature until a reasonable by-eye fit was found. The by-eye fit results were given as starting parameters for the MCMC $\chi^{2}$ minimisation routine. The $\chi^{2}$ minimisation was performed on lines that are not blended and have minimal contributions from the wings of neighboring lines. Blending species were identified by checking the line position for other lines of known hot core / corino species with A$_{\rm ij}$ > 1$\times$10$^{-6}$ and E$_{\rm up}$ of 0--1000~K. The column density was given as a free parameter over two orders of magnitude centered on the by-eye fit column density and the excitation temperature was a free parameter from 50 -- 350 K. For SMM1-a, $\Delta V$ was a free parameter from 1.0 -- 4.0 km s$^{-1}$, and the source velocity was a free parameter between 6.0 -- 9.0 km s$^{-1}$. The source size was assumed to be equal to the beam size and taken to be 1$\farcs$2, resulting in a beam filling factor of 0.5. A background continuum temperature ($T_{\rm bg}$) of 5.2~K was used.

For the analysis of IRAS~16293B, $\Delta V$ and $V_{\rm LSR}$ were fixed to 1.0 km s$^{-1}$ and 2.7 km s$^{-1}$, respectively. The source size was assumed to be equal to the beam size at 0$\farcs$5, while a background continuum temperature of $T_{\rm bg}$ = 21~K was used.

\section{Results}
\label{sec:results}

\subsection{SMM1}

In the spectra of SMM1-a, multiple unblended rotational lines of the isomers HOCH$_{2}$CN and CH$_{3}$NCO~$\nu$=0 are found. Moment 0 maps (the spatial mapping of the integrated line intensity of a single rotational line) of both species show that most emission originates from SMM1-a, see Fig. \ref{fig:mom0_smm1}. The identified lines towards SMM1-a are presented in Figs. \ref{fig:lines_HOCH2CN} and \ref{fig:lines_CH3NCO_v=0}. For HOCH$_{2}$CN, this is the second independent interstellar detection of this molecule \citep[the first detection of HOCH$_{2}$CN was presented towards IRAS~16293B by][]{zeng2019}, while the first detection of CH$_{3}$NCO towards SMM1-a is part of only a handful of detections of this species towards other interstellar sources. 

Several other species are identified in the SMM1-a spectra as well. Rotational lines of acetonitrile (CH$_{3}$CN~$\nu_{8}$=1), cyanamide (NH$_{2}$CN), ethanol (CH$_{3}$CH$_{2}$OH), the anti-conformer of ethylene glycol (a-(CH$_{2}$OH)$_{2}$), deuterated formaldehyde (D$_{2}$CO), isocyanic acid (HN$^{12}$CO and HN$^{13}$CO), methylformate (CH$_{3}$OCHO), and methanol ($^{12}$CH$_{3}$OH and CH$_{3}^{18}$OH) are detected. Rotational lines of the C$_{2}$H$_{3}$NO isomers methyl fulmiate (CH$_{3}$CNO) and methyl cyanate (CH$_{3}$OCN) were not identified in the spectra. The molecules aminoacetonitrile (NH$_{2}$CH$_{2}$CN), ketenimine (CH$_{2}$CNH), and formylcyanide (HC(O)CN) are searched for, but not identified. Spectra are presented in Appendix \ref{ap:supporting_SMM1-a} and spectroscopic parameters of transitions are provided in Table \ref{tab:SMM1_lines}. Following the procedure detailed in Sect. \ref{sec:analysis_method}, the best-fit parameters of these species are determined. For undetected species, upper limit column densities are determined by assuming $T_{\rm ex}$ = 200~K, which is chosen as a representative excitation temperature from the molecules that are detected. The fit parameters are presented in Table \ref{tab:best_fits}. The main isotopologues of CH$_{3}$OH and HNCO are optically thick and their column densities are therefore determined from minor isotopologues. This is done by multiplying with the local interstellar $^{12}$C/$^{13}$C ratio of 52.5$\pm$15.4 \citep{yan2019} and $^{16}$O/$^{18}$O ratio of 560 \citep{wilson1999}.

\subsection{IRAS~16293B}

\begin{figure*}[h]
\includegraphics[width=\hsize]{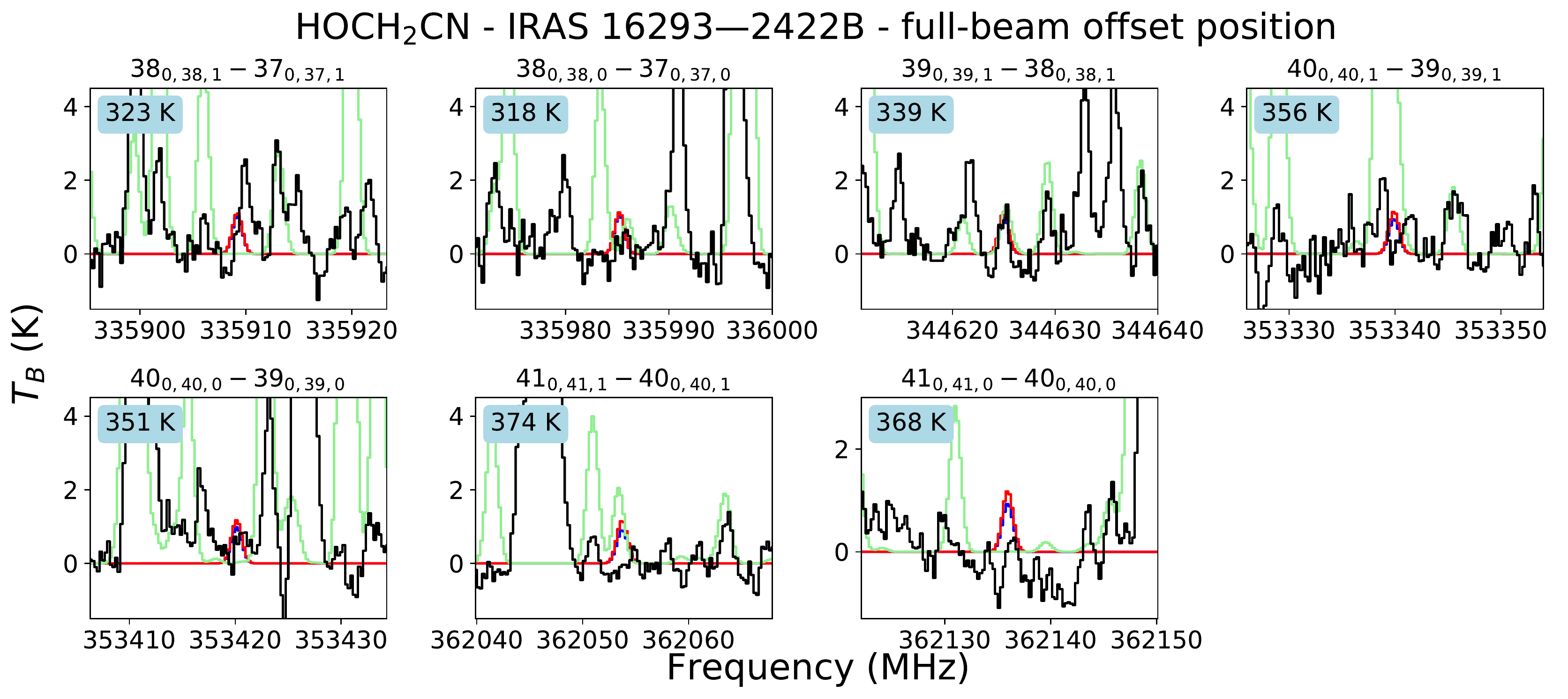}
\caption{Overview of a-type ($J'_{0,J'}$ -- $J''_{0,J''}$) HOCH$_{2}$CN transitions covered by the PILS spectrum towards the full-beam offset position of IRAS~16293B, illustrating the non-detection of these lines in this spectrum. The observed spectrum is plotted in black and synthetic spectra for a column density of 1.0$\times$10$^{15}$ cm$^{-2}$ and excitation temperatures of 150 (blue) and 300~K (red) are overplotted. The synthetic spectrum of the entire molecular inventory determined with PILS data towards this position is plotted in green. The quantum numbers of the transition are indicated at the top of each panel and the upper state energy is given in the top left of each panel.}
\label{fig:PILS_1.0_HOCH2CN_a-type}
\end{figure*}

\begin{figure*}[h]
\includegraphics[width=\hsize]{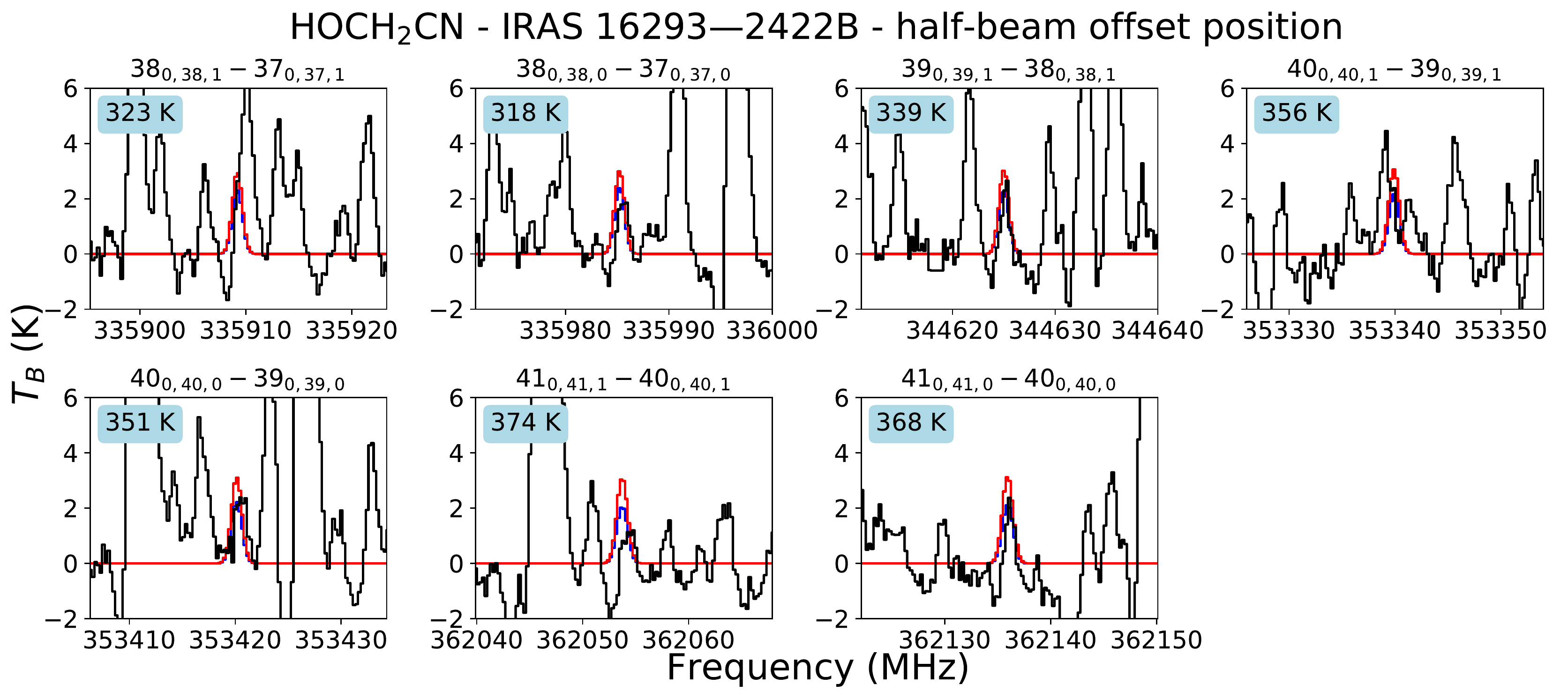}
\caption{Overview of a-type ($J'_{0,J'}$ -- $J''_{0,J''}$) HOCH$_{2}$CN transitions covered by the PILS spectrum towards the half-beam offset position of IRAS~16293B, illustrating the detection of a number of these lines. The observed spectrum is plotted in black and synthetic spectra for a column density of 3.0$\times$10$^{15}$ cm$^{-2}$ and excitation temperatures of 150 (blue) and 300~K (red) are overplotted. The quantum numbers of the transition are indicated at the top of each panel and the upper state energy is given in the top left of each panel.}
\label{fig:PILS_0.5_HOCH2CN_a-type}
\end{figure*}

Because the chemical inventory of IRAS~16293B is well characterised with results from the PILS survey, this data set is used to search for HOCH$_{2}$CN and related species to make an unbiased chemical comparison with SMM1-a. While HOCH$_{2}$CN was identified towards IRAS~16293B by \citet{zeng2019}, this detection cannot be confirmed in the PILS spectrum at the full-beam offset position (the position commonly used for the molecular analysis of IRAS~16293B with PILS data, see Fig.\ref{fig:mom0_16293B}). Figure \ref{fig:PILS_1.0_HOCH2CN_a-type} shows the HOCH$_{2}$CN a-type transitions ($J'_{0,J'}$ -- $J''_{0,J''}$) at this position covered by the PILS spectral range with a synthetic glycolonitrile spectrum at $N_{\rm T}$ = 1.0$\times$10$^{15}$ cm$^{-2}$ and $T_{\rm ex}$ = 150 and 300~K. A synthetic spectrum of the previously identified molecules towards IRAS~16293B at the full-beam offset position is added in green. Note that this synthetic spectrum only includes molecules listed in previous publications and does not include HOCH$_{2}$CN. The molecules and parameters used in this fit are listed in Table \ref{tab:params_16293_fit}. When a rotational spectrum is experimentally measured, a-type transitions are usually the first and most accurately determined transitions. The assignment of these a-type transitions, therefore, is key to claim an unambiguous identification. However, of the seven a-type transitions covered by the PILS survey, only the $39_{0,39,1}$ -- $38_{0,38,1}$ transition at 344625 MHz is possibly detected, although this feature has a contribution of a HONO and CH$_{2}$DOH transition \citep[][]{jorgensen2016,jorgensen2018,coutens2019}, which also can fully reproduce this line. Appendix \ref{ap:PILS} presents all the HOCH$_{2}$CN transitions at the full-beam offset position that are largely unblended and have $A_{\rm ij}$ $\geq$ 1$\times$10$^{-3}$ s$^{-1}$. At the full-beam offset position, a large number ($\sim$40) of glycolonitrile transitions that are present in the synthetic spectrum are not seen in the observed spectrum. We note that due to the line-richness of the source the baseline subtraction is challenging and in some cases, it can be oversubtracted. This can explain why certain HOCH$_{2}$CN lines are not clearly observed, as the baseline at these positions dips. Furthermore, some transitions seem to be better reproduced with low excitation temperatures, whereas others require warmer temperatures. This could indicate that two gas components are traced, as also shown by \citet{zeng2019}, but at this position these components are hard to distinguish. Therefore, HOCH$_{2}$CN can only tentatively be identified towards the full-beam offset position, with a column density of $\sim$1.0$\times$10$^{15}$ cm$^{-2}$ at $T_{\rm ex}$ = 150 and 300~K.

At the half-beam offset position (see Fig. \ref{fig:mom0_16293B}), which is closer to the continuum peak of IRAS~16293B, HOCH$_{2}$CN can be identified. At this position, at least four a-type transitions are clearly detected, while the three other lines suffer from line blending or absorption features, see Fig. \ref{fig:PILS_0.5_HOCH2CN_a-type}. These lines can approximately be fitted with synthetic spectra of $N_{\rm T}$ = 3.0$\times$10$^{15}$ cm$^{-2}$ for $T_{\rm ex}$ = 150 and 300~K. In appendix \ref{ap:PILS} the remaining HOCH$_{2}$CN transitions at the half-beam offset position are shown, with the same selection criteria for the full-beam offset position.

In Fig. \ref{fig:mom0_16293B} the moment 0 map of the HOCH$_{2}$CN 36$_{4,32}$--35$_{4,31}$ transition towards IRAS~16293B is shown. This map shows that HOCH$_{2}$CN emission towards IRAS~16293B is compact. This explains the non-detection of glycolonitrile towards the full-beam offset position, as this position misses most of the HOCH$_{2}$CN emission. At the same time, this map demonstrates why \citet{zeng2019} could detect HOCH$_{2}$CN towards IRAS~16293B, since these authors use a larger observational beam of 1$\farcs$6 beam (with an assumed source size of 0$\farcs$5), which covers the entire emitting area.

Because the chemical inventory at the full-beam offset position of the PILS data is best characterised, the tentative detection of HOCH$_{2}$CN towards this position, with a column density of $\sim$1.0$\times$10$^{15}$ cm$^{-2}$), is used for further analysis in this paper. However, the identification of HOCH$_{2}$CN towards the half-beam offset position in combination with the moment 0 map support the detection of HOCH$_{2}$CN towards IRAS~16923B by \citet{zeng2019}.

The related species CH$_{2}$CNH, CH(O)CN, and NH$_{2}$CH$_{2}$CN are also searched for in the PILS dataset, but clear and unblended lines are not identified. For these species upper limit column densities are determined at the full-beam offset position at excitation temperatures of 100 and 300~K. The upper limit column densities and abundances of all species are presented in Table \ref{tab:PILS_species}.

\section{Discussion}
\label{sec:discussion}

In this work, several molecules are detected towards SMM1-a and analysed. Most notable are the detection of the C$_{2}$H$_{3}$NO isomers CH$_{3}$NCO (methyl isocyanate) and HOCH$_{2}$CN (glycolonitrile). For HOCH$_{2}$CN, this is only its second interstellar detection. CH$_{3}$NCO has been detected in multiple interstellar sources, but this is the first detection towards SMM1 and therefore also the first detection towards an intermediate-mass source. These new detections serve as additional evidence for a large and diverse reservoir of prebiotic molecules in star- and planet-forming regions, which can contribute to the emergence of biomolecules on planetary bodies. Of the C$_{2}$H$_{3}$NO isomers, CH$_{3}$NCO is energetically the most favorable, followed by HOCH$_{2}$CN, which has a higher relative energy of 12.1 -- 18.6 kcal mol$^{-1}$ (0.5 -- 0.8 eV molecule$^{-1}$ or 5800 -- 9300 K molecule$^{-1}$), depending on the level of theory used \citep{fourre2020}. In a thermodynamic equilibrium, lower energy or more stable products are favored and in such a scenario, CH$_{3}$NCO is expected to be more abundant than HOCH$_{2}$CN by a factor of at least 1$\times$10$^{9}$ (assuming a temperature of 300~K). The fact that CH$_{3}$NCO and HOCH$_{2}$CN are found to be equally abundant is, therefore, evidence that the formation of these molecules is rather driven by kinetics. 

To better understand the interstellar chemistry of HOCH$_{2}$CN and CH$_{3}$NCO, their abundances and those of several other species are compared for SMM1-a, IRAS~16293B, and several other sources for which at least some of these species are detected. For this comparison, it is important to emphasise observational differences. For example, SMM1 (D $\approx$436 pc) is located further away than IRAS~16293 \citep[D $\approx$141 pc,][]{dzib2018} and the beam size used in this work (1$\farcs$2) is larger than that of the PILS survey (0$\farcs$5). Therefore, the chemical inventory of SMM1-a and the involved chemical processes are investigated on a much larger spatial scale of roughly 500 au, compared to about 70 au for IRAS~16293B. The SMM1-a observations can cover a larger range of physical environments (e.g. also the envelope or outflow), and potentially a larger temperature gradient. At the same time, the higher luminosity of SMM1 results in a larger area where hot core conditions are present, thus compensating for the lower spatial resolution. Furthermore, the SMM1-a spectrum is extracted towards the continuum peak, but for IRAS~16293B a full-beam offset position from the continuum peak is used. Therefore, when comparing molecular ratios between the two sources, they may not only be affected by different physical conditions but also due to the way the sources were observed. Observational parameters of SMM1-a, IRAS~16293B, and other sources used for comparisons are listed in Table \ref{tab:obs_params}.

\begin{figure*}[h]
\includegraphics[width=1\hsize]{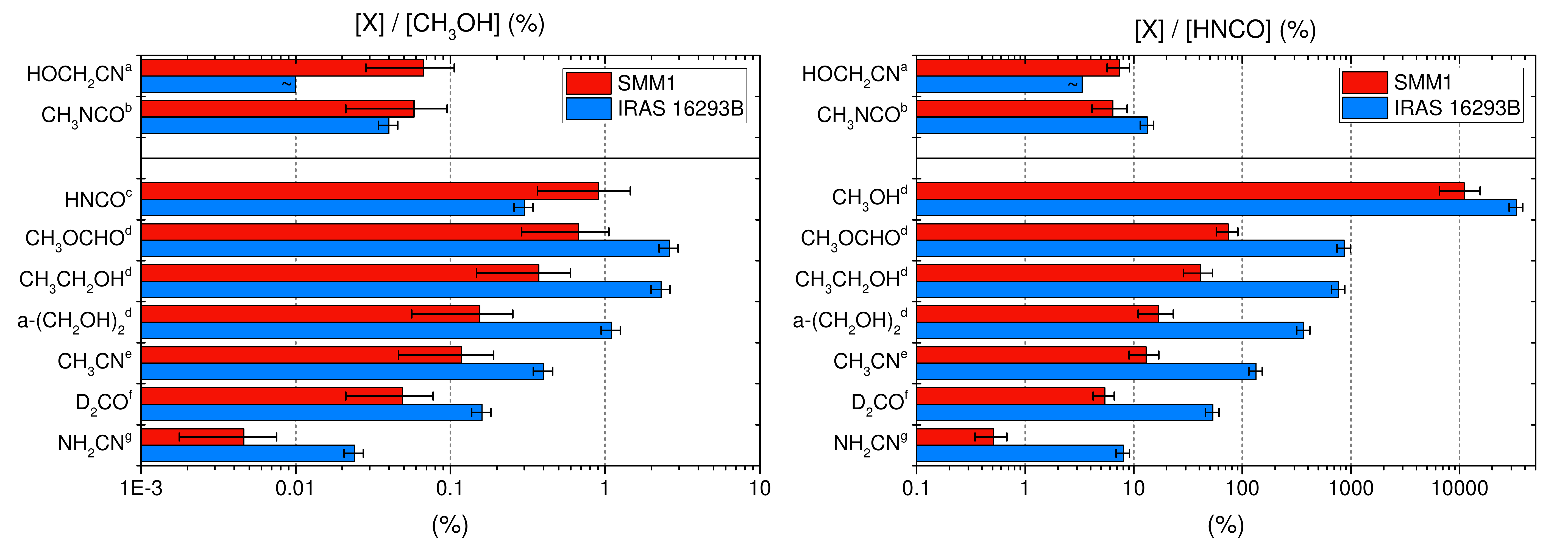}
\caption{Ratios of [X]/[CH$_{3}$OH] towards SMM1-a (red) and IRAS~16293B (blue) in decreasing order of SMM1-a abundance. The ``$\sim$'' symbol indicates that these HOCH$_{2}$CN ratios have been determined with the column density of the tentative HOCH$_{2}$CN detection towards the IRAS~16293B full-beam offset position. For IRAS~16293B, column densities derived towards the full-beam offset position from the following PILS publications are used: $^{a}$This work, $^{b}$\citet{ligterink2017}, $^{c}$\citet{coutens2016}, $^{d}$\citet{jorgensen2016,jorgensen2018}, $^{e}$\citet{calcutt2018}, $^{f}$\citet{persson2018}, $^{g}$\citet{coutens2018}.} 
\label{fig:X_over}
\end{figure*}

HOCH$_{2}$CN is thus far only detected in two sources, SMM1-a and IRAS~16293B \citep{zeng2019}, and therefore a chemical comparison is limited to these two objects. To use molecular ratios that are unbiased by observational parameters, only data of the chemical inventory of IRAS~16293B obtained with PILS survey data at the full-beam offset position is used for the source comparison. This means that the tentative column density of HOCH$_{2}$CN in IRAS~16293B is used. Figure \ref{fig:X_over} shows the abundances of molecules detected in this work to CH$_{3}$OH and HNCO in SMM1-a and IRAS~16293B. For IRAS~16293B, the analysis performed in this work is combined with results from \citet{jorgensen2016,coutens2016,ligterink2017,coutens2018,persson2018,calcutt2018,jorgensen2018}. For both the [X]/[CH$_{3}$OH] and [X]/[HNCO] ratios, all the oxygen-bearing molecules, CH$_{3}$CN, and NH$_{2}$CN are found to be more abundant in IRAS~16293B than in SMM1-a. For [X]/[CH$_{3}$OH] its ratios are generally a factor of a few lower in SMM1-a, while for the [X]/[HNCO] ratios the difference is usually more than a factor of ten. 

\subsection{SMM1-a: A HOCH$_{2}$CN-rich source}
\label{sec:hoch2cn_form}

While general trends are found in the [X]/[CH$_{3}$OH] and [X]/[HNCO] ratios displayed in Fig. \ref{fig:X_over}, three molecules deviate from the general trend. The abundance of CH$_{3}$NCO is found to be approximately equal in both sources, while HOCH$_{2}$CN and HNCO are more abundant in SMM1-a compared to IRAS~16293B. In particular, for HOCH$_{2}$CN a large difference is seen in its ratios to CH$_{3}$OH, which can be more than an order of magnitude different between the two sources.

\begin{figure}[h]
\includegraphics[width=1\hsize]{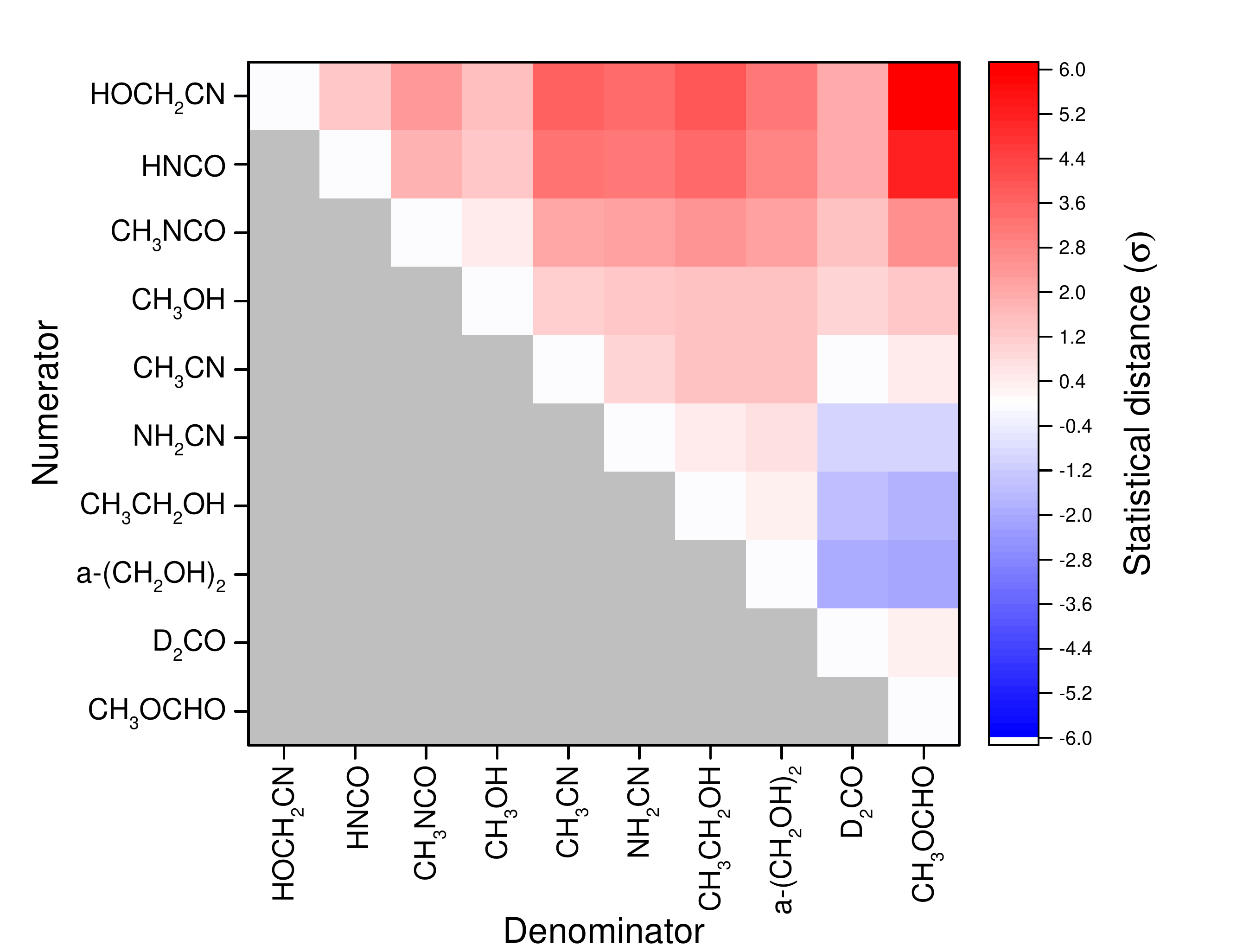}
\caption{Statistical distance between molecular ratios in SMM1 and IRAS~16293B, given in $\sigma$. Larger $\sigma$ values indicate a larger difference between the two sources for a given ratio. Positive values indicate that a ratio is higher in SMM1-a, while negative values indicate that a ratio is lower in SMM1-a. In particular, all ratios of HOCH$_{2}$CN are found to be higher in SMM1-a than in IRAS~16239B.}
\label{fig:statistical_difference}
\end{figure}

To gain further insight in how the chemical compositions of SMM1-a and IRAS~16293B differ, the statistical distance of molecular ratios are plotted in Fig. \ref{fig:statistical_difference}. The statistical distance indicates how significant the difference in a molecular ratio between SMM1-a and IRAS~16293B is \citep[see][and Appendix \ref{sec:stat_distance}]{manigand2020}. Greater values indicate a greater difference in the molecular ratios between the two sources. A positive value indicates that the ratio of SMM1-a is greater than that of IRAS~16293B and vice versa. Fig. \ref{fig:statistical_difference} highlights that HOCH$_{2}$CN and HNCO are more abundant in SMM1-a ($\sigma$ ranging from 3 -- 6) and CH$_{3}$NCO is moderately more abundant in SMM1-a ($\sigma$ $\geq$ 2). 

The statistical distance results have several implications. Since the abundances of both HOCH$_{2}$CN and HNCO are enhanced in SMM1-a, this may indicate a relationship between these two species. For CH$_{3}$NCO a much less significant enhancement is seen, which can imply that both C$_{2}$H$_{3}$NO isomers form via different chemical reactions or under different physical conditions. However, it is important to stress that an abundance correlation does not always imply a formational link between species \citep{belloche2020}.

The statistical distances of ratios involving CH$_{3}$CN, NH$_{2}$CN, a-(CH$_{2}$OH), and CH$_{3}$CH$_{2}$OH show that there is almost no variation in these molecules between SMM1-a and IRAS~16293B. This is interesting because some of these molecules can form in reactions involving radicals from which HOCH$_{2}$CN  also can be formed, such as CN, CH$_{2}$OH, and CH$_{2}$CN. This hints that either HOCH$_{2}$CN is not formed from these radicals, is not linked to the reaction networks that form the four aforementioned species, or HOCH$_{2}$CN forms from the same radicals, but under different physical conditions. 

Finally, it is interesting to note that ratios of D$_{2}$CO and CH$_{3}$OCHO to NH$_{2}$CN, CH$_{3}$CH$_{2}$OH, and a-(CH$_{2}$OH)$_{2}$ seem to be a bit more abundant in SMM1-a than in IRAS~16293B. This is particularly significant for the case of D$_{2}$CO since H$_{2}$CO is involved in the Strecker-like formation of HOCH$_{2}$CN, see reaction \ref{eq:cn-h2co}. A higher abundance of H$_{2}$CO may indicate that the Strecker-like reaction can more efficiently take place. However, care needs to be taken with this interpretation, since the D$_{2}$CO spectral lines in the SMM1-a spectrum are blended and thus there is a large uncertainty on its column density. At the same time, the D/H ratio of H$_{2}$CO is not known in SMM1-a, which introduces another source of uncertainty. To investigate if there is a correlation between HOCH$_{2}$CN and H$_{2}$CO, both species should be identified towards more sources and H$_{2}$CO should be observed through its minor $^{13}$C and $^{18}$O isotopes instead of the deuterated species. For now, however, the Strecker-like synthesis of HOCH$_{2}$CN in the ISM can neither be confirmed nor ruled out.

\subsection{Interstellar formation of CH$_{3}$NCO}
\label{sec:ch3nco_form}

\begin{figure}[h]
\includegraphics[width=1.1\hsize]{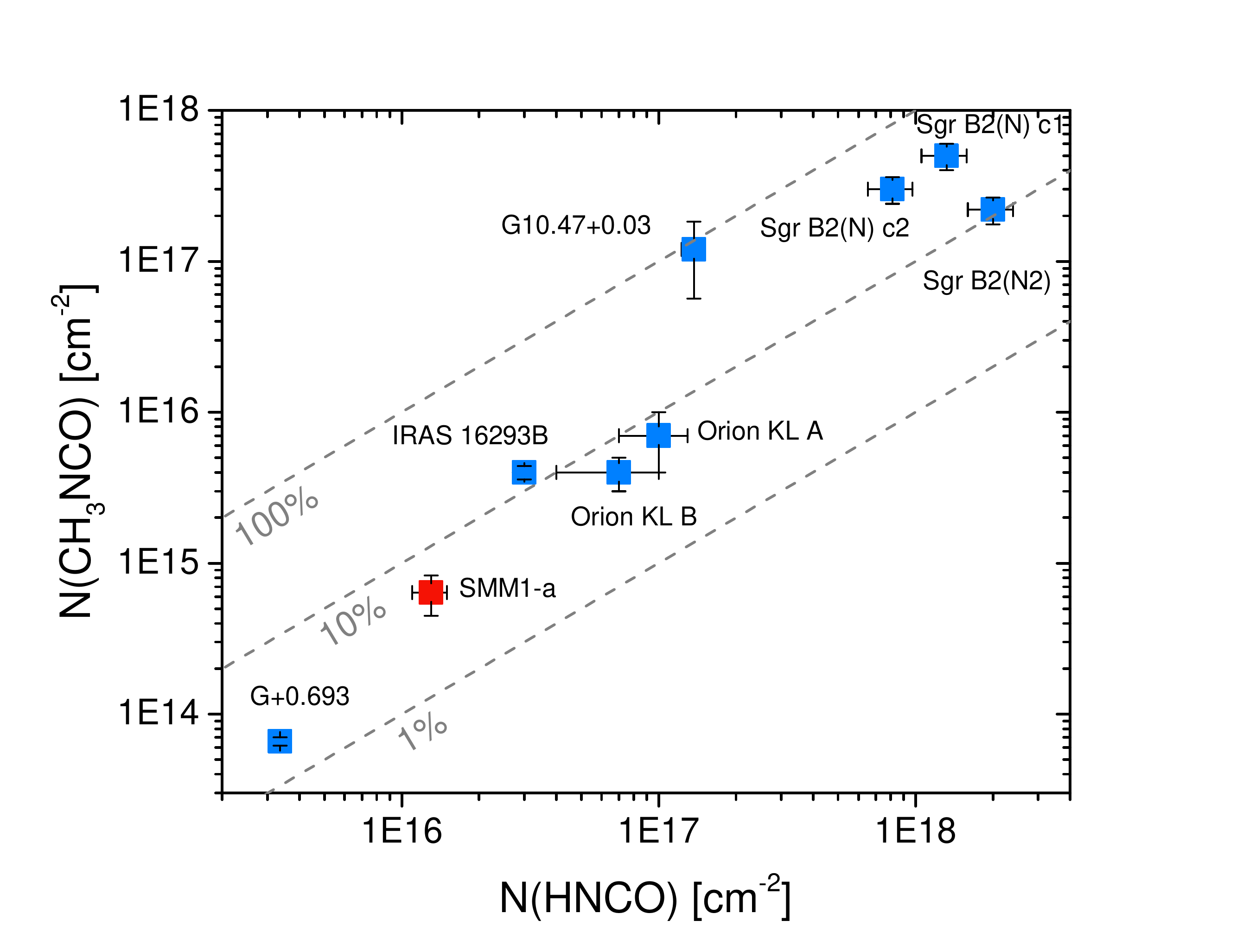}
\caption{Ratios of [CH$_{3}$NCO]/[HNCO] ratios towards SMM1-a and various other sources. Column densities from the following publications are used: \citet{belloche2013}, \citet{cernicharo2016}, \citet{ligterink2017}, \citet{belloche2017}, \citet{zeng2018}, \citet{gorai2020}, and this work.}
\label{fig:CH3NCO_ratios}
\end{figure}

Since its first detection in the ISM, the formation of CH$_{3}$NCO has been hypothesised to be linked to that of HNCO, see Eq. \ref{eq:nco_ch3}. To investigate the interstellar chemical relationship between these two species, their gas-phase ratios towards interstellar sources are plotted in Fig. \ref{fig:CH3NCO_ratios}. The result of the SMM1-a analysis is complemented by data from the low-mass protostar IRAS~16239B \citep{ligterink2017}, the quiescent giant molecular cloud G+0.693 \citep{zeng2018}, the high-mass star-forming region Orion KL \citep{cernicharo2016}, the high-mass hot molecular core G10.47+0.03 \citep{gorai2020}, and the galactic center source Sagittarius B2(N) \citep[Sgr B2(N),][]{belloche2013,cernicharo2016,belloche2017}. The majority of these sources is categorised as hot cores or corinos. In these sources, thermal desorption of molecules from ice mantles plays an important role to chemically enrich the gas surrounding the protostar, in particular around the desorption temperature of water ice ($\sim$100~K). The exception is the source G+0.693, which is a molecular cloud. Molecules observed in the gas of this cloud are assumed to be the result of gas-phase formation reactions or non-thermal desorption from ice mantles of dust grains.

The ratios of CH$_{3}$NCO and HNCO are generally similar at [CH$_{3}$NCO] / [HNCO] = $\sim$10\% and vary only by a factor of a few, in particular when the lowest and highest ratio are omitted. The lowest ratio, found towards G+0.693, arises in a source with a very different physical structure and can therefore not directly be compared with the hot core and corino sources. The highest ratio is found towards G10.47+0.03, but the analysis of HNCO in this source is likely performed on optically thick HNCO lines and the analysis seems to underestimate the HNCO column density in this source \citep[see Fig. 4 in][]{gorai2020}. Removing these results from the analysis, a correlation between CH$_{3}$NCO and HNCO is found, which may indicate a chemical link between CH$_{3}$NCO and HNCO. Furthermore, this correlation spans a variety of sources of different masses and over four order of magnitude in luminosities. This results in different physical conditions, such as gas and dust temperature and radiation fields for each source. Therefore, the lack of source-to-source variation in [CH$_{3}$NCO]/[HNCO] ratio hints that the abundances of these species are set at an early stage of star formation. Formation of CH$_{3}$NCO via reaction \ref{eq:nco_ch3} in the ice mantles of dust grains during the dark cloud stage or while the ice mantles are warmed up in the hot core/corino stage are a plausible scenario.

\subsection{On the formation of -CN and -NCO molecules}

Figure \ref{fig:statistical_difference} shows that HOCH$_{2}$CN and HNCO abundances are enhanced in SMM1-a compared to IRAS~16293B. At the same time, CH$_{3}$NCO is only marginally enhanced and CH$_{3}$CN and NH$_{2}$CN abundances show little difference between the two sources. Why the abundances of HOCH$_{2}$CN and HNCO are enhanced in SMM1-a and those of CH$_{3}$NCO, CH$_{3}$CN, and NH$_{2}$CN are not is not straightforward to explain. However, it is likely that the various -CN and -NCO molecules form in different chemical reactions and physical conditions.

As this work shows, CH$_{3}$NCO probably forms at an early stage of star-formation, as do CH$_{3}$CN and NH$_{2}$CN. All three species are suggested to form in radical-radical addition reactions in ice mantles. These reactions take place during the dark cloud stage in cold ($\sim$10~K) ice mantles and significantly speed up when the ice mantle temperature increases to $\sim$30~K and radicals become mobile \citep{garrod2008,coutens2018}. Some reactions can compete for the same radical, such as CH$_{3}$CN and NH$_{2}$CN, which both compete for the -CN radical.   

The fact that HOCH$_{2}$CN and HNCO are enhanced in abundance in SMM1-a, can indicate a link between these species. Both molecules can be formed from HCN in ice mantles of interstellar dust grains. HOCH$_{2}$CN can be formed in the Strecker-like reaction when HCN is converted to CN$^{-}$. HNCO and the related anion OCN$^{-}$ are formed when HCN:H$_{2}$O mixtures are processed with energetic UV photons or protons \citep{gerakines2004}. These reactions are aided by high grain temperatures ($\gg$30~K, but below the water sublimation temperature) for a prolonged time and high fluxes of photons and energetic particles. If these conditions are met in SMM1, they can explain the higher abundances of HOCH$_{2}$CN and HNCO compared to IRAS~16239B and present a formational link between some -CN and -NCO molecules. However, only circumstantial evidence can be presented for such conditions in SMM1, which is based on the fact that SMM1 hosts multiple protostellar sources and outflows, which can warm and irradiate the cloud \citep[e.g.][]{choi2009,dionatos2014,hull2017,tychoniec2019}. 

To gain further insight into the reactions that form -CN and -NCO molecules, unbiased observations of these species towards a multitude of sources spanning different physical conditions are needed. Not only the isomers HOCH$_{2}$CN and CH$_{3}$NCO should be targeted for these observations, but also species like CH$_{3}$CN, HNCO, NH$_{2}$CN, C$_{2}$H$_{3}$CN, and C$_{2}$H$_{5}$CN. Laboratory, theoretical and modeling efforts should be focused on understanding the formation of these species. In particular, the formation of HOCH$_{2}$CN via pathways other than the thermal Strecker-like synthesis needs to be studied and a better understanding of the formation of CH$_{3}$CN is required.

\section{Conclusions}
\label{sec:conclusion}

This publication presents the simultaneous detection of the C$_{2}$H$_{3}$NO isomers methyl isocyanate (CH$_{3}$NCO) and glycolonitrile (HOCH$_{2}$CN). Both species are identified towards the intermediate-mass Class 0 protostar Serpens SMM1-a. This is only the second interstellar detection of glycolonitrile, while for methyl isocyanate it is the first detection towards an intermediate-mass protostar. Additionally, CH$_{3}$OH, HNCO, CH$_{3}$OCHO, CH$_{3}$CH$_{2}$OH, a-(CH$_{2}$OH)$_{2}$, D$_{2}$CO, and NH$_{2}$CN are detected. CH$_{2}$CNH, CH(O)CN, and NH$_{2}$CH$_{2}$CN, molecules that are related to HOCH$_{2}$CN, are searched for but not identified. Data from the PILS survey towards IRAS~16293B are analysed in search for HOCH$_{2}$CN and this molecule is identified in a spectrum extracted at a half-beam offset position from the continuum peak of IRAS~16239B. The molecules CH$_{2}$CNH, CH(O)CN, and NH$_{2}$CH$_{2}$CN are not identified towards this source.

The detection of CH$_{3}$NCO and HOCH$_{2}$CN towards SMM1-a is additional evidence of a large interstellar reservoir of prebiotic molecules. Delivery of these molecules to planetary surfaces may contribute to the formation of biomolecules on these objects. The column densities and abundances of CH$_{3}$NCO ($N_{\rm T}$ =  6.4$\times$10$^{14}$ cm$^{-2}$ and [CH$_{3}$NCO]/[CH$_{3}$OH] = 5.3$\times$10$^{-4}$) and HOCH$_{2}$CN ($N_{\rm T}$ = 7.4$\times$10$^{14}$ cm$^{-2}$ and [HOCH$_{2}$CN]/[CH$_{3}$OH] = 6.2$\times$10$^{-4}$) are found to be equal within their error bars. Since HOCH$_{2}$CN is the least energetically favorable of the two isomers, thermodynamics predicts that CH$_{3}$NCO should be more abundant. The equal ratio between both molecules is therefore evidence that the formation of these molecules is driven by kinetics.  

The comparison of molecular ratios between SMM1-a and IRAS~16293B show that HOCH$_{2}$CN and HNCO are significantly more abundant in the former source. The molecular ratios of HOCH$_{2}$CN hint that the formation of this molecule does not heavily depend on solid-state radical-radical addition reactions, such as HOCH$_{2}$ + CN and HO + CH$_{2}$CN. Formation via the thermal Strecker-like reaction [X$^{+}$CN$^{-}$] + H$_{2}$CO in ice mantles cannot be confirmed nor ruled out based on the current data but may be a prominent formation pathway.

To investigate the possibility that CH$_{3}$NCO formation is related to HNCO, the ratios of these molecules in SMM1-a and other sources are analysed. These ratios are found to be uniform throughout all sources at [CH$_{3}$NCO]/[HNCO] = $\sim$10 \%. This indicates that there is a chemical link between both species, but also that its ratios is already set at an early stage of star-formation. Presumably CH$_{3}$NCO forms via the radical-radical reaction CH$_{3}$ + NCO in ice mantles during the dark cloud stage. 

It is difficult to establish a chemical link between -CN and -NCO molecules. Some may be related, such as HOCH$_{2}$CN and HNCO, which can both form from HCN at elevated ($\gg$30~K) grain temperatures and in relatively high radiation fields. Continued observational, laboratory, and theoretical studies of -CN and -NCO molecules are required to gain further insight into their formation and links in their chemistry.

\begin{acknowledgements}
We thank E.G. B{\o}gelund, S.F. Wampfler, M.N. Drozdovskaya, B. Kulterer, and B.A. McGuire for helpful discussions on the observations, spectroscopy, and the chemistry of the C$_{2}$H$_{3}$NO isomers. The authors acknowledge assistance from Allegro, the European ALMA Regional Center node in the Netherlands. We thank the anonymous referee for their thorough review of this manuscript and helpful comments. This paper makes use of the following ALMA data: ADS/JAO.ALMA\#2018.1.00836.S and ADS/JAO.ALMA\#2013.1.00278.S. ALMA is a partnership of ESO (representing its member states), NSF (USA) and NINS (Japan), together with NRC (Canada), MOST and ASIAA (Taiwan), and KASI (Republic of Korea), in cooperation with the Republic of Chile. The Joint ALMA Observatory is operated by ESO, AUI/NRAO and NAOJ. NFWL is supported by the Swiss National Science Foundation
(SNSF) Ambizione grant 193453. JKJ is supported by the European Research Council (ERC) under the European Union's Horizon 2020 research and innovation programme through ERC Consolidator Grant ``S4F'' (grant agreement No~646908). AC acknowledges financial support from the Agence Nationale de la Recherche (grant ANR-19-ERC7-0001-01).    
\end{acknowledgements}

\bibliographystyle{aa}
\bibliography{bibliography}

\begin{appendix}

\section{Spectroscopic data}
\label{ap:spec_data}

In this paper, the CDMS and JPL spectroscopic databases are the primary sources of molecular line lists. In the following table, an overview of the analysed molecules, their identifier and catalog, and the most important publications in literature on which these entries are based is given.

\begin{table*}[h]
\caption{}             
\label{tab:spec_data}      
\centering          
\begin{tabular}{l c c c c}     
\hline\hline  
Molecule & ID & catalog & entry date & reference \\ 
\hline                    
D$_{2}$CO & 32502 & CDMS & Jan 2016 & \citet{bocquet1999} \\
& & & & \citet{zakharenko2015} \\
$^{12}$CH$_{3}$OH & 32504 & CDMS & May 2016 & \citet{xu2008} \\
CH$_{3}^{18}$OH & 34504 & CDMS & Sep 2020 & \citet{fisher2007} \\
CH$_{3}$CN, $\nu_{8}$=1 & 41509 & CDMS & Nov 2016 & \citet{muller2015} \\
& & & & \citet{koivusaari1992} \\ 
NH$_{2}$CN & 42003 & JPL & Jan 1991 & \citet{read1986} \\
HN$^{12}$CO & 43511 & CDMS & May 2009 & \cite{kukolich1971a} \\
& & & &\cite{hocking1975} \\
& & & &\cite{niedenhoff1995} \\
& & & &\cite{lapinov2007} \\
HN$^{13}$CO & 44008 & JPL & Jul 1987 & \cite{hocking1975} \\
CH$_{3}$CH$_{2}$OH & 46524 & CDMS & Nov 2016 & \citet{pearson2008} \\
& & & & \citet{muller2016} \\
CH$_{3}$NCO, $\nu$=0 & 57505 & CDMS & Mar 2016 &\citet{cernicharo2016} \\
CH$_{3}$NCO, $\nu$=1 & 57506 & CDMS & Mar 2016 & \citet{cernicharo2016} \\
HOCH$_{2}$CN & 57512 & CDMS & Mar 2017 & \citet{margules2017} \\
CH$_{3}$OCHO & 60003 & JPL & Apr 2009 & \citet{ilyushin2009} \\
a-(CH$_{2}$OH)$_{2}$ & 62503 & CDMS & Sep 2003 & \citet{christen1995} \\
& & & & \citet{christen2003} \\
\hline             
\end{tabular}
\end{table*}

\section{Supporting information for the SMM1-a analysis}
\label{ap:supporting_SMM1-a}

\begin{figure*}
\includegraphics[width=\hsize]{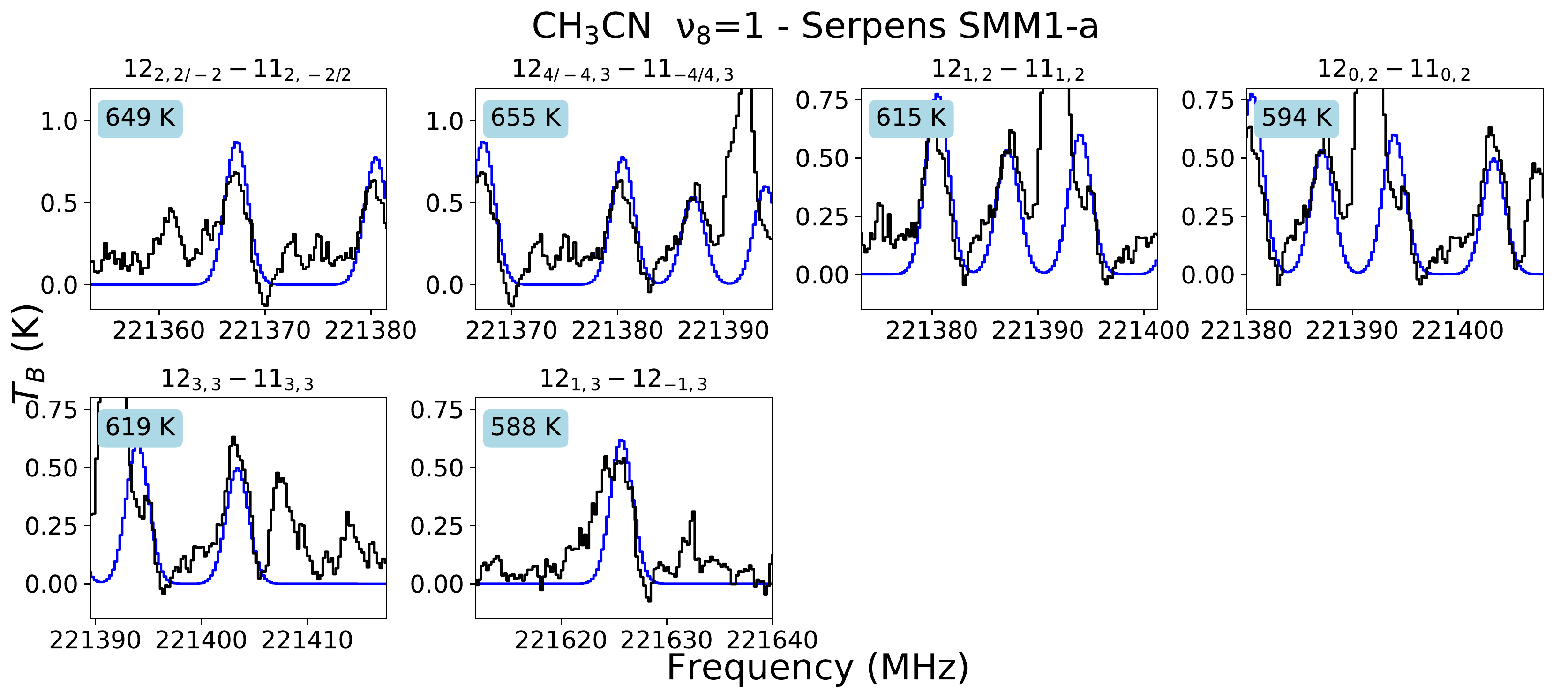}
\caption{Identified lines of CH$_{3}$CN~$\nu_{8}$=1 towards SMM1. The observed spectrum is plotted in black, with the best-fit synthetic spectrum overplotted in blue ($N_{\rm T}$ = (1.3$\pm$0.3)$\times$10$^{15}$ cm$^{-2}$, $T_{\rm ex}$ = 190$\pm$25~K). The transition is indicated at the top of each panel and the upper state energy is given in the top left of each panel.}
\label{fig:lines_CH3CN_v8=1}
\end{figure*}

\begin{figure*}
\includegraphics[width=\hsize]{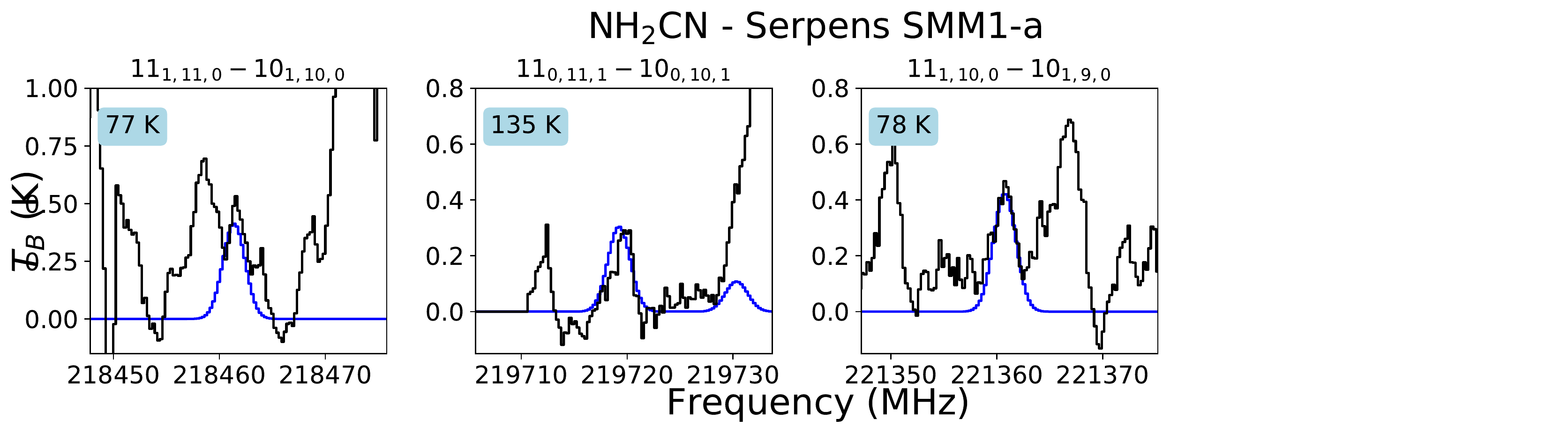}
\caption{Identified lines of NH$_{2}$CN towards SMM1. The observed spectrum is plotted in black, with the best-fit synthetic spectrum overplotted in blue ($N_{\rm T}$ = (5.1$\pm$1.3)$\times$10$^{13}$ cm$^{-2}$, $T_{\rm ex}$ = 190$\pm$40~K). The transition is indicated at the top of each panel and the upper state energy is given in the top left of each panel.}
\label{fig:lines_NH2CN}
\end{figure*}

\begin{figure*}
\includegraphics[width=\hsize]{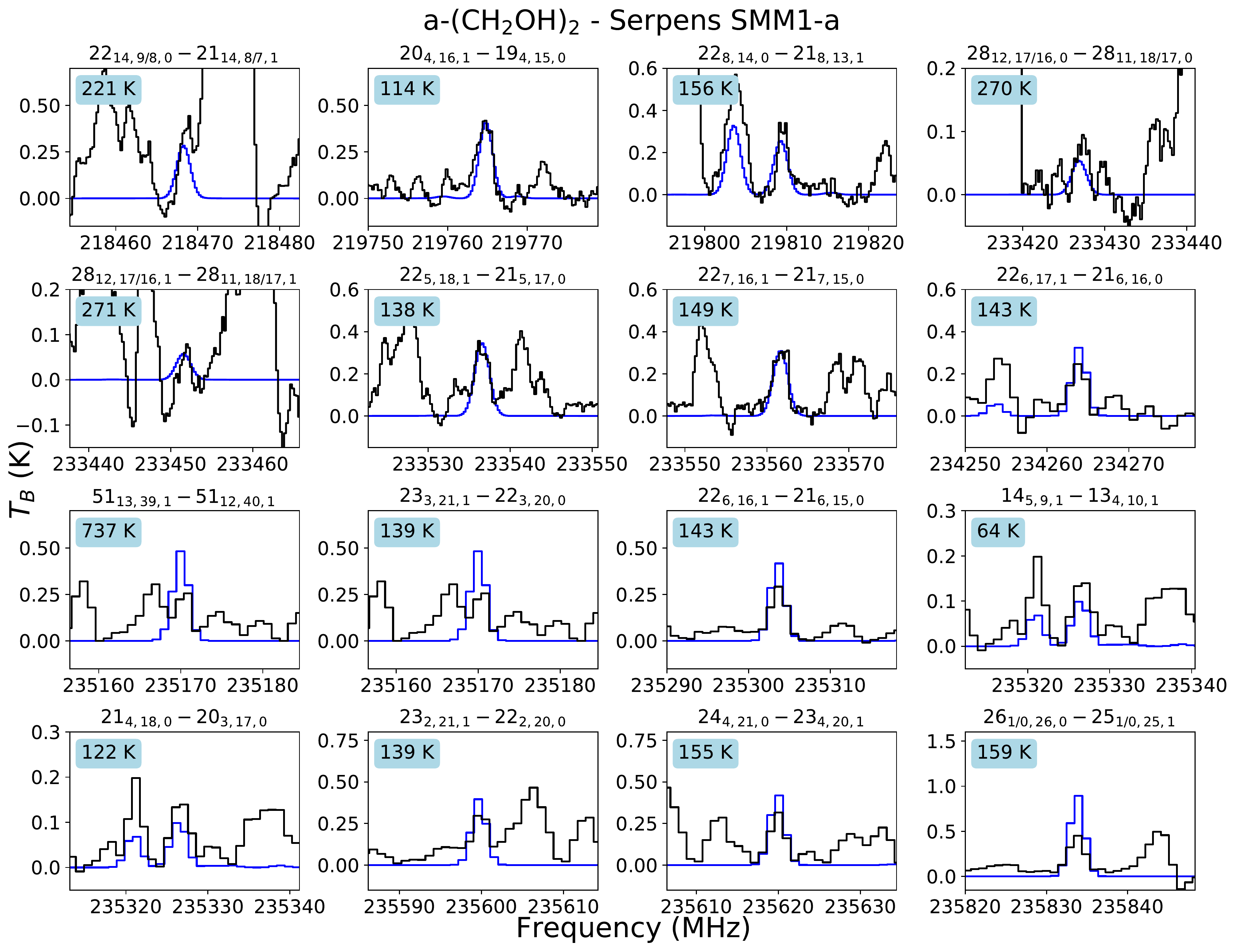}
\caption{Identified lines of a-(CH$_{2}$OH)$_{2}$ towards SMM1. The observed spectrum is plotted in black, with the best-fit synthetic spectrum overplotted in blue ($N_{\rm T}$ = (1.7$\pm$0.5)$\times$10$^{15}$ cm$^{-2}$, $T_{\rm ex}$ = 195$\pm$70~K). The transition is indicated at the top of each panel and the upper state energy is given in the top left of each panel.}
\label{fig:lines_(CH2OH)2}
\end{figure*}

\begin{figure*}
\includegraphics[width=\hsize]{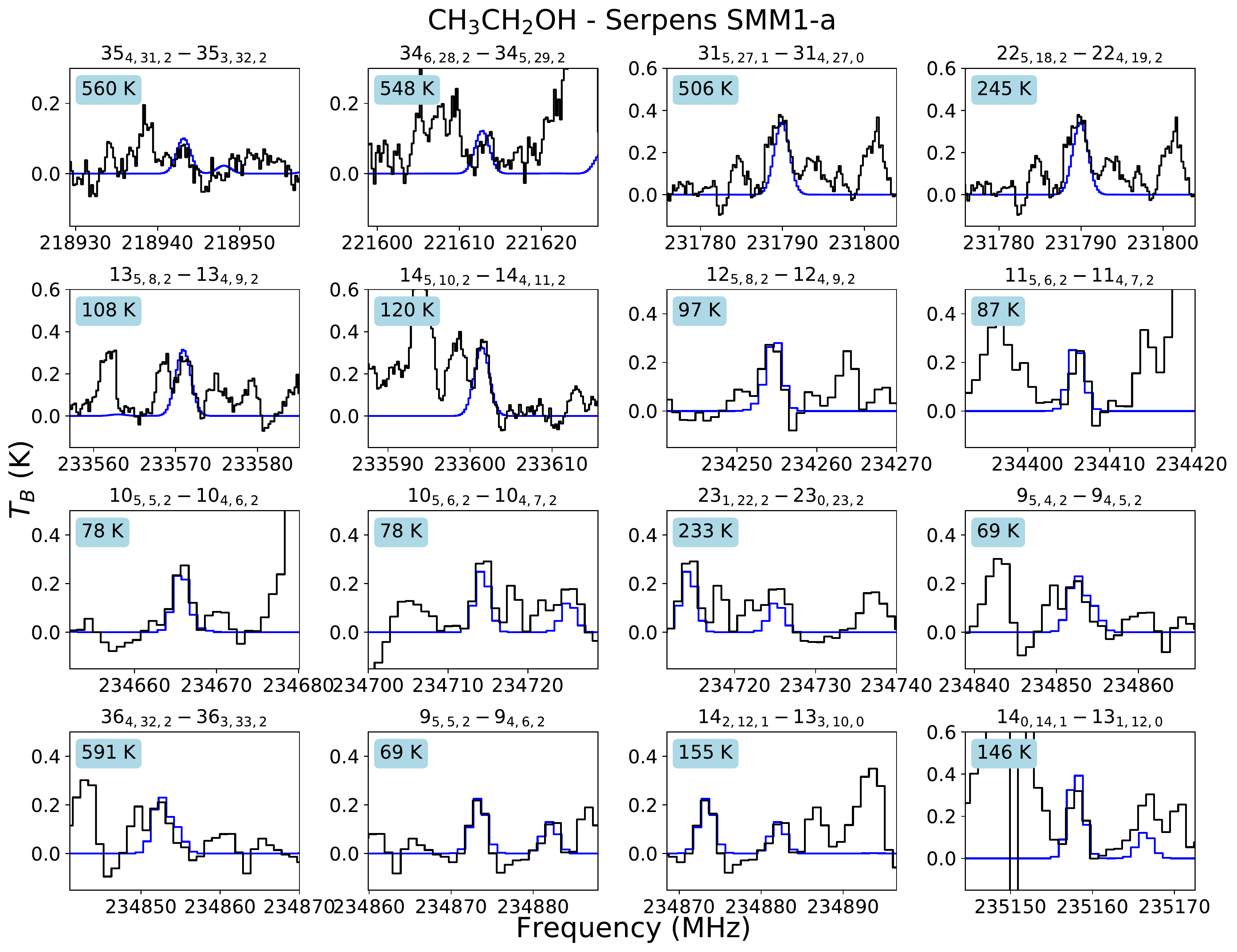}
\caption{Identified lines of CH$_{3}$CH$_{2}$OH towards SMM1. The observed spectrum is plotted in black, with the best-fit synthetic spectrum overplotted in blue ($N_{\rm T}$ = (4.1$\pm$0.9)$\times$10$^{15}$ cm$^{-2}$, $T_{\rm ex}$ = 210$\pm$25~K). The transition is indicated at the top of each panel and the upper state energy is given in the top left of each panel.}
\label{fig:lines_CH3CH2OH}
\end{figure*}

\begin{figure*}
\includegraphics[width=\hsize]{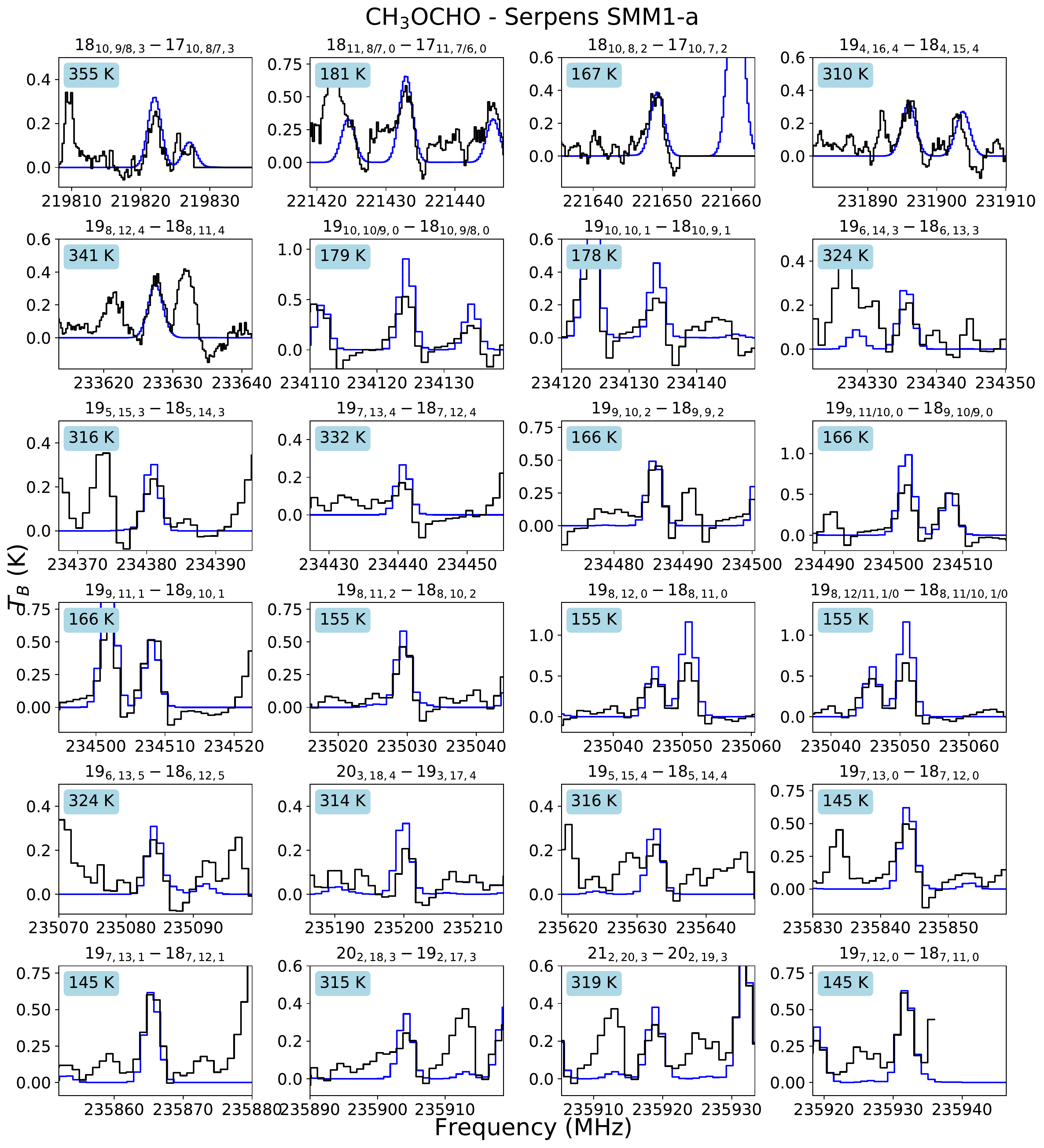}
\caption{Identified lines of CH$_{3}$OCHO towards SMM1. The observed spectrum is plotted in black, with the best-fit synthetic spectrum overplotted in blue ($N_{\rm T}$ = (7.4$\pm$0.7)$\times$10$^{15}$ cm$^{-2}$, $T_{\rm ex}$ = 215$\pm$20~K). The transition is indicated at the top of each panel and the upper state energy is given in the top left of each panel.}
\label{fig:lines_CH3OCHO}
\end{figure*}

\begin{figure*}
\includegraphics[width=\hsize]{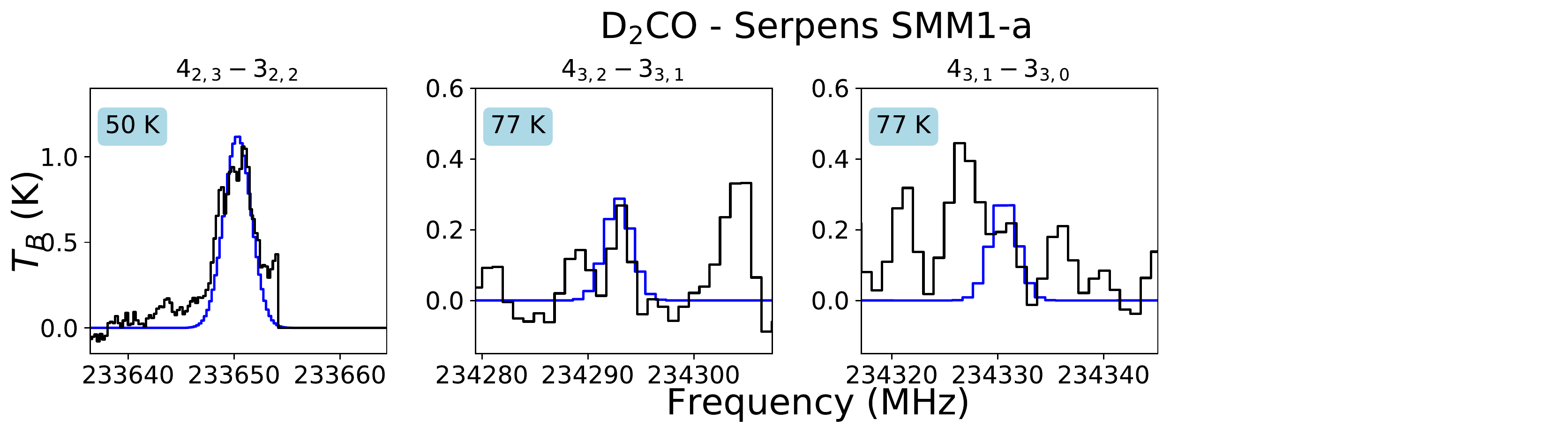}
\caption{Identified lines of D$_{2}$CO towards SMM1. The observed spectrum is plotted in black, with the best-fit synthetic spectrum overplotted in blue ($N_{\rm T}$ = (5.4$\pm$0.5)$\times$10$^{14}$ cm$^{-2}$, $T_{\rm ex}$ = [200]~K). The transition is indicated at the top of each panel and the upper state energy is given in the top left of each panel.}
\label{fig:lines_D2CO}
\end{figure*}

\begin{figure*}
\includegraphics[width=\hsize]{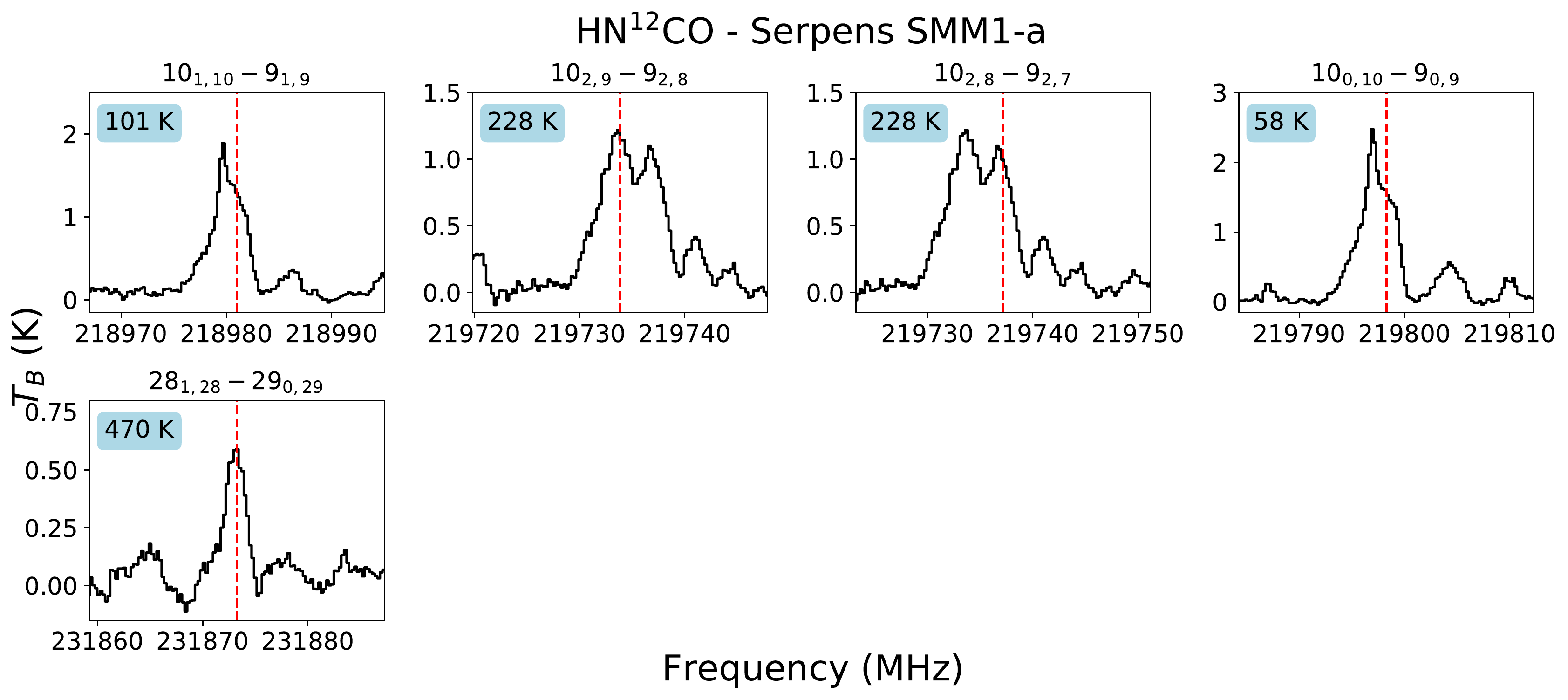}
\caption{Identified lines of HN$^{12}$CO towards SMM1. The observed spectrum is plotted in black and the line position is indicated by the red dotted line. Because these lines are optically thick, not synthetic fit is given. The transition is indicated at the top of each panel and the upper state energy is given in the top left of each panel.}
\label{fig:lines_HN-12-CO}
\end{figure*}

\begin{figure*}
\includegraphics[width=\hsize]{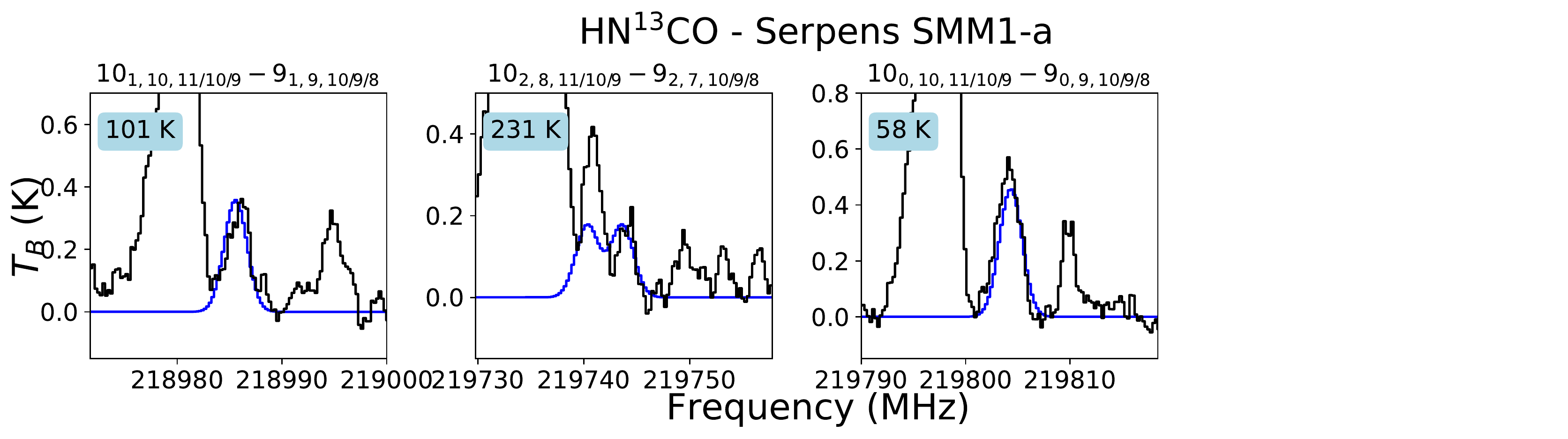}
\caption{Identified lines of HN$^{13}$CO towards SMM1. The observed spectrum is plotted in black, with the best-fit synthetic spectrum overplotted in blue ($N_{\rm T}$ = (1.9$\pm$0.3)$\times$10$^{14}$ cm$^{-2}$, $T_{\rm ex}$ = 190$\pm$30~K). The transition is indicated at the top of each panel and the upper state energy is given in the top left of each panel.}
\label{fig:lines_HN-13-CO}
\end{figure*}

\begin{figure*}
\includegraphics[width=\hsize]{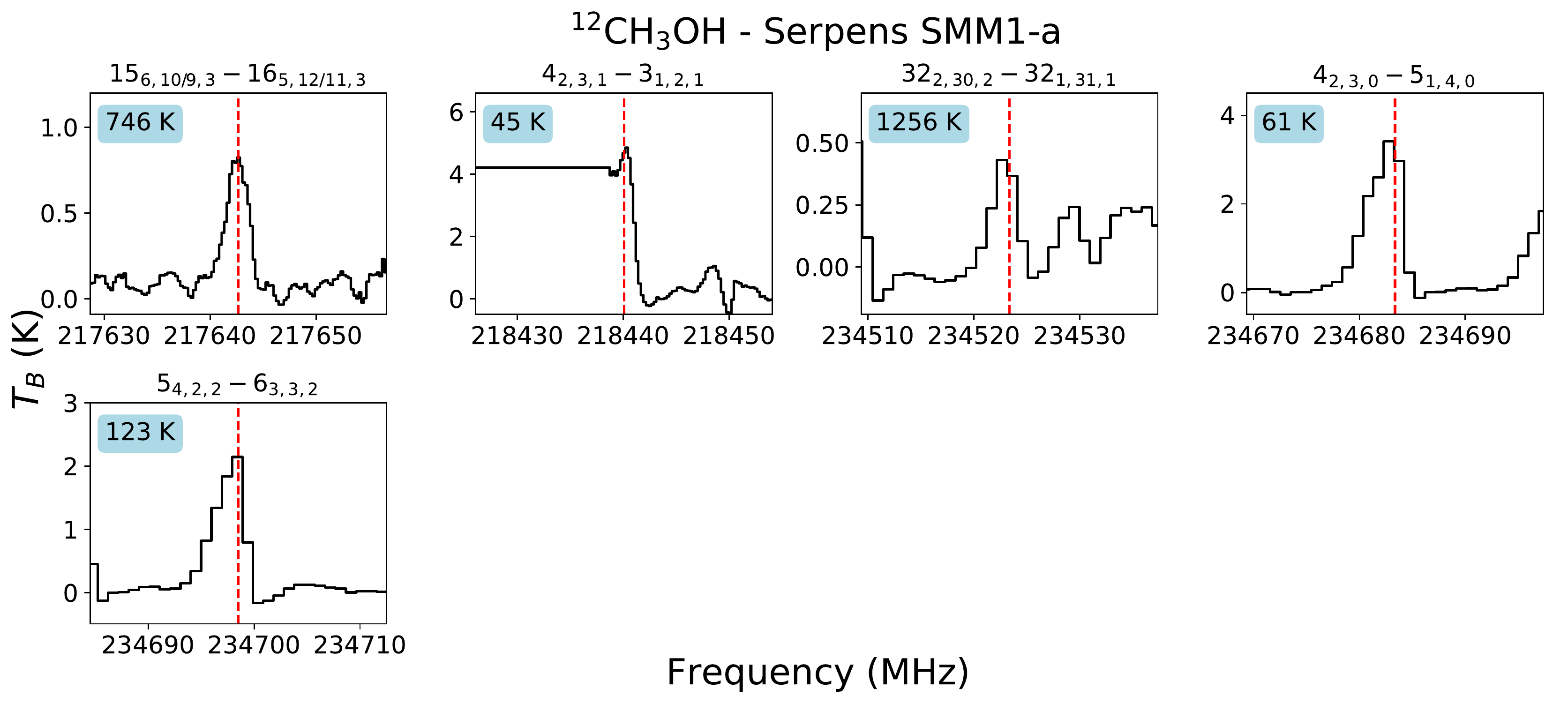}
\caption{Identified lines of $^{12}$CH$_{3}$OH towards SMM1. The observed spectrum is plotted in black and the line position is indicated by the red dotted line. Because these lines are optically thick, not synthetic fit is given. The transition is indicated at the top of each panel and the upper state energy is given in the top left of each panel.}
\label{fig:lines_CH3-12-OH}
\end{figure*}

\begin{figure*}
\includegraphics[width=\hsize]{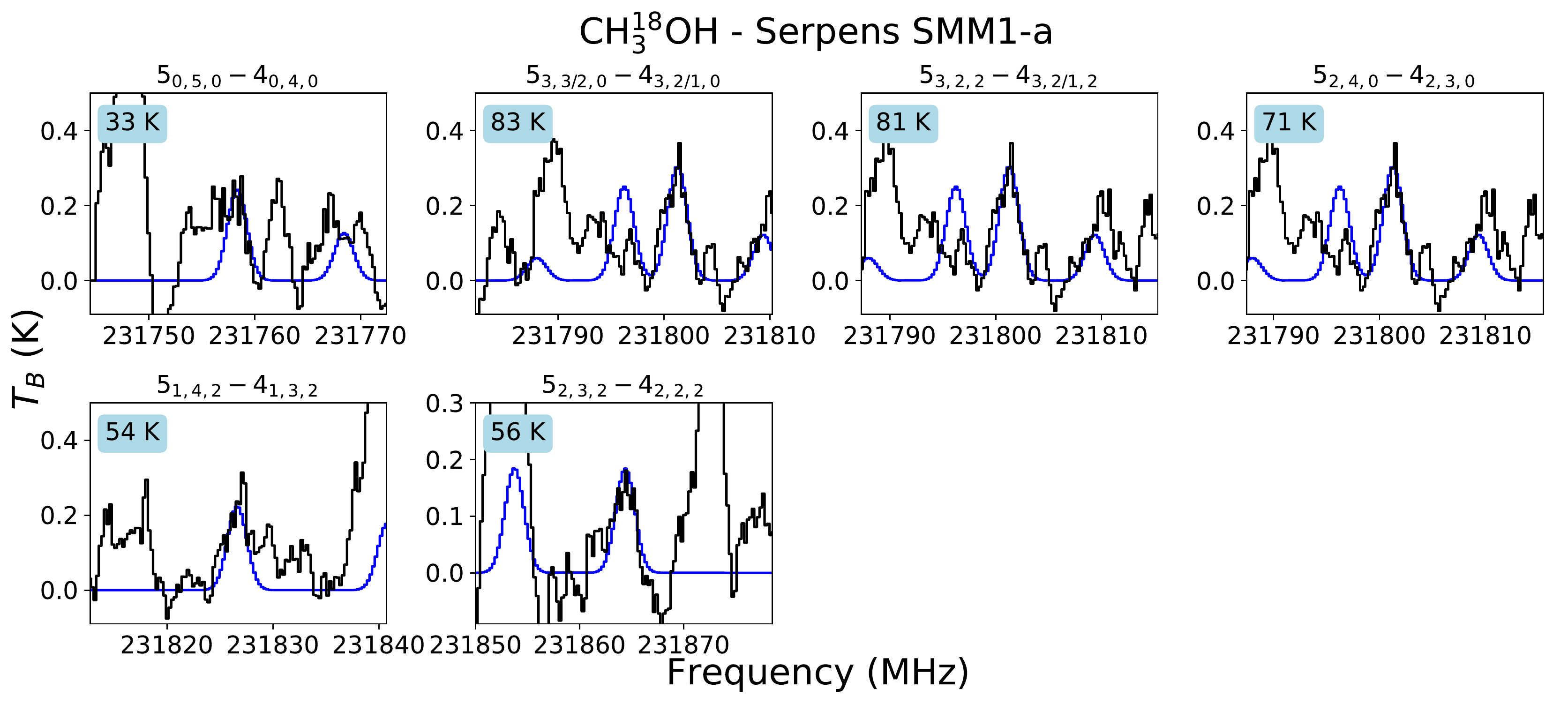}
\caption{Identified lines of CH$_{3}^{18}$OH towards SMM1. The observed spectrum is plotted in black, with the best-fit synthetic spectrum overplotted in blue ($N_{\rm T}$ = (2.0$\pm$0.7)$\times$10$^{15}$ cm$^{-2}$, $T_{\rm ex}$ = 250$\pm$60~K). The transition is indicated at the top of each panel and the upper state energy is given in the top left of each panel.}
\label{fig:lines_CH3-18-OH}
\end{figure*}

\longtab[1]{
\begin{longtable}{l c c l c c}
\caption{Spectral information of molecules detected towards SMM1-a}\\
\hline
\hline
Molecule & Database entry & Transition & Frequency & $E_{\rm up}$ & $A_{\rm ij}$ \\  
& & $J,K_{\rm a},K_{\rm c},(F)$ & (MHz) & (K) & s$^{-1}$ \\
\hline
\endfirsthead
\caption{Continued.} \\
\hline
Molecule & Database entry & Transition & Frequency & $E_{\rm up}$ & $A_{\rm ij}$ \\  
& & $J,K_{\rm a},K_{\rm c},(F)$ & (MHz) & (K) & s$^{-1}$ \\
\hline
\endhead
\hline
\endfoot
\hline
\endlastfoot
D$_{2}$CO & 32502 &  4 2 3 - 3 2 2 & 233 650.441 (0.0500) & 49.63 & 2.69$\times$10$^{-4}$ \\
& CDMS & 4 3 2 - 3 3 1 & 234 293.361 (0.0500) & 76.62 & 1.58$\times$10$^{-4}$ \\
& & 4 3 1 - 3 3 0 & 234 331.062 (0.0500) & 76.63 & 1.58$\times$10$^{-4}$ \\
\hline
$^{12}$CH$_{3}$OH & 32504 & 15 6 9 3 - 16 5 11 3  & 217 642.677 (0.0220) & 746 & 1.89$\times$10$^{-5}$ \\
 & CDMS & 15 6 10 3 - 16 5 12 3 & 217 642.678 (0.0220) & 746 & 1.89$\times$10$^{-5}$ \\
 & & 4 2 3 1 - 3 1 2 1 & 218 440.063 (0.0130) & 45 & 5.69$\times$10$^{-5}$ \\
 & & 32 2 30 2 - 32 1 31 1  & 234 523.365 (0.1070) & 1256 & 8.95$\times$10$^{-5}$ \\
 & & 4 2 3 0 - 5 1 4 0 & 234 683.370 (0.0120) & 61 & 1.87$\times$10$^{-5}$ \\
 & & 5 4 2 2 - 6 3 3 2 & 234 698.519 (0.0150) & 123 & 6.34$\times$10$^{-6}$ \\
 \hline
CH$_{3}^{18}$OH & 34504 & 5 0 5 0 - 4 0 4 0 & 231 758.446 (0.0300) & 33 & 5.33$\times$10$^{-5}$ \\
 & CDMS & 5 3 3 0 - 4 3 2 0 & 231 796.218 (0.0300) & 83 & 3.41$\times$10$^{-5}$ \\
 & & 5 3 2 0 - 4 3 1 0 & 231 796.521 (0.0300) & 83 & 3.41$\times$10$^{-5}$ \\
 & & 5 3 2 2 - 4 3 1 2 & 231 801.304 (0.0300) & 81 & 3.42$\times$10$^{-5}$ \\ 
 & & 5 2 4 0 - 4 2 3 0 & 231 801.466 (0.0300) & 71 & 4.53$\times$10$^{-5}$ \\
 & & 5 1 4 2 - 4 1 3 2 & 231 826.744 (0.0300) & 54 & 5.33$\times$10$^{-5}$ \\
 & & 5 2 3 2 - 4 2 2 2 & 231 864.501 (0.0300) & 56 & 4.41$\times$10$^{-5}$ \\
\hline
CH$_{3}$CN,~$\nu_{8}$=1 & 41509 & 12 2 2 - 11 -2 2 & 221 367.450 (0.0011) & 649 & 8.98$\times$10$^{-4}$ \\
& CDMS & 12 -2 2 - 11 2 2 & 221 367.450 (0.0011) & 649 & 8.98$\times$10$^{-4}$ \\
& & 12 4 3 - 11 -4 3 & 221 380.608 (0.0011) & 655 & 8.21$\times$10$^{-4}$ \\
& & 12 -4 3 - 11 4 3 & 221 380.608 (0.0011) & 655 & 8.21$\times$10$^{-4}$ \\
& & 12 1 2 - 11 1 2 & 221 387.271 (0.0011) & 615 & 9.17$\times$10$^{-4}$ \\
& & 12 0 2 - 11 0 2 & 221 394.085 (0.0011) & 594 & 9.24$\times$10$^{-4}$ \\
& & 12 3 3 - 11 3 3 & 221 403.521 (0.0011) & 619 & 8.66$\times$10$^{-4}$ \\
& & 12 1 3 - 11 -1 3 & 221 625.840 (0.0011) & 588 & 9.20$\times$10$^{-4}$ \\
\hline
NH$_{2}$CN & 42003 & 11 1 11 0 - 10 1 10 0 & 218 461.795 (0.0200) & 77 & 1.08$\times$10$^{-3}$ \\
& JPL & 11 0 11 1 - 10 0 10 1 & 219 719.651 (0.0200) & 135 & 1.08$\times$10$^{-3}$ \\
& & 11 1 10 0 - 10 1 9 0 & 221 361.160 (0.0200) & 78 & 1.12$\times$10$^{-3}$ \\
\hline
HN$^{12}$CO & 43511 & 10 1 10 - 9 1 9 & 218 985.696 (0.0192) & 101 & 1.48$\times$10$^{-4}$ \\
 & CDMS & 10 2 9 - 9 2 8 & 219 733.850 (0.0300) & 228 & 1.35$\times$10$^{-4}$ \\
 & & 10 2 8 - 9 2 7 & 219 737.193 (0.0300) & 228 & 1.35$\times$10$^{-4}$ \\
 & & 10 0 10 - 9 0 9 & 219 798.274 (0.0040) & 58 & 1.47$\times$10$^{-4}$ \\
 & & 28 1 28 - 29 0 29 & 231 873.255	(0.0064) & 470 & 6.68$\times$10$^{-5}$ \\
\hline
HN$^{13}$CO & 44008 & 10 1 10 9 - 9 1 9 9 & 218 984.716 (0.1053) & 101 & 1.63$\times$10$^{-6}$ \\
 & & 10 1 10 11 - 9 1 9 10 & 218 985.697 (0.0192) & 101 & 1.48$\times$10$^{-4}$ \\
 & & 10 1 10 10 - 9 1 9 9 & 218 985.705 (0.0191) & 101 & 1.46$\times$10$^{-4}$ \\
 & & 10 1 10 9 - 9 1 9 8 & 218 985.706 (0.0191) & 101 & 1.46$\times$10$^{-4}$ \\
 & & 10 1 10 10 - 9 1 9 10 & 218 986.596 (0.0191) & 101 & 1.48$\times$10$^{-6}$ \\
 & & 10 1 10 10 - 9 1 9 10 & 218 986.596 (0.0191) & 101 & 1.48$\times$10$^{-6}$ \\
 & & 10 2 9 9 - 9 2 8 9 & 219 739.762 (0.0962) & 231 & 1.60$\times$10$^{-6}$ \\
 & & 10 2 9 11 - 9 2 8 10 & 219 740.451 (0.0274) & 231 & 1.45$\times$10$^{-4}$ \\
 & & 10 2 9 9 - 9 2 8 8 & 219 740.456 (0.0274) & 231 & 1.43$\times$10$^{-4}$ \\
 & & 10 2 9 10 - 9 2 8 9 & 219 740.471 (0.0274) & 231 & 1.43$\times$10$^{-4}$ \\
 & & 10 2 9 10 - 9 2 8 10 & 219 741.095 (0.0885) & 231 & 1.45$\times$10$^{-6}$ \\
 & & 10 2 8 9 - 9 2 7 9 & 219 743.054 (0.0966) & 231 & 1.60$\times$10$^{-6}$ \\
 & & 10 2 8 11 - 9 2 7 10 & 219 743.742 (0.0288) & 231 & 1.45$\times$10$^{-4}$ \\
 & & 10 2 8 9 - 9 2 7 8 & 219 743.747 (0.0288) & 231 & 1.43$\times$10$^{-4}$ \\
 & & 10 2 8 10 - 9 2 7 9 & 219 743.762 (0.0288) & 231 & 1.43$\times$10$^{-4}$ \\
 & & 10 2 8 10 - 9 2 7 10 & 219 744.386 (0.0890) & 231 & 1.45$\times$10$^{-6}$ \\
 & & 10 0 10 9 - 9 0 9 9 & 219 803.645 (0.1070) & 58 & 1.67$\times$10$^{-6}$ \\
 & & 10 0 10 11 - 9 0 9 10 & 21 9804.439 (0.0171) & 58 & 1.51$\times$10$^{-4}$ \\
 & & 10 0 10 10 - 9 0 9 9 & 219 804.442 (0.0171) & 58 & 1.50$\times$10$^{-4}$ \\
 & & 10 0 10 9 - 9 0 9 8 & 219 804.446 (0.0171) & 58 & 1.49$\times$10$^{-4}$ \\
 & & 10 0 10 10 - 9 0 9 10 & 219 805.163 (0.0965) & 58 & 1.51$\times$10$^{-6}$ \\ 
\hline
CH$_{3}$CH$_{2}$OH & 46524 & 35 4 31 2 - 35 3 32 2 & 218 943.289	(0.0068) & 560 & 7.38$\times$10$^{-5}$ \\
& CDMS & 34 6 28 2 - 34 5 29 2 & 221 612.904	(0.0080) & 548 & 8.84$\times$10$^{-5}$ \\
& & 31 5 27 1 - 31 4 27 0 & 231 789.850	(0.0120) & 506 & 2.24$\times$10$^{-5}$ \\
& & 22 5 18 2 - 22 4 19 2 & 231 790.056 (0.0035) & 245 & 8.46$\times$10$^{-5}$ \\
& & 13 5 8 2 - 13 4 9 2 & 233 571.051 (0.0032) & 108 & 7.54$\times$10$^{-5}$ \\
& & 14 5 10 2 - 14 4 11 2 & 233 601.554 (0.0032) & 120 & 7.71$\times$10$^{-5}$ \\
& & 12 5 8 2 - 12 4 9 2 & 234 255.240 (0.0032) & 97 & 7.39$\times$10$^{-5}$ \\
& & 11 5 6 2 - 11 4 7 2 & 234 406.433 (0.0033) & 87 & 7.17$\times$10$^{-5}$ \\
& & 10 5 5 2 - 10 4 6 2 & 234 666.142 (0.0035) & 78 & 6.89$\times$10$^{-5}$ \\
& & 10 5 6 2 - 10 4 7 2 & 234 714.782 (0.0035) & 78 & 6.90$\times$10$^{-5}$ \\
& & 23 1 22 2 - 23 0 23 2 & 234 725.620	(0.0090) & 233 & 3.17$\times$10$^{-5}$ \\
& & 9 5 4 2 - 9 4 5 2 & 234 852.862 (0.0038) & 69 & 6.53$\times$10$^{-5}$ \\
& & 36 4 32 2 - 36 3 33 2 & 234 855.010 (0.0073) & 591 & 8.70$\times$10$^{-5}$ \\
& & 9 5 5 2 - 9 4 6 2 & 234 873.873 (0.0038) & 69 & 6.54$\times$10$^{-5}$ \\
& & 14 2 12 1 - 13 3 10 0 & 234 882.537 (0.0105) & 155 & 3.64$\times$10$^{-5}$ \\
& & 14 0 14 1 - 13 1 12 0 & 235 158.494 (0.0058) & 146 & 1.12$\times$10$^{-4}$ \\
\hline
CH$_{3}$NCO, $\nu$=0 & 57505 & 25 -1 0 2 - 24 -1 0 2 & 217 595.174 (0.0500) & 194 & 4.84$\times$10$^{-4}$ \\
& CDMS & 25 0 0 2 - 24 0 0 2 & 217 595.174 (0.0500) & 188 & 4.85$\times$10$^{-4}$ \\
& & 25 2 0 1 - 24 2 0 1 & 217 652.088 (0.0500) & 171 & 4.82$\times$10$^{-4}$ \\
& & 24 3 0 1 - 23 3 0 1 & 217 701.086 (0.0500) & 191 & 4.40$\times$10$^{-4}$ \\
& & 25 1 0 -3 - 24 1 0 -3 & 218 002.461 (0.0500) & 258 & 4.93$\times$10$^{-4}$ \\
& & 25 0 0 -3 - 24 0 0 -3 & 218 014.630 (0.0500) & 251 & 4.84$\times$10$^{-4}$ \\
& & 25 1 0 3 - 24 1 0 3 & 218 069.900 (0.0500) & 257 & 4.93$\times$10$^{-4}$ \\
& & 25 1 24 0 - 24 1 23 0 & 218 541.803 (0.0500) & 142 & 4.94$\times$10$^{-4}$ \\
& & 27 -1 0 1 - 26 -1 0 1 & 231 793.783 (0.0500) & 175 & 6.02$\times$10$^{-4}$ \\
& & 27 2 26 0 - 26 2 25 0 & 234 088.125 (0.0500) & 181 & 6.05$\times$10$^{-4}$ \\
& & 27 0 0 2 - 26 0 0 2 & 234 932.492 (0.0500) & 210 & 6.11$\times$10$^{-4}$ \\
& & 27 -3 0 2 - 26 -3 0 2 & 235 663.096 (0.0500) & 264 & 6.06$\times$10$^{-4}$ \\
& & 27 2 0 3 - 26 2 0 3 & 235 801.163 (0.0500) & 296 & 6.12$\times$10$^{-4}$ \\
& & 27 2 0 -3 - 26 2 0 -3 & 235 803.211 (0.0500) & 297 & 6.12$\times$10$^{-4}$ \\
\hline
HOCH$_{2}$CN & 57512 & 24 2 23 1 - 23 2 22 1 & 218 994.156	(0.0009) & 143 & 3.23$\times$10$^{-4}$ \\
& CDMS & 24 8 17 1 - 23 8 16 1 & 221 334.546	(0.0009) & 227 & 2.98$\times$10$^{-4}$ \\
& & 24 8 16 1 - 23 8 15 1 & 221 334.546	(0.0009) & 227 & 2.98$\times$10$^{-4}$ \\
& & 24 9 15 1 - 23 9 14 1 & 221 344.331	(0.0010) & 251 & 2.88$\times$10$^{-4}$ \\
& & 24 9 16 1 - 23 9 15 1 & 221 344.331	(0.0010) & 251 & 2.88$\times$10$^{-4}$ \\
& & 24 10 14 1 - 23 10 13 1 & 221 372.125	(0.0010) & 277 & 2.77$\times$10$^{-4}$ \\
& & 24 10 15 1 - 23 10 14 1 & 221 372.125	(0.0010) & 277 & 2.77$\times$10$^{-4}$ \\
& & 24 6 19 1 - 23 6 18 1 & 221 406.933	(0.0009) & 188 & 3.15$\times$10$^{-4}$ \\
& & 24 6 18 1 - 23 6 17 1 & 221 407.170	(0.0009) & 188 & 3.15$\times$10$^{-4}$ \\
& & 24 11 13 1 - 23 11 12 1 & 221 413.638	(0.0010) & 306 & 2.65$\times$10$^{-4}$ \\
& & 24 11 14 1 - 23 11 13 1 & 221 413.638	(0.0010) & 306 & 2.65$\times$10$^{-4}$ \\
& & 24 12 12 1 - 23 12 11 1 & 221 466.304	(0.0011) & 338 & 2.52$\times$10$^{-4}$ \\
& & 24 12 13 1 - 23 12 12 1 & 221 466.304	(0.0011) & 338 & 2.52$\times$10$^{-4}$ \\
& & 24 3 22 1 - 23 3 21 1 & 221 466.604	(0.0009) & 151 & 3.32$\times$10$^{-4}$ \\
& & 24 8 17 0 - 23 8 16 0 & 221 480.199	(0.0009) & 222 & 2.95$\times$10$^{-4}$ \\
& & 24 8 16 0 - 23 8 15 0 & 221 480.199	(0.0009) & 222 & 2.95$\times$10$^{-4}$ \\
& & 24 9 15 0 - 23 9 14 0 & 221 488.785	(0.0010) & 245 & 2.86$\times$10$^{-4}$ \\
& & 24 9 16 0 - 23 9 15 0 & 221 488.785	(0.0010) & 245 & 2.86$\times$10$^{-4}$ \\
& & 24 7 18 0 - 23 7 17 0 & 221 497.997	(0.0009) & 201 & 3.04$\times$10$^{-4}$ \\
& & 24 7 17 0 - 23 7 16 0 & 221 498.002	(0.0009) & 201 & 3.04$\times$10$^{-4}$ \\
& & 24 10 14 0 - 23 10 13 0 & 221 515.800	(0.0010) & 272 & 2.75$\times$10$^{-4}$ \\
& & 24 10 15 0 - 23 10 14 0 & 221 515.800	(0.0010) & 272 & 2.75$\times$10$^{-4}$ \\
& & 24 13 11 1 - 23 13 10 1 & 221 528.501	(0.0011) & 372 & 2.38$\times$10$^{-4}$ \\
& & 24 13 12 1 - 23 13 11 1 & 221 528.501	(0.0011) & 372 & 2.38$\times$10$^{-4}$ \\
& & 24 5 20 1 - 23 5 19 1 & 221 533.058	(0.0009) & 173 & 3.22$\times$10$^{-4}$ \\
& & 24 5 19 1 - 23 5 18 1 & 221 542.097	(0.0009) & 173 & 3.22$\times$10$^{-4}$ \\
& & 24 11 13 0 - 23 11 12 0 & 221 556.802	(0.0010) & 301 & 2.63$\times$10$^{-4}$ \\
& & 24 11 14 0 - 23 11 13 0 & 221 556.802	(0.0010) & 301 & 2.63$\times$10$^{-4}$ \\
& & 24 6 19 0 - 23 6 18 0 & 221 557.660	(0.0009) & 183 & 3.12$\times$10$^{-4}$ \\
& & 24 6 18 0 - 23 6 17 0 & 221 557.911	(0.0009) & 183 & 3.12$\times$10$^{-4}$ \\
& & 24 12 12 0 - 23 12 11 0 & 221 609.140	(0.0010) & 332 & 2.50$\times$10$^{-4}$ \\
& & 24 12 13 0 - 23 12 12 0 & 221 609.140	(0.0010) & 332 & 2.50$\times$10$^{-4}$ \\
& & 25 2 23 1 - 24 2 22 1 & 234 584.932	(0.0300) & 157 & 4.36$\times$10$^{-4}$ \\
& & 25 2 23 0 - 24 2 22 0 & 235 112.110	(0.0300) & 152 & 3.94$\times$10$^{-4}$ \\
\hline
CH$_{3}$OCHO & 60003 & 18 10 9 3 - 17 10 8 3 & 219 822.126 (0.1000) & 355 & 1.11$\times$10$^{-4}$ \\
 & JPL & 18 10 8 3 - 17 10 7 3 & 219 822.126 (0.1000) & 355 & 1.11$\times$10$^{-4}$ \\
 & & 18 11 7 0 - 17 11 6 0 & 221 433.019 (0.1000) & 181 & 1.03$\times$10$^{-4}$ \\
 & & 18 11 8 0 - 17 11 7 0 & 221 433.019 (0.1000) & 181 & 1.03$\times$10$^{-4}$ \\
 & & 18 10 8 2 - 17 10 7 2 & 221 649.411 (0.1000) & 167 & 1.14$\times$10$^{-4}$ \\
 & & 19 4 16 4 - 18 4 15 4 & 231 896.060 (0.1000) & 310 & 1.79$\times$10$^{-4}$ \\
 & & 19 8 12 4 - 18 8 11 4 & 233 627.478 (0.1000) & 341 & 1.59$\times$10$^{-4}$ \\
 & & 19 10 10 0 - 18 10 9 0 & 234 124.883 (0.1000) & 179 & 1.40$\times$10$^{-4}$ \\
 & & 19 10 9 0 - 18 10 8 0 & 234 124.883 (0.1000) & 179 & 1.40$\times$10$^{-4}$ \\
 & & 19 10 10 1 - 18 10 9 1 & 234 134.600 (0.0500) & 178 & 1.40$\times$10$^{-4}$ \\
 & & 19 6 14 3 - 18 6 13 3 & 234 336.107 (0.1000) & 324 & 1.75$\times$10$^{-4}$ \\
 & & 19 5 15 3 - 18 5 14 3 & 234 381.269 (0.1000) & 316 & 1.80$\times$10$^{-4}$ \\
 & & 19 7 13 4 - 18 7 12 4 & 234 441.264 (0.1000) & 332 & 1.68$\times$10$^{-4}$ \\
 & & 19 9 10 2 - 18 9 9 2 & 234 486.395 (0.1000) & 166 & 1.51$\times$10$^{-4}$ \\
 & & 19 9 11 0 - 18 9 10 0 & 234 502.241	(0.0009) & 166 & 1.51$\times$10$^{-4}$ \\
 & & 19 9 10 0 - 18 9 9 0 & 234 502.432	(0.0009) & 166 & 1.51$\times$10$^{-4}$ \\
 & & 19 9 11 1 - 18 9 10 1 & 234 508.614 (0.1000) & 166 & 1.51$\times$10$^{-4}$ \\
 & & 19 8 11 2 - 18 8 10 2 & 235 029.952 (0.1000) & 155 & 1.61$\times$10$^{-4}$ \\
 & & 19 8 12 0 - 18 8 11 0 & 235 046.493 (0.1000) & 155 & 1.61$\times$10$^{-4}$ \\
 & & 19 8 11 0 - 18 8 10 0 & 235 051.378 (0.1000) & 155 & 1.61$\times$10$^{-4}$ \\
 & & 19 8 12 1 - 18 8 11 1 & 235 051.378 (0.1000) & 155 & 1.61$\times$10$^{-4}$ \\
 & & 19 6 13 5 - 18 6 12 5 & 235 084.738 (0.1000) & 324 & 1.76$\times$10$^{-4}$ \\
 & & 20 3 18 4 - 19 3 17 4 & 235 200.422 (0.1000) & 314 & 1.90$\times$10$^{-4}$ \\
 & & 19 5 15 4 - 18 5 14 4 & 235 633.058 (0.1000) & 316 & 1.81$\times$10$^{-4}$ \\
 & & 19 7 13 0 - 18 7 12 0 & 235 844.544 (0.1000) & 145 & 1.71$\times$10$^{-4}$ \\
 & & 19 7 13 1 - 18 7 12 1 & 235 865.969 (0.1000) & 145 & 1.67$\times$10$^{-4}$ \\
 & & 20 2 18 3 - 19 2 17 3 & 235 904.655 (0.1000) & 315 & 1.91$\times$10$^{-4}$ \\
 & & 21 2 20 3 - 20 2 19 3 & 235 919.352 (0.1000) & 319 & 1.94$\times$10$^{-4}$ \\
 & & 19 7 12 0 - 18 7 11 0 & 235 932.379 (0.1000) & 145 & 1.72$\times$10$^{-4}$ \\
\hline
a-(CH$_{2}$OH)$_{2}$ & 62503 & 22 14 8 0 - 21 14 7 1 & 218 468.381 (0.0044) & 221 & 1.51$\times$10$^{-4}$ \\
& CDMS & 22 14 9 0 - 21 14 8 1 & 218 468.381 (0.0044) & 221 & 1.51$\times$10$^{-4}$ \\
& & 20 4 16 1 - 19 4 15 0 & 219 764.925 (0.0040) & 114 & 2.45$\times$10$^{-4}$ \\
& & 22 8 14 0 - 21 8 13 1 & 219 809.406 (0.0026) & 156 & 2.24$\times$10$^{-4}$ \\
& & 28 12 16 0 - 28 11 17 0 & 233 426.949 (0.0036) & 270 & 3.11$\times$10$^{-5}$ \\
& & 28 12 17 0 - 28 11 18 0 & 233 427.018 (0.0036) & 270 & 3.10$\times$10$^{-5}$ \\
& & 28 12 16 1 - 28 11 17 1 & 233 451.651 (0.0036) & 271 & 3.32$\times$10$^{-5}$ \\
& & 28 12 17 1 - 28 11 18 1 & 233 451.723 (0.0036) & 271 & 3.32$\times$10$^{-5}$ \\
& & 22 5 18 1 - 21 5 17 0 & 233 536.696 (0.0025) & 138 & 2.93$\times$10$^{-4}$ \\
& & 22 7 16 1 - 21 7 15 0 & 233 561.784 (0.0024) & 149 & 2.79$\times$10$^{-4}$ \\
& & 22 6 17 1 - 21 6 16 0 & 234 264.446 (0.0026) & 143 & 2.84$\times$10$^{-4}$ \\
& & 51 13 39 1 - 51 12 40 1 & 235 170.476 (0.0288) & 737 & 4.73$\times$10$^{-5}$ \\
& & 23 3 21 1 - 22 3 20 0 & 235 170.573 (0.0023) & 139 & 2.91$\times$10$^{-4}$ \\
& & 22 6 16 1 - 21 6 15 0 & 235 304.050 (0.0026) & 143 & 2.90$\times$10$^{-4}$ \\
& & 14 5 9 1 - 13 4 10 1 & 235 326.413	(0.0021) & 64 & 2.39$\times$10$^{-5}$ \\
& & 21 4 18 0 - 20 3 17 0 & 235 327.161	(0.0027) & 122 & 6.30$\times$10$^{-5}$ \\
& & 23 2 21 1 - 22 2 20 0 & 235 600.178 (0.0022) & 139 & 3.28$\times$10$^{-4}$ \\
& & 24 4 21 0 - 23 4 20 1 & 235 620.372 (0.0027) & 155 & 2.88$\times$10$^{-4}$ \\
& & 26 1 26 0 - 25 1 25 1 & 235 834.239 (0.0040) & 159 & 3.22$\times$10$^{-4}$ \\
& & 26 0 26 0 - 25 0 25 1 & 235 834.327 (0.0040) & 159 & 3.22$\times$10$^{-4}$ \\
\end{longtable}
\label{tab:SMM1_lines}
}

\section{Analysis of PILS data}
\label{ap:PILS}

\longtab[1]{
\begin{longtable}{l c c c c c}
\caption{Molecules and parameters used for the IRAS~16293B synthetic spectrum}\\
\hline
\hline
Molecule	&	Name	&	Tag	&	Database	&	$N_{\rm T}$	&	$T_{\rm ex}$	\\
	&		&		&		&	(cm$^{-2}$)	&	(K)	\\
\hline
\endfirsthead
\caption{Continued.} \\
\hline
Molecule	&	Name	&	Tag	&	Database	&	$N_{\rm T}$	&	$T_{\rm ex}$	\\
	&		&		&		&	(cm$^{-2}$)	&	(K)	\\
\hline
\endhead
\hline
\endfoot
\hline
\endlastfoot
CCH	&	Ethynyl radical	&	25501	&	CDMS	&	3.00$\times$10$^{13}$	&	120	\\
HCN	&	Hydrogen cyanide	&	27501	&	CDMS	&	5.00$\times$10$^{16}$	&	120	\\
HNC	&	Hydrogen isocyanide	&	27502	&	CDMS	&	5.00$\times$10$^{16}$	&	120	\\
H$^{13}$CN	&	Hydrogen cyanide	&	28501	&	CDMS	&	2.00$\times$10$^{14}$	&	300	\\
CO	&	Carbon monoxide	&	28503	&	CDMS	&	1.00$\times$10$^{20}$	&	100	\\
HC$^{15}$N	&	Hydrogen cyanide	&	28506	&	CDMS	&	2.00$\times$10$^{14}$	&	300	\\
DNC	&	Hydrogen cyanide	&	28508	&	CDMS	&	7.00$\times$10$^{14}$	&	300	\\
$^{13}$CO	&	Carbon monoxide	&	29501	&	CDMS	&	3.10$\times$10$^{19}$	&	100	\\
C$^{17}$O	&	Carbon monoxide	&	29503	&	CDMS	&	8.00$\times$10$^{16}$	&	100	\\
H$^{13}$C$^{15}$	&	Hydrogen cyanide	&	29512	&	CDMS	&	2.00$\times$10$^{14}$	&	300	\\
HNCH$_{2}$	&	Methanimine	&	29518	&	CDMS	&	8.00$\times$10$^{14}$	&	100	\\
NO	&	Nitrogen oxide	&	30008	&	JPL	&	2.00$\times$10$^{16}$	&	100	\\
H$_{2}$CO	&	Formaldehyde	&	30501	&	CDMS	&	1.80$\times$10$^{18}$	&	105	\\
C$^{18}$O	&	Carbon monoxide	&	30502	&	CDMS	&	1.00$\times$10$^{17}$	&	100	\\
DCO$^{+}$	&	Formyl radical	&	30510	&	CDMS	&	3.00$\times$10$^{12}$	&	29	\\
CH$_{3}$NH$_{2}$	&	Methylamine	&	31008	&	JPL	&	5.30$\times$10$^{14}$&	100	\\
HDCO	&	Formaldehyde	&	31501	&	CDMS	&	1.30$\times$10$^{17}$	&	105	\\
H$_{2}^{13}$CO	&	Formaldehyde	&	31503	&	CDMS	&	3.60$\times$10$^{16}$	&	105	\\
D$_{2}$CO	&	Formaldehyde	&	32502	&	CDMS	&	1.60$\times$10$^{16}$	&	105	\\
H$_{2}$C$^{18}$O	&	Formaldehyde	&	32503	&	CDMS	&	2.50$\times$10$^{15}$	&	105	\\
CH$_{3}$OH	&	Methanol	&	32504	&	CDMS	&	2.00$\times$10$^{19}$	&	300	\\
CH$_{2}$DOH	&	Methanol	&	33004	&	JPL	&	7.10$\times$10$^{17}$	&	300	\\
$^{13}$CH$_{3}$OH	&	Methanol	&	33502	&	CDMS	&	4.00$\times$10$^{16}$	&	300	\\
NH$_{2}$OH	&	Hydroxylamine	&	33503	&	CDMS	&	3.70$\times$10$^{14}$	&	100	\\
D$_{2}^{13}$CO	&	Formaldehyde	&	33506	&	CDMS	&	2.20$\times$10$^{14}$	&	105	\\
HDC$^{18}$O	&	Formaldehyde	&	33510	&	CDMS	&	1.40$\times$10$^{14}$	&	105	\\
H$_{2}$S	&	Hydrogen sulfide	&	34502	&	CDMS	&	1.00$\times$10$^{18}$	&	125	\\
CH$_{3}^{18}$OH	&	Methanol	&	34504	&	CDMS	&	2.00$\times$10$^{16}$	&	300	\\
HDS	&	Hydrogen sulfide	&	35502	&	CDMS	&	2.00$\times$10$^{16}$	&	125	\\
HD$^{34}$S	&	Hydrogen sulfide	&	37503	&	CDMS	&	1.00$\times$10$^{15}$	&	125	\\
c-C$_{3}$H$_{2}$	&	Cyclopropenylidene	&	38508	&	CDMS	&	2.00$\times$10$^{14}$	&	100	\\
CH$_{3}$CCH	&	Propyne	&	40502	&	CDMS	&	6.80$\times$10$^{15}$	&	100	\\
CH$_{3}$CN	&	Acetonitrile	&	41505	&	CDMS	&	4.00$\times$10$^{16}$	&	120	\\
CH$_{3}$CN $\nu_{8}$=1	&	Acetonitrile	&	41509	&	CDMS	&	4.00$\times$10$^{16}$	&	120	\\
CH$_{3}$NC	&	Methyl isocyanide	&	41514	&	CDMS	&	2.00$\times$10$^{14}$	&	150	\\
H$_{2}$CCO	&	Ketene	&	42501	&	CDMS	&	4.80$\times$10$^{16}$	&	125	\\
HNCNH	&	Carbodiimide	&	42506	&	CDMS	&	2.40$\times$10$^{16}$	&	300	\\
$^{13}$CH$_{3}$CN	&	Acetonitrile	&	42508	&	CDMS	&	3.30$\times$10$^{14}$	&	130	\\
CH$_{3}^{13}$CN	&	Acetonitrile	&	42509	&	CDMS	&	3.00$\times$10$^{14}$	&	130	\\
CH$_{3}$C$^{15}$N	&	Acetonitrile	&	42510	&	CDMS	&	8.70$\times$10$^{13}$	&	130	\\
CH$_{2}$DCN	&	Acetonitrile	&	42511	&	CDMS	&	5.60$\times$10$^{14}$	&	130	\\
H$_{2}$C$^{13}$CO	&	Ketene	&	43505	&	CDMS	&	7.10$\times$10$^{14}$	&	125	\\
H$_{2}^{13}$CCO	&	Ketene	&	43506	&	CDMS	&	7.10$\times$10$^{14}$	&	125	\\
HDC$_{2}$O	&	Ketene	&	43507	&	CDMS	&	2.00$\times$10$^{15}$	&	125	\\
HNCO	&	Isocyanic acid	&	43511	&	CDMS	&	3.70$\times$10$^{16}$	&	300	\\
CHD$_{2}$CN	&	Acetonitrile	&	43514	&	CDMS	&	1.20$\times$10$^{14}$	&	130	\\
H$_{2}$N$^{13}$CN	&	Cyanamide	&	43515	&	CDMS	&	3.00$\times$10$^{13}$	&	300	\\
N$_{2}$O	&	Nitrous oxide	&	44004	&	JPL	&	5.00$\times$10$^{16}$	&	100	\\
DNCO	&	Isocyanic acid	&	44006	&	JPL	&	3.00$\times$10$^{14}$	&	300	\\
HN$^{13}$CO	&	Isocyanic acid	&	44008	&	JPL	&	4.00$\times$10$^{14}$	&	300	\\
CS	&	Carbon monosulfide	&	44501	&	CDMS	&	1.00$\times$10$^{16}$	&	125	\\
c-C$_{2}$H$_{4}$O	&	Ethylene oxide	&	44504	&	CDMS	&	4.10$\times$10$^{15}$	&	125	\\
SiO	&	Silicon monoxide	&	44505	&	CDMS	&	7.00$\times$10$^{13}$	&	300	\\
s-H$_{2}$CCHOH	&	Vinylalcohol	&	44506	&	CDMS	&	1.20$\times$10$^{17}$	&	125	\\
a-H$_{2}$CCHOH	&	Vinylalcohol	&	44507	&	CDMS	&	1.20$\times$10$^{17}$	&	125	\\
C$^{33}$S	&	Carbon monosulfide	&	45502	&	CDMS	&	1.00$\times$10$^{14}$	&	125	\\
NH$_{2}$CHO	&	Formamide	&	45512	&	CDMS	&	1.00$\times$10$^{16}$	&	300	\\
CH$_{3}$CDO	&	Acetaldehyde	&	45524	&	CDMS	&	7.40$\times$10$^{15}$	&	125	\\
CH$_{2}$DCHO	&	Acetaldehyde	&	45525	&	CDMS	&	6.20$\times$10$^{15}$	&	125	\\
C$^{34}$S	&	Carbon monosulfide	&	46501	&	CDMS	&	3.00$\times$10$^{14}$	&	125	\\
t-HCOOH	&	Formic acid	&	46506	&	CDMS	&	5.08$\times$10$^{16}$	&	300	\\
H$_{2}$CS	&	Thioformaldehyde	&	46509	&	CDMS	&	1.50$\times$10$^{15}$	&	125	\\
NH$_{2}^{13}$CHO	&	Formamide	&	46512	&	CDMS	&	1.00$\times$10$^{14}$	&	300	\\
CH$_{3}$OCH$_{3}$	&	Dimethyl ether	&	46514	&	CDMS	&	3.00$\times$10$^{17}$	&	125	\\
NH$_{2}$CDO	&	Formamide	&	46520	&	CDMS	&	1.40$\times$10$^{14}$	&	300	\\
cis-NHDCHO	&	Formamide	&	46521	&	CDMS	&	1.40$\times$10$^{14}$	&	300	\\
trans-NHDCHO	&	Formamide	&	46522	&	CDMS	&	1.20$\times$10$^{14}$	&	300	\\
C$_{2}$H$_{5}$OH	&	Ethanol	&	46524	&	CDMS	&	2.30$\times$10$^{17}$	&	300	\\
HONO	&	Nitrous acid	&	47007	&	JPL	&	9.00$\times$10$^{14}$	&	100	\\
t-H$^{13}$COOH	&	Formic acid	&	47503	&	CDMS	&	8.30$\times$10$^{14}$	&	300	\\
HDCS	&	Thioformaldehyde	&	47504	&	CDMS	&	1.50$\times$10$^{14}$	&	125	\\
CH$_{3}^{13}$CH$_{2}$OH	&	Ethanol	&	47511	&	CDMS	&	4.60$\times$10$^{14}$	&	300	\\
$^{13}$CH$_{3}$CH$_{2}$OH	&	Ethanol	&	47512	&	CDMS	&	4.60$\times$10$^{14}$	&	300	\\
CH$_{3}$CH$_{2}$OD	&	Ethanol	&	47515	&	CDMS	&	5.75$\times$10$^{14}$	&	300	\\
CH$_{3}$CHDOH	&	Ethanol	&	47516	&	CDMS	&	1.15$\times$10$^{15}$	&	300	\\
a-CH$_{2}$DCH$_{2}$OH	&	Ethanol	&	47517	&	CDMS	&	1.34$\times$10$^{15}$	&	300	\\
s-CH$_{2}$DCH$_{2}$OH	&	Ethanol	&	47518	&	CDMS	&	6.51$\times$10$^{14}$	&	300	\\
SO	&	Sulfur monoxide	&	48501	&	CDMS	&	5.00$\times$10$^{14}$	&	125	\\
C$^{36}$S	&	Carbon monosulfide	&	48503	&	CDMS	&	2.00$\times$10$^{13}$	&	125	\\
CH$_{3}$SH	&	Methyl mercaptan 	&	48510	&	CDMS	&	5.50$\times$10$^{15}$	&	125	\\
CH$_{3}^{35}$Cl	&	Chloromethane	&	50007	&	JPL	&	3.10$\times$10$^{13}$	&	125	\\
HC$_{3}$N	&	Cyanoacetylene	&	51501	&	CDMS	&	1.40$\times$10$^{14}$	&	100	\\
CH$_{3}^{37}$Cl	&	Chloromethane	&	52009	&	JPL	&	2.20$\times$10$^{14}$	&	125	\\
C$_{2}$H$_{3}$CN	&	Vinyl cyanide	&	53515	&	CDMS	&	4.80$\times$10$^{14}$	&	110	\\
C$_{2}$H$_{5}$CN	&	Ethyl cyanide	&	55502	&	CDMS	&	1.50$\times$10$^{15}$	&	160	\\
CH$_{3}$NCO	&	Methyl isocyanate	&	57505	&	CDMS	&	3.00$\times$10$^{15}$	&	300	\\
CH$_{3}$C(O)CH$_{3}$	&	Acetone	&	58003	&	JPL	&	3.40$\times$10$^{16}$	&	125	\\
CH$_{3}$CH$_{2}$CHO	&	Propanal	&	58505	&	CDMS	&	1.48$\times$10$^{15}$	&	125	\\
CH$_{3}$OCHO	&	Methylformate	&	60003	&	JPL	&	2.60$\times$10$^{17}$	&	300	\\
HOCH$_{2}$CHO	&	Glycolaldehyde	&	60501	&	CDMS	&	3.40$\times$10$^{16}$	&	300	\\
OCS	&	Carbonyl sulfide	&	60503	&	CDMS	&	2.00$\times$10$^{16}$	&	125	\\
OCS $\nu_{2}$=1	&	Carbonyl sulfide	&	60504	&	CDMS	&	2.00$\times$10$^{17}$	&	125	\\
CH$_{3}$COOH	&	Acetic acid	&	60523	&	CDMS	&	3.00$\times$10$^{15}$	&	300	\\
O$^{13}$CS	&	Carbonyl sulfide	&	61502	&	CDMS	&	5.00$\times$10$^{15}$	&	125	\\
OC$^{33}$S	&	Carbonyl sulfide	&	61503	&	CDMS	&	3.00$\times$10$^{15}$	&	100	\\
HOCH$_{2}^{13}$CHO	&	Glycolaldehyde	&	61513	&	CDMS	&	4.46$\times$10$^{14}$	&	300	\\
HO$^{13}$CH$_{2}$CHO	&	Glycolaldehyde	&	61514	&	CDMS	&	4.46$\times$10$^{14}$	&	300	\\
CH$_{3}$O$^{13}$CHO	&	Glycolaldehyde	&	61515	&	CDMS	&	6.30$\times$10$^{15}$	&	300	\\
DOCH$_{2}$CHO	&	Glycolaldehyde	&	61516	&	CDMS	&	4.86$\times$10$^{14}$	&	300	\\
HOCHDCHO	&	Glycolaldehyde	&	61517	&	CDMS	&	1.27$\times$10$^{15}$	&	300	\\
HOCH$_{2}$CDO	&	Glycolaldehyde	&	61518	&	CDMS	&	6.25$\times$10$^{14}$	&	300	\\
a-(CH$_{2}$OH)$_{2}$	&	Ethylene glycol	&	62503	&	CDMS	&	1.37$\times$10$^{16}$	&	300	\\
s-(CH$_{2}$OH)$_{2}$	&	Ethylene glycol	&	62504	&	CDMS	&	3.62$\times$10$^{16}$	&	300	\\
OC$^{34}$S	&	Carbonyl sulfide	&	62505	&	CDMS	&	1.50$\times$10$^{16}$	&	125	\\
$^{18}$OCS	&	Carbonyl sulfide	&	62506	&	CDMS	&	7.00$\times$10$^{14}$	&	125	\\
SO$_{2}$	&	Sulfur dioxide	&	64502	&	CDMS	&	1.50$\times$10$^{15}$	&	125	\\
$^{34}$O$_{2}$	&	Sulfur dioxide	&	66501	&	CDMS	&	4.00$\times$10$^{14}$	&	125	\\
\end{longtable}
\label{tab:params_16293_fit}
}

\begin{figure*}
\includegraphics[width=\hsize]{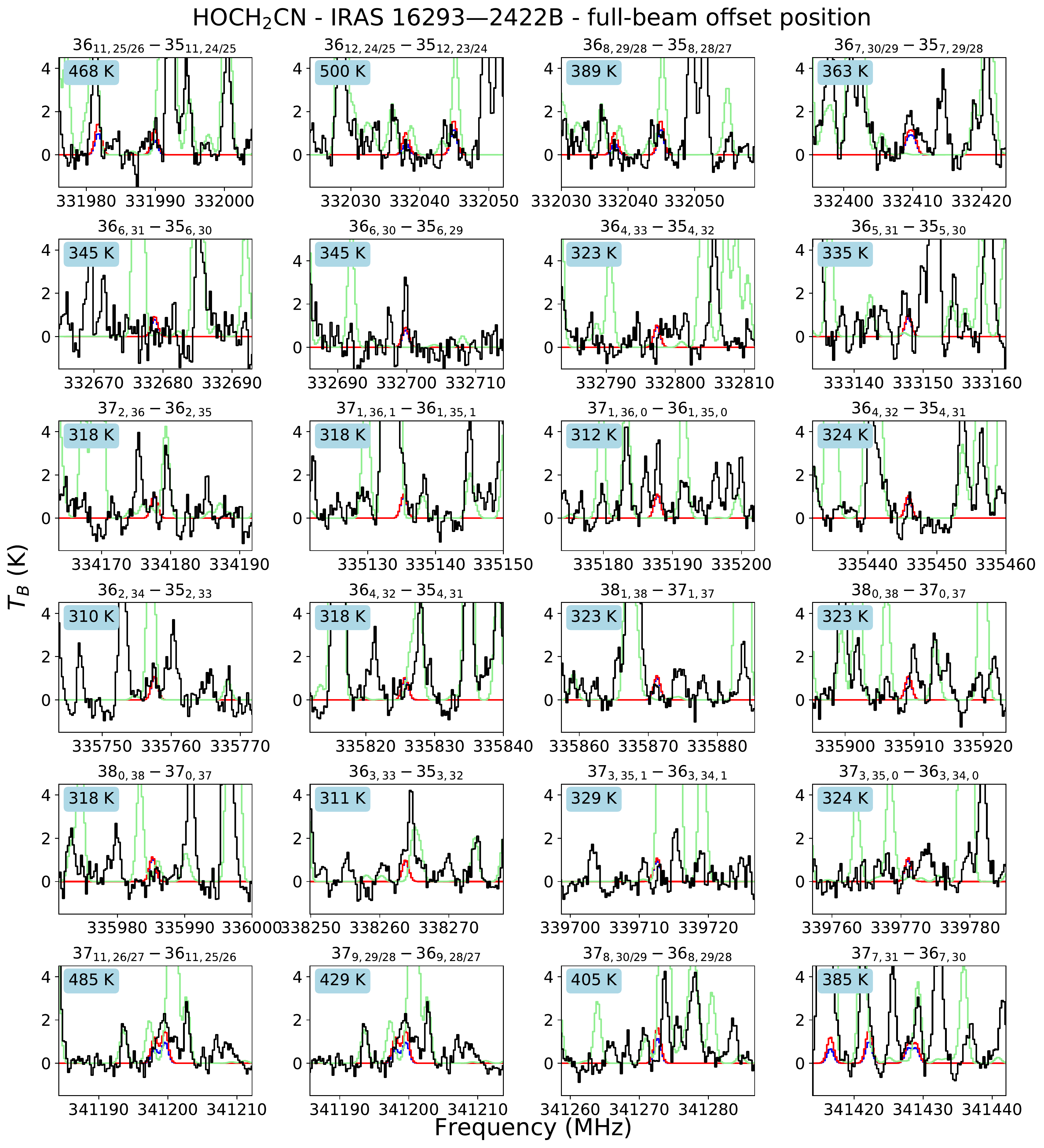}
\caption{Spectral lines of HOCH$_{2}$CN in the PILS spectrum towards IRAS~16293B at the full-beam offset position. The observed spectrum is plotted in black, with synthetic spectra overplotted ($N_{\rm T}$ = [1.0$\times$10$^{15}$] cm$^{-2}$, $T_{\rm ex}$ = 150, blue, and 300~K, red). The synthetic spectrum of the entire molecular inventory determined with PILS data towards this position is plotted in green. All covered transitions with $A_{\rm ij}$ $\geq$ 1.0$\times$10$^{-3}$ s$^{-1}$ that are not blended are shown. The transition is indicated at the top of each panel and the upper state energy is given in the top left of each panel. HOCH$_{2}$CN is not detected in the full-beam offset position spectrum towards IRAS~16293B.}
\label{fig:PILS_1.0_HOCH2CN-1}
\end{figure*}

\begin{figure*}
\includegraphics[width=\hsize]{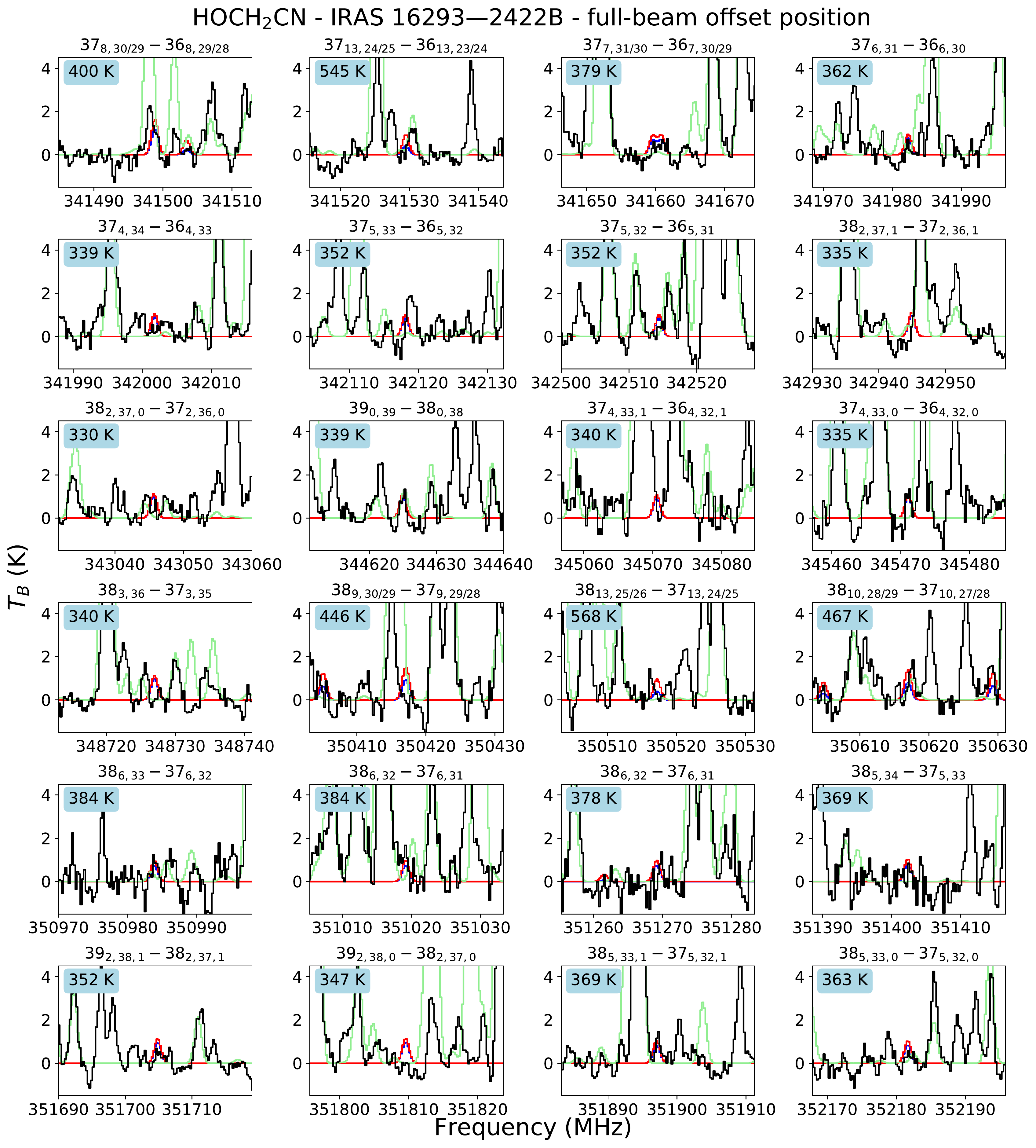}
\caption{Same as Fig. \ref{fig:PILS_1.0_HOCH2CN-1}}
\label{fig:PILS_1.0_HOCH2CN-2}
\end{figure*}

\begin{figure*}
\includegraphics[width=\hsize]{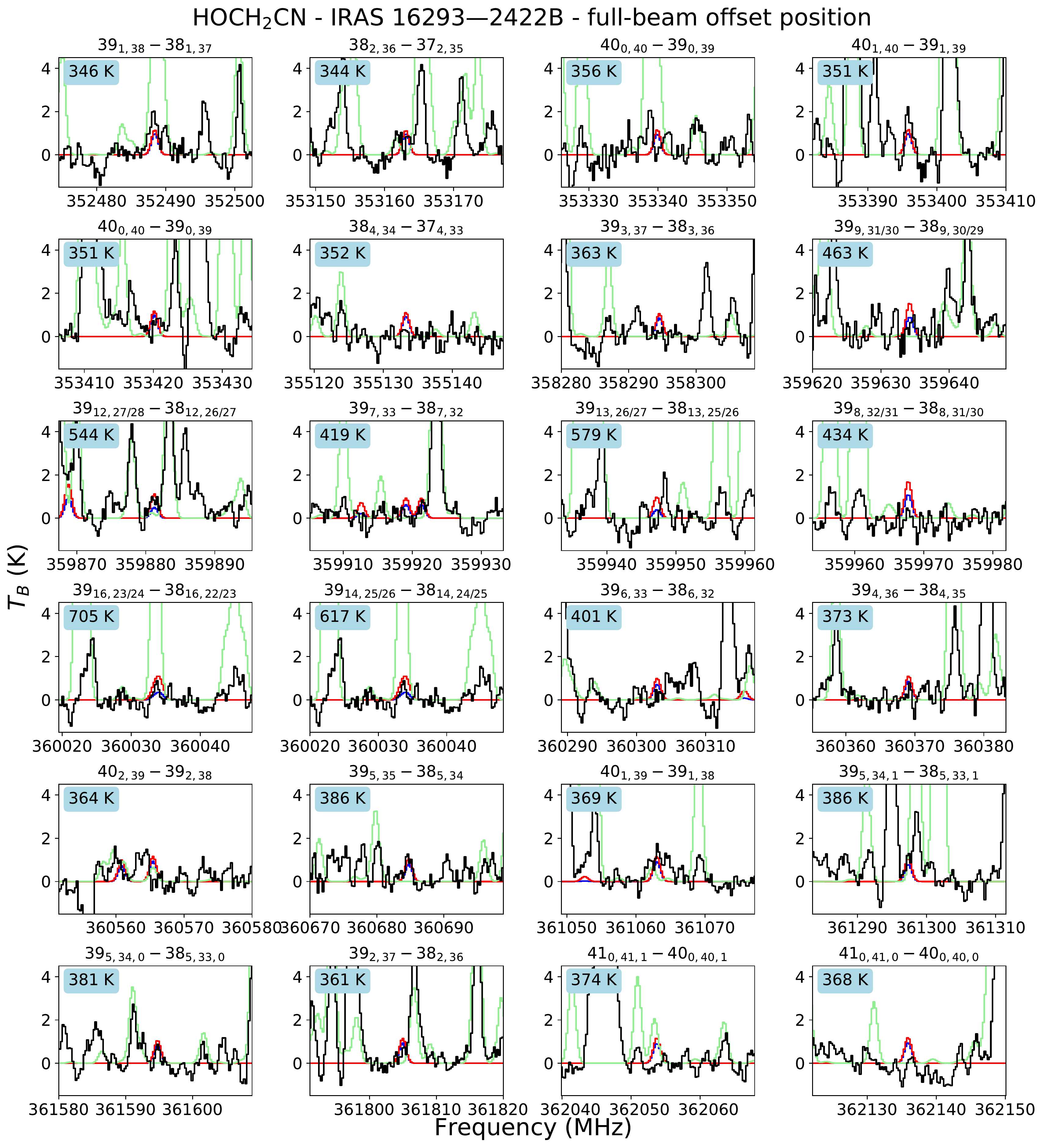}
\caption{Same as Fig. \ref{fig:PILS_1.0_HOCH2CN-1}}
\label{fig:PILS_1.0_HOCH2CN-3}
\end{figure*}

\begin{figure*}
\includegraphics[width=\hsize]{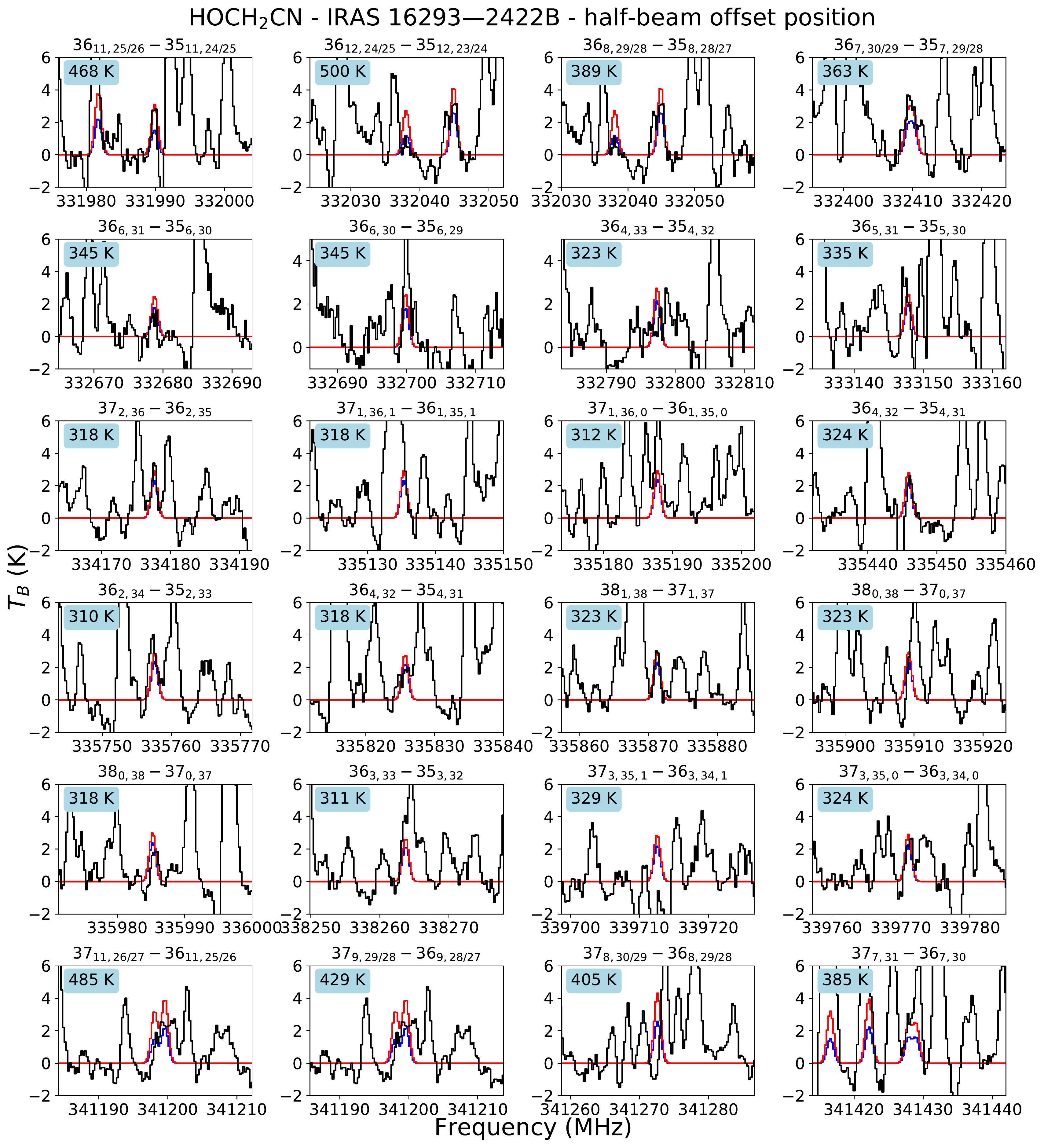}
\caption{Spectral lines of HOCH$_{2}$CN in the PILS spectrum towards IRAS~16293B at the half-beam offset position. The observed spectrum is plotted in black, with synthetic spectra overplotted ($N_{\rm T}$ = [3.0$\times$10$^{15}$] cm$^{-2}$, $T_{\rm ex}$ = 150, blue, and 300~K, red). All covered transitions with $A_{\rm ij}$ $\geq$ 1.0$\times$10$^{-3}$ s$^{-1}$ that are not blended are shown. The transition is indicated at the top of each panel and the upper state energy is given in the top left of each panel. HOCH$_{2}$CN is detected in the half-beam offset position spectrum towards IRAS~16293B.}
\label{fig:PILS_0.5_HOCH2CN-1}
\end{figure*}

\begin{figure*}
\includegraphics[width=\hsize]{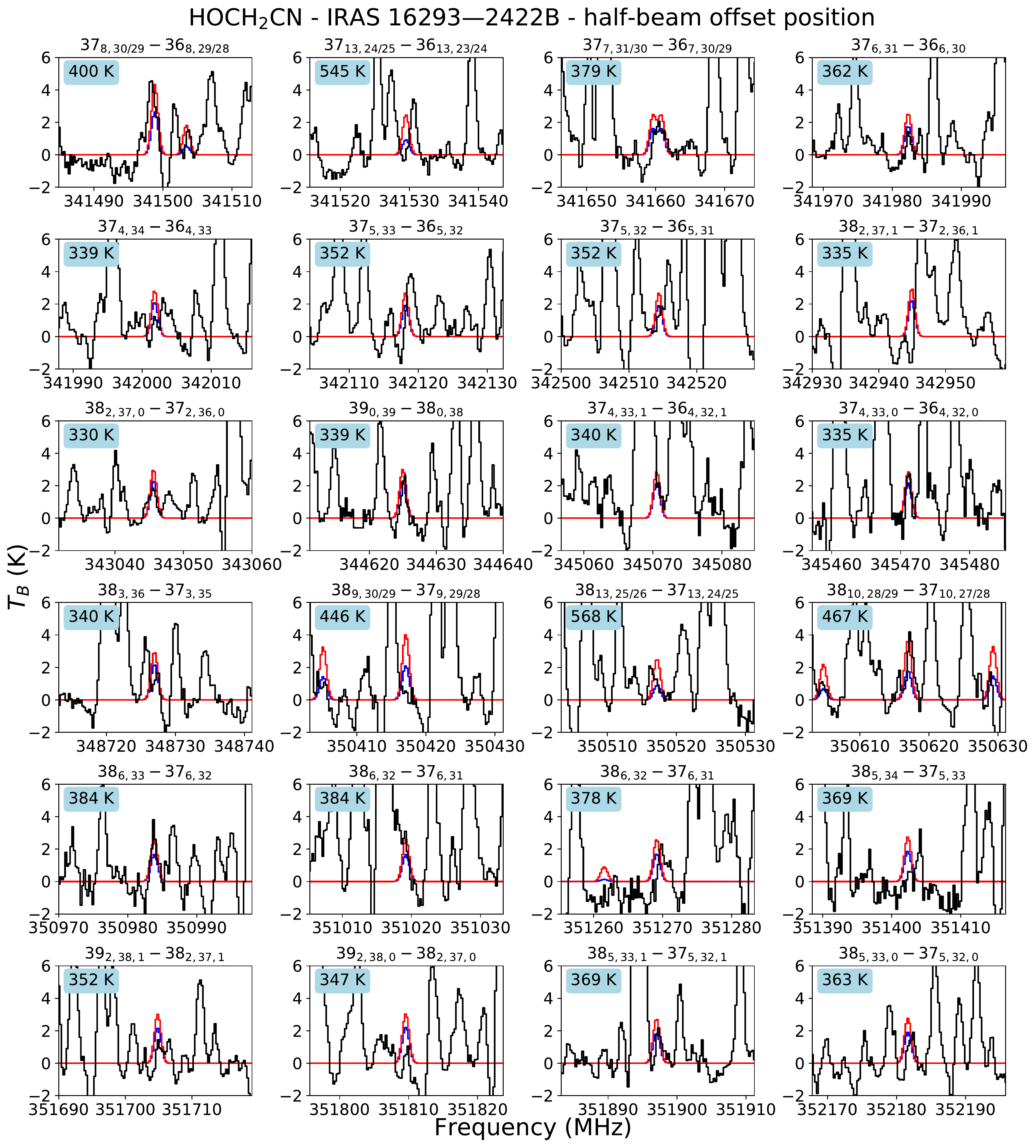}
\caption{Same as Fig. \ref{fig:PILS_0.5_HOCH2CN-1}}
\label{fig:PILS_0.5_HOCH2CN-2}
\end{figure*}

\begin{figure*}
\includegraphics[width=\hsize]{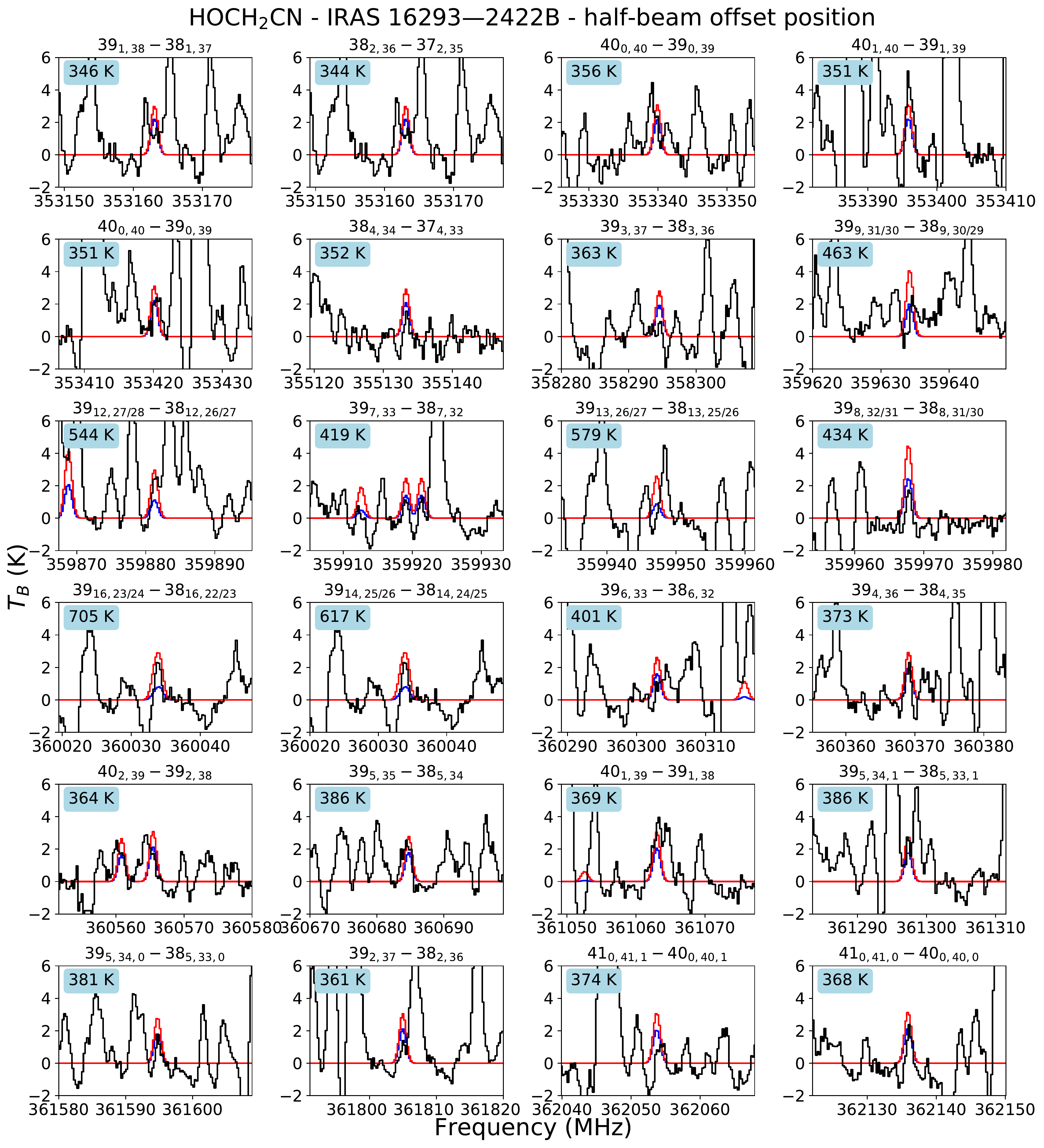}
\caption{Same as Fig. \ref{fig:PILS_0.5_HOCH2CN-1}}
\label{fig:PILS_0.5_HOCH2CN-3}
\end{figure*}

\longtab[1]{
\begin{longtable}{ccccc}
\caption{HOCH$_{2}$CN lines in the PILS data towards IRAS~16293B}\\
\hline
\hline
Transition & Frequency & $E_{\rm up}$ & $A_{\rm ij}$ & Blending species  \\  
$J,K_{\rm a},K_{\rm c},F$ & (MHz) & (K) & s$^{-1}$ & \\ 
\hline
\endfirsthead
\caption{Continued.} \\
\hline
Transition & Frequency & $E_{\rm up}$ & $A_{\rm ij}$ & Blending species  \\  
$J,K_{\rm a},K_{\rm c},F$ & (MHz) & (K) & s$^{-1}$ & \\ 
\hline
\endhead
\hline
\endfoot
\hline
\endlastfoot 
36 11 25 1 - 35 11 24 1 	&	331 990.003	&	468	&	1.03$\times$10$^{-3}$	& g-(CH$_{2}$OH)$_{2}$ \\ 
36 11 26 1 - 35 11 25 1 	&	331 990.003	&	468	&	1.03$\times$10$^{-3}$	& g-(CH$_{2}$OH)$_{2}$ \\
36 12 24 1 - 35 12 23 1 	&	332 038.123	&	500	&	1.01$\times$10$^{-3}$	& CH$_{3}$OCHO \\
36 12 25 1 - 35 12 24 1 	&	332 038.123	&	500	&	1.01$\times$10$^{-3}$	& CH$_{3}$OCHO \\
36 8 29 1 - 35 8 28 1 	&	332 045.021	&	389	&	1.08$\times$10$^{-3}$	& g-(CH$_{2}$OH)$_{2}$ \\
36 8 28 1 - 35 8 27 1 	&	332 045.044	&	389	&	1.08$\times$10$^{-3}$	& g-(CH$_{2}$OH)$_{2}$ \\
36 7 30 0 - 35 7 29 0 	&	332 409.418	&	363	&	1.09$\times$10$^{-3}$	& -- \\
36 7 29 0 - 35 7 28 0 	&	332 410.281	&	363	&	1.09$\times$10$^{-3}$	& -- \\
36 6 31 0 - 35 6 30 0 	&	332 678.904	&	345	&	1.10$\times$10$^{-3}$	& -- \\
36 6 30 0 - 35 6 29 0 	&	332 699.957	&	345	&	1.10$\times$10$^{-3}$	& -- \\
36 4 33 1 - 35 4 32 1 	&	332 797.489	&	323	&	1.14$\times$10$^{-3}$	& -- \\
36 5 31 1 - 35 5 30 1 	&	333 147.976	&	335	&	1.13$\times$10$^{-3}$	& -- \\
37 2 36 1 - 36 2 35 1 	&	334 177.762	&	318	&	1.15$\times$10$^{-3}$	& CH$_{3}$OCHO, CH$_{3}$O$^{13}$CHO, C$_{2}$H$_{5}$CN, HO$^{13}$CH$_{2}$CHO  \\
37 1 36 1 - 36 1 35 1 	&	335 135.448	&	318	&	1.16$\times$10$^{-3}$	& CH$_{3}$OH \\
37 1 36 0 - 36 1 35 0 	&	335 187.943	&	313	&	1.16$\times$10$^{-3}$	& -- \\
36 4 32 1 - 35 4 31 1 	&	335 446.009	&	324	&	1.17$\times$10$^{-3}$	& -- \\
36 2 34 1 - 35 2 33 1 	&	335 757.704	&	310	&	1.15$\times$10$^{-3}$	& g-(CH$_{2}$OH)$_{2}$\\
36 4 32 0 - 35 4 31 0 	&	335 825.850	&	318	&	1.15$\times$10$^{-3}$	& g-(CH$_{2}$OH)$_{2}$, CH$_{3}$C(O)CH$_{3}$, CH$_{3}$OCHO \\
38 1 38 1 - 37 1 37 1 	&	335 871.388	&	323	&	1.17$\times$10$^{-3}$	& -- \\
38 0 38 1 - 37 0 37 1 	&	335 909.319	&	323	&	1.17$\times$10$^{-3}$	& -- \\
38 0 38 0 - 37 0 37 0 	&	335 985.342	&	318	&	1.17$\times$10$^{-3}$	& CH$_{3}$CH$_{2}$OH, g-(CH$_{2}$OH)$_{2}$\\
36 3 33 0 - 35 3 32 0 	&	338 263.867	&	311	&	1.07$\times$10$^{-3}$	& CH$_{3}$OCHO \\
37 3 35 1 - 36 3 34 1 	&	339 712.712	&	329	&	1.21$\times$10$^{-3}$	& NH$_{2}$CHO \\
37 3 35 0 - 36 3 34 0 	&	339 771.178	&	324	&	1.20$\times$10$^{-3}$	& -- \\
37 11 26 1 - 36 11 25 1 	&	341 198.194	&	484	&	1.13$\times$10$^{-3}$	& -- \\
37 11 27 1 - 36 11 26 1 	&	341 198.194	&	484	&	1.13$\times$10$^{-3}$	& -- \\
37 9 29 1 - 36 9 28 1 	&	341 199.695	&	429	&	1.16$\times$10$^{-3}$	& g-(CH$_{2}$OH)$_{2}$ \\
37 9 28 1 - 36 9 27 1 	&	341 199.696	&	429	&	1.16$\times$10$^{-3}$	& g-(CH$_{2}$OH)$_{2}$ \\
37 8 30 1 - 36 8 29 1 	&	341 272.746	&	405	&	1.18$\times$10$^{-3}$	& HOCH$_{2}$CHO, CH$_{3}$CDO\\
37 8 29 1 - 36 8 28 1 	&	341 272.781	&	405	&	1.18$\times$10$^{-3}$	& HOCH$_{2}$CHO, CH$_{3}$CDO\\
37 7 31 1 - 36 7 30 1 	&	341 427.962	&	385	&	1.20$\times$10$^{-3}$	& CH$_{3}$CHO \\
37 8 30 0 - 36 8 29 0 	&	341 498.917	&	400	&	1.17$\times$10$^{-3}$	& CH$_{3}$OCHO \\
37 8 29 0 - 36 8 28 0 	&	341 498.954	&	400	&	1.17$\times$10$^{-3}$	& CH$_{3}$OCHO \\
37 13 24 0 - 36 13 23 0 	&	341 529.626	&	545	&	1.08$\times$10$^{-3}$	& a-(CH$_{2}$OH)$_{2}$ \\
37 13 25 0 - 36 13 24 0 	&	341 529.626	&	545	&	1.08$\times$10$^{-3}$	& a-(CH$_{2}$OH)$_{2}$ \\
37 7 31 0 - 36 7 30 0 	&	341 659.715	&	379	&	1.18$\times$10$^{-3}$	& -- \\
37 7 30 0 - 36 7 29 0 	&	341 660.941	&	379	&	1.18$\times$10$^{-3}$	& -- \\
37 6 31 0 - 36 6 30 0 	&	341 982.421	&	362	&	1.20$\times$10$^{-3}$	& CH$_{3}$O$^{13}$CHO \\
37 4 34 1 - 36 4 33 1 	&	342 001.917	&	339	&	1.23$\times$10$^{-3}$	& -- \\
37 5 33 1 - 36 5 32 1 	&	342 118.267	&	352	&	1.23$\times$10$^{-3}$	& HOCH$_{2}$CHO \\
37 5 32 1 - 36 5 31 1 	&	342 514.548	&	352	&	1.23$\times$10$^{-3}$	& CH$_{3}$CDO \\
38 2 37 1 - 37 2 36 1 	&	342 945.040	&	335	&	1.25$\times$10$^{-3}$	& H$_{2}$CS, CH$_{3}$O$^{13}$CHO \\
38 2 37 0 - 37 2 36 0 	&	343 045.762	&	330	&	1.24$\times$10$^{-3}$	& CH$_{3}$CDO \\
39 0 39 1 - 38 0 38 1 	&	344 625.126	&	339	&	1.27$\times$10$^{-3}$	& HONO, CH$_{2}$DOH, CH$_{3}$CHO \\
37 4 33 1 - 36 4 32 1 	&	345 070.674	&	340	&	1.27$\times$10$^{-3}$	& CH$_{3}$OCHO \\
37 4 33 0 - 36 4 32 0 	&	345 471.229	&	335	&	1.25$\times$10$^{-3}$	& CH$_{3}$OCHO \\
38 3 36 0 - 37 3 35 0 	&	348 727.094	&	340	&	1.30$\times$10$^{-3}$	& -- \\
38 9 30 1 - 37 9 29 1 	&	350 417.202	&	446	&	1.27$\times$10$^{-3}$	& a-(CH$_{2}$OH)$_{2}$ \\
38 9 29 1 - 37 9 28 1 	&	350 417.203	&	446	&	1.27$\times$10$^{-3}$	& a-(CH$_{2}$OH)$_{2}$ \\
38 13 25 1 - 37 13 24 1 	&	350 517.318	&	568	&	1.19$\times$10$^{-3}$	& -- \\
38 13 26 1 - 37 13 25 1 	&	350 517.318	&	568	&	1.19$\times$10$^{-3}$	& -- \\
38 10 28 0 - 37 10 27 0 	&	350 617.089	&	467	&	1.24$\times$10$^{-3}$	& CH$_{3}$O$^{13}$CHO \\
38 10 29 0 - 37 10 28 0 	&	350 617.089	&	467	&	1.24$\times$10$^{-3}$	& CH$_{3}$O$^{13}$CHO \\
38 6 33 1 - 37 6 32 1 	&	350 983.970	&	384	&	1.32$\times$10$^{-3}$	& CH$_{3}$OCHO, CH$_{3}$CDO \\
38 6 32 1 - 37 6 31 1 	&	351 019.318	&	384	&	1.32$\times$10$^{-3}$	& H$_{2}$C$^{13}$CO, CH$_{3}$OCHO \\
38 6 32 0 - 37 6 31 0 	&	351 269.282	&	378	&	1.30$\times$10$^{-3}$	& -- \\
38 5 34 1 - 37 5 33 1 	&	351 402.459	&	369	&	1.33$\times$10$^{-3}$	& -- \\
39 2 38 1 - 38 2 37 1 	&	351 704.972	&	352	&	1.35$\times$10$^{-3}$	& -- \\
39 2 38 0 - 38 2 37 0 	&	351 809.704	&	346	&	1.34$\times$10$^{-3}$	& -- \\
38 5 33 1 - 37 5 32 1 	&	351 897.151	&	369	&	1.34$\times$10$^{-3}$	& -- \\
38 5 33 0 - 37 5 32 0 	&	352 181.763	&	363	&	1.32$\times$10$^{-3}$	& CH$_{3}$O$^{13}$CHO \\
39 1 38 0 - 38 1 37 0 	&	352 488.531	&	346	&	1.35$\times$10$^{-3}$	& CH$_{3}^{18}$OH \\
38 2 36 1 - 37 2 35 1 	&	353 163.191	&	344	&	1.35$\times$10$^{-3}$	& -- \\
40 0 40 1 - 39 0 39 1 	&	353 340.026	&	356	&	1.37$\times$10$^{-3}$	& CH$_{3}$OCHO \\
40 1 40 0 - 39 1 39 0 	&	353 395.992	&	351	&	1.37$\times$10$^{-3}$	& -- \\
40 0 40 0 - 39 0 39 0 	&	353 420.250	&	351	&	1.37$\times$10$^{-3}$	& -- \\
38 4 34 0 - 37 4 33 0 	&	355 133.419	&	352	&	1.36$\times$10$^{-3}$	& -- \\
39 3 37 1 - 38 3 36 1 	&	358 294.680	&	363	&	1.33$\times$10$^{-3}$	& -- \\
39 9 31 1 - 38 9 30 1 	&	359 634.337	&	463	&	1.37$\times$10$^{-3}$	& -- \\
39 9 30 1 - 38 9 29 1 	&	359 634.339	&	463	&	1.37$\times$10$^{-3}$	& -- \\
39 12 27 0 - 38 12 26 0 	&	359 881.373	&	545	&	1.30$\times$10$^{-3}$	& -- \\
39 12 28 0 - 38 12 27 0 	&	359 881.373	&	545	&	1.30$\times$10$^{-3}$	& -- \\
39 7 33 1 - 38 7 32 1 	&	359 919.218	&	419	&	1.41$\times$10$^{-3}$	& -- \\
39 13 26 0 - 38 13 25 0 	&	359 947.354	&	579	&	1.28$\times$10$^{-3}$	& -- \\
39 13 27 0 - 38 13 26 0 	&	359 947.354	&	579	&	1.28$\times$10$^{-3}$	& -- \\
39 8 32 0 - 38 8 31 0 	&	359 967.849	&	434	&	1.38$\times$10$^{-3}$	& -- \\
39 8 31 0 - 38 8 30 0 	&	359 967.931	&	434	&	1.38$\times$10$^{-3}$	& -- \\
39 16 23 1 - 38 16 22 1 	&	360 033.494	&	705	&	1.21$\times$10$^{-3}$	& t-HCOOH \\
39 16 24 1 - 38 16 23 1 	&	360 033.494	&	705	&	1.21$\times$10$^{-3}$	& t-HCOOH \\
39 14 25 0 - 38 14 24 0 	&	360 034.284	&	617	&	1.25$\times$10$^{-3}$	& t-HCOOH \\
39 14 26 0 - 38 14 25 0 	&	360 034.284	&	617	&	1.25$\times$10$^{-3}$	& t-HCOOH \\
39 6 33 1 - 38 6 32 1 	&	360 303.101	&	401	&	1.43$\times$10$^{-3}$	& -- \\
39 4 36 1 - 38 4 35 1 	&	360 369.224	&	373	&	1.45$\times$10$^{-3}$	& -- \\
40 2 39 0 - 39 2 38 0 	&	360 565.572	&	364	&	1.44$\times$10$^{-3}$	& -- \\
39 5 35 1 - 38 5 34 1 	&	360 684.867	&	386	&	1.44$\times$10$^{-3}$	& -- \\
40 1 39 1 - 39 1 38 1 	&	361 063.161	&	369	&	1.46$\times$10$^{-3}$	& -- \\
39 5 34 1 - 38 5 33 1 	&	361 297.474	&	386	&	1.45$\times$10$^{-3}$	& CH$_{3}$CH$_{2}$OH \\
39 5 34 0 - 38 5 33 0 	&	361 594.894	&	381	&	1.43$\times$10$^{-3}$	& -- \\
39 2 37 1 - 38 2 36 1 	&	361 805.086	&	361	&	1.46$\times$10$^{-3}$	& CH$_{3}$OCHO \\
41 0 41 1 - 40 0 40 1 	&	362 053.851	&	374	&	1.47$\times$10$^{-3}$	& CH$_{3}$C(O)CH$_{3}$ \\
41 0 41 0 - 40 0 40 0 	&	362 136.053	&	368	&	1.47$\times$10$^{-3}$	& -- \\               
\end{longtable}
\tablefoot{List of identified and unidentified HOCH$_{2}$CN transitions with A$_{\rm ij}$ $\geq$ 1.0$\times$10$^{-3}$ s$^{-1}$ towards IRAS~16293B in the PILS data set.}
}

\section{Source parameters}

\begin{table*}[h]
\caption{Physical parameters of interstellar sources used for abundance comparison.}             
\label{tab:obs_params}      
\centering          
\begin{tabular}{c c c c c c c}     
\hline\hline                           
Source & Telescope & Distance & Luminosity & Beam size & physical size & Reference \\
& & (pc) & $L_{\odot}$ & ($\arcsec$ $\times$ $\arcsec$) & au & \\
\hline                    
IRAS~16293--2422B & ALMA & 141 & 3$^{a}$ & 0.5$\times$0.5 & 70 & \citet{jorgensen2016} \\
Serpens SMM1-a & ALMA & 436 & 100$^{b}$ & 1.3$\times$1.0 & 500 & this work \\
Orion KL & ALMA & 414 & 1$\times$10$^{5}$ & 1.8$\times$1.8 & 750 & \citet{cernicharo2016} \\
Sgr B2(N2) & ALMA & 8300 & 4.7$\times$10$^{6}$ & 1.6$\times$1.2 -- 2.9$\times$1.5 & 1.3$\times$10$^{4}$ & \citet{belloche2017} \\
G10.47+0.03 & ALMA & 8550 & 5$\times$10$^{5}$ & 2.0$\times$1.4 -- 2.4$\times$1.6 & 1.6$\times$10$^{4}$ & \citet{gorai2020} \\
Sgr B2(N) & IRAM 30m & 8300 & 4.7$\times$10$^{6}$ & 3.2$\times$2.8 -- 12.2$\times$4.4 & 3.0$\times$10$^{4}$ & \citet{cernicharo2016} \\
G+0.693 & IRAM 30m \& GBT & 8300$^{c}$ & -- & 9$\times$9 -- 55$\times$55 &  $\geq$7.5$\times$10$^{4}$ & \citet{zeng2019} \\
\hline              
\end{tabular}  
\tablefoot{$^{a}$Luminosity determined from a modeling investigation by \citet{jacobsen2018}. $^{b}$Luminosity for the entire SMM1 source. $^{c}$The distance to G+0.693 is assumed to be the same as Sgr B2.}  
\end{table*}

\section{Statistical distance}
\label{sec:stat_distance}

The statistical distance of molecular ratios between SMM1-a and IRAS~16293B is calculated according to the following equation:
\begin{equation}
   	S_{\rm X/Y} = \frac{\left(\frac{N_{\rm X}}{N_{\rm Y}}\right)_{\rm SMM1-a} - \left(\frac{N_{\rm X}}{N_{\rm Y}}\right)_{\rm IRAS~16293B}}{\sqrt{\sigma^2_{\rm SMM1-a}+\sigma^2_{\rm IRAS~16293B}}},
    \label{eq:stat_distance}
\end{equation}
where $N_{\rm x}$ and $N_{\rm y}$ are the column densities of two different molecules and $\sigma$ is the uncertainty on the column density ratio $\left(\frac{N_{\rm X}}{N_{\rm Y}}\right)$. The value of $S_{\rm X/Y}$ is given in $\sigma$ and indicates the significance of the difference, with greater values implying that there is a more significant difference between the two sources. In this equation, positive values indicate that $\left(\frac{N_{\rm X}}{N_{\rm Y}}\right)$ is more abundant in SMM1-a than in IRAS~16293B, while negative values indicate the opposite.  


\end{appendix}
\end{document}